\documentclass[preprint,aps,prd,nofootinbib,floatfix]{revtex4-2}

\usepackage[T1]{fontenc}
\usepackage[utf8]{inputenc}
\usepackage{graphicx}
\usepackage{amsmath,amssymb,bm}
\usepackage{booktabs}
\usepackage{array}
\usepackage{multirow}
\usepackage{placeins}
\usepackage{needspace}
\usepackage{hyperref}
\usepackage{xcolor}

\graphicspath{{figures/}}
\hypersetup{
  colorlinks=true,
  linkcolor=blue,
  citecolor=blue,
  urlcolor=blue
}
\newcommand{\Br}{\mathop{\mathrm{Br}}}
\newcommand{\GeV}{\mathrm{GeV}}
\newcommand{\Lag}{\mathcal{L}}

\newcommand{\hc}{\mathrm{h.c.}}
\newcommand{\diag}{\mathrm{diag}}

\begin{document}

\title{Top-quark flavor-changing neutral currents and charged-lepton flavor violation in a generational three-Higgs-doublet model}

\author{Hao-Ran Ma$^{1,2,3,\ast,\dagger}$}

\author{Ti-Bin Hou$^{1,2,3,\ast,\ddagger}$}

\author{Yu-Ju Peng$^{1,2,3}$}

\author{Jin-Lei Yang$^{1,2,3,\S}$}

\author{Tai-Fu Feng$^{1,2,3,4,\Vert}$}

\affiliation{$^1$Department of Physics, Hebei University, Baoding 071002, China}
\affiliation{$^2$Hebei Key Laboratory of High-precision Computation and Application of Quantum Field Theory, Baoding 071002, China}
\affiliation{$^3$Hebei Research Center of the Basic Discipline for Computational Physics, Baoding 071002, China}
\affiliation{$^4$Department of Physics, Chongqing University, Chongqing 401331, China}

\begin{abstract}
In this work, we systematically investigate top-quark flavor-changing neutral-current decays, charged-lepton flavor-violating processes, and coherent $\mu-e$ conversion in the CP-conserving generational three-Higgs-doublet model (G3HDM), including the relevant tree-level and one-loop contributions.  The 4839-point sample passes conservative sufficient vacuum-stability conditions, necessary perturbative-unitarity preselection cuts, Higgs-mass and signal-strength constraints, electroweak precision data, the UTfit 2025 neutral-meson-mixing ranges, $\overline B\to X_s\gamma$, and $B_s\to\mu^+\mu^-$.  In the scanned region, $\Br(t\to ch_1)\leq1.04\times10^{-5}$ and $\Br(h_1\to\mu\tau)\leq4.83\times10^{-4}$, while the maximal conversion rates are $\mathrm{CR}(\mu\mathrm{Al}\to e\mathrm{Al})\approx4.95\times10^{-15}$ and $\mathrm{CR}(\mu\mathrm{Au}\to e\mathrm{Au})\approx1.05\times10^{-14}$.  The dominant collider signals are mainly driven by tree-level flavor-changing neutral-Higgs couplings, which determine the relative strengths of these processes and induce the pronounced enhancement of $t\to ch_1$ and $h_1\to\mu\tau$. Meanwhile, the upper range of the predicted $\mu-e$ conversion rate in aluminum lies within the sensitivity reach of next-generation $\mu-e$ conversion experiments.
\end{abstract}

\maketitle
\begingroup
\renewcommand{\thefootnote}{\fnsymbol{footnote}}
\footnotetext[1]{These authors contributed equally to this work.}
\footnotetext[2]{Contact author: \href{mailto:haoranma2024@163.com}{haoranma2024@163.com}}
\footnotetext[3]{Contact author: \href{mailto:tibinhou2312@163.com}{tibinhou2312@163.com}}
\footnotetext[4]{Contact author: \href{mailto:jlyang@hbu.edu.cn}{jlyang@hbu.edu.cn}}
\renewcommand{\thefootnote}{\ensuremath{\Vert}}
\footnotetext{Contact author: \href{mailto:fengtf@hbu.edu.cn}{fengtf@hbu.edu.cn}}
\endgroup
\clearpage

\section{Introduction}

The discovery of a scalar boson with a mass close to $125~\GeV$ and the continuing agreement of its measured couplings with the Standard Model (SM) have established the single-doublet description as an excellent low-energy approximation~\cite{ATLAS:2022vkf,CMS:2022dwd,ParticleDataGroup:2024cfk}.  At the same time, neither the SM nor the measured Higgs couplings explain the large hierarchies among fermion masses and mixing angles.  Multi-Higgs-doublet theories offer a direct way to connect electroweak symmetry breaking to flavor, because different vacuum expectation values (vevs) and different Yukawa matrices may participate in generating the observed fermion spectrum~\cite{Branco:2011iw,Altmannshofer:2024jyv}.

In a generic multi-doublet theory, several Yukawa matrices contribute to one fermion mass matrix.  Their simultaneous diagonalization is not guaranteed, and neutral scalars then mediate flavor-changing neutral currents (FCNCs) at tree level.  The Glashow--Weinberg condition of natural flavor conservation avoids this problem by allowing fermions of a given electric charge to couple to only one scalar doublet~\cite{Glashow:1976nt}; alignment, minimal-flavor-violation, and controlled generic-Yukawa alternatives have also been developed~\cite{Altmannshofer:2012azMFV2HDM,Crivellin:2013wna,Penuelas:2017ikk}.  A general three-Higgs-doublet model (3HDM) need not satisfy this condition, but completely arbitrary full-rank Yukawa matrices usually produce excessive flavor violation unless their off-diagonal entries are small, the additional scalars are heavy, or cancellations are arranged.

Three-doublet scalar sectors exhibit a particularly rich symmetry, vacuum, alignment, and collider structure~\cite{Keus:2013hya,Keus:2014jha,Grzadkowski:2009bt,Ahriche:2015mea,Hartmann:2014ppa,Cordero:2017owj,Aranda:2019vda,Khater:2021wcx,Kuncinas:2022whn}.  Their basis geometry and symmetry classifications have been developed systematically~\cite{Ivanov:2010ww,Ivanov:2010wz,Ivanov:2012ry,Nishi:2007nh}, while recent studies illustrate CP, dark-sector, and phenomenological variants~\cite{Pilaftsis:2016erj,Das:2019yad,Boto:2021qgu,Das:2022gbm,Boto:2023nyi,Darvishi:2019dbh,Darvishi:2021txa}.  These works also make clear that theoretical consistency is more involved in a 3HDM than in a two-doublet model: boundedness, the vacuum structure, and the coupled-channel scattering spectrum must be treated separately.

A complementary line of work relates Higgs flavor violation directly to fermion-mass generation~\cite{Altmannshofer:2015esa,Altmannshofer:2016oaq,Blechman:2010cs,Das:1995df,Botella:2016krk,Ghosh:2015gpa,Egana-Ugrinovic:2019dqu}.  Flavorful and flavor-locked two-doublet constructions provide predictive precedents for such an organization~\cite{Altmannshofer:2016zrn,Altmannshofer:2017uvs,Altmannshofer:2018bch}, and rare top decays are especially clean probes of their neutral-scalar flavor structure~\cite{Altmannshofer:2019ogm}.  More recently, flavor-violating Higgs and top decays have been analyzed in the Type 1B flavorful two-Higgs-doublet model with a twist, including one-loop effects and low-energy flavor constraints~\cite{Hou:2026Type1B}.  For comparison, top-quark FCNC decays have also been studied in gauge-sector extensions such as the B--L supersymmetric model and a flavor-dependent $U(1)_X$ model, where the flavor-changing mechanisms differ from the rank-one multi-doublet structure considered here~\cite{Yang:2018fvw,Ge:2024fdx}.

The generational three-Higgs-doublet model (G3HDM) introduced in Ref.~\cite{AltmannshoferToner:2025} and subsequently applied to rare Higgs decays in Ref.~\cite{Ma:2026G3HDMbs} takes a more structured route.  Each scalar doublet couples through a rank-one Yukawa matrix and dominantly provides the mass of one fermion generation.  A hierarchy $v_1\ll v_2\ll v_3$ can therefore account for part of the fermion-mass hierarchy.  The rank-one Yukawa matrices are not, in general, aligned in flavor space.  Their misalignment generates the Cabibbo--Kobayashi--Maskawa (CKM) matrix and also induces tree-level neutral-scalar FCNCs.  The flavor violation is thus neither removed by natural flavor conservation nor described by three arbitrary full-rank matrices: it is organized by a small number of generational texture parameters and by the misalignment of the physical scalar with the electroweak vacuum direction.

In this work we study top-quark FCNC decays and charged-lepton flavor violation (LFV) in the CP-conserving G3HDM.  We focus on $t\to qh_1$ with $q=u,c$, $h_1\to uc$, radiative charged-lepton decays, three-body charged-lepton decays, and coherent $\mu$--$e$ conversion.  Unlike an ordinary 3HDM with independent full-rank FCNC Yukawa matrices, the G3HDM uses rank-one matrices, hierarchical vevs, and generation-tagged doublets to reduce flavor violation to a few texture factors multiplying common scalar-misalignment quantities.  The resulting cross-channel correlations characterize the generational construction.


The paper is organized as follows.  Section~II summarizes the scalar and Yukawa sectors of the G3HDM and emphasizes its distinction from a generic FCNC 3HDM.  Section~III defines the observables, scan, and applied constraints.  Section~IV presents the numerical results and identifies the common tree-level driver of $t\to ch_1$ and $h_1\to\mu\tau$.  We conclude in Sec.~V.

\section{The Generational Three-Higgs-Doublet Model}
\label{sec:G3HDM}

In this work, we consider a G3HDM, in which the SM Higgs sector is extended by two additional scalar $SU(2)_L$ doublets.  The model is motivated by the possibility that fermion mass hierarchies (or part thereof) originate not only from hierarchical Yukawa couplings, but also from a hierarchical pattern of electroweak symmetry breaking.  In particular, the three Higgs doublets may be arranged such that they provide masses predominantly to the first, second, and third generations of SM fermions, respectively, and a hierarchy among their vacuum expectation values can then partially address the SM flavor puzzle~\cite{AltmannshoferToner:2025,Ma:2026G3HDMbs}.

\subsection{The scalar sector of the G3HDM}
\label{subsec:G3HDM_scalar}

\paragraph{Field content and scalar potential.}
The scalar sector contains three Higgs doublets $\Phi_a$ ($a=1,2,3$), transforming under the SM gauge group $SU(3)_c\times SU(2)_L\times U(1)_Y$ as
\begin{equation}
 \Phi_a\sim(1,2,\tfrac12),\qquad a=1,2,3.
 \label{eq:Phi_rep}
\end{equation}
We assume that the renormalizable Higgs potential respects, to a good approximation, a softly broken $U(1)^3$ symmetry, with each $U(1)$ factor acting on a single Higgs doublet.  The resulting potential can be written as~\cite{AltmannshoferToner:2025,Ma:2026G3HDMbs,Faro:2019vcd,Boto:2022uwv}
\begin{align}
 V={}&\sum_{a=1}^3m_{aa}^2\Phi_a^\dagger\Phi_a
 -\left[m_{12}^2\Phi_1^\dagger\Phi_2
 +m_{13}^2\Phi_1^\dagger\Phi_3
 +m_{23}^2\Phi_2^\dagger\Phi_3+\hc\right]
 \nonumber\\
 &+\lambda_1(\Phi_1^\dagger\Phi_1)^2
 +\lambda_2(\Phi_2^\dagger\Phi_2)^2
 +\lambda_3(\Phi_3^\dagger\Phi_3)^2
 \nonumber\\
 &+\lambda_4(\Phi_1^\dagger\Phi_1)(\Phi_2^\dagger\Phi_2)
 +\lambda_5(\Phi_1^\dagger\Phi_1)(\Phi_3^\dagger\Phi_3)
 +\lambda_6(\Phi_2^\dagger\Phi_2)(\Phi_3^\dagger\Phi_3)
 \nonumber\\
 &+\lambda_7(\Phi_1^\dagger\Phi_2)(\Phi_2^\dagger\Phi_1)
 +\lambda_8(\Phi_1^\dagger\Phi_3)(\Phi_3^\dagger\Phi_1)
 +\lambda_9(\Phi_2^\dagger\Phi_3)(\Phi_3^\dagger\Phi_2).
 \label{eq:potential}
\end{align}
Here the diagonal mass parameters $m_{aa}^2$ and all quartic couplings $\lambda_i$ are taken to be real, while the off-diagonal mass parameters $m_{ab}^2$ ($a\neq b$) softly break the approximate $U(1)^3$ symmetry and can in general be complex, potentially inducing CP violation in the Higgs sector.

\paragraph{Electroweak symmetry breaking and VEV parametrization.}
We assume that electroweak symmetry breaking proceeds as usual, $SU(2)_L\times U(1)_Y\to U(1)_{\rm em}$, with the vacuum aligned such that $U(1)_{\rm em}$ remains unbroken.  In a component-field decomposition, one may parametrize
\begin{equation}
 \Phi_a=\begin{pmatrix}
 \phi_a^+\\[1mm]
 (v_a+\rho_a+i\eta_a)/\sqrt{2}
 \end{pmatrix},\qquad a=1,2,3,
 \label{eq:doublets}
\end{equation}
where $v_a$ denote the vacuum expectation values (VEVs).  We work in a phase convention in which the VEVs are real and label the fields such that $v_1\ll v_2\ll v_3$.  The electroweak scale is fixed by
\begin{equation}
 v^2\equiv v_1^2+v_2^2+v_3^2\simeq(246~\GeV)^2.
 \label{eq:vsum}
\end{equation}
A convenient parametrization trades $(v_1,v_2,v_3)$ for $(v,\beta,\beta')$,
\begin{align}
 v_1&=v\cos\beta', &
 v_2&=v\sin\beta'\cos\beta, &
 v_3&=v\sin\beta'\sin\beta,
 \label{eq:vevs}
\end{align}
implying
\begin{equation}
 \tan\beta'=\frac{\sqrt{v_2^2+v_3^2}}{v_1},\qquad
 \tan\beta=\frac{v_3}{v_2}.
 \label{eq:tanbeta}
\end{equation}

\paragraph{Tadpole conditions.}
Working with real VEVs generally implies nontrivial relations among the imaginary parts of the soft mass parameters.  The minimization conditions yield
\begin{equation}
 \operatorname{Im}m_{13}^2=-\frac{v_2}{v_3}\operatorname{Im}m_{12}^2,
 \qquad
 \operatorname{Im}m_{23}^2=\frac{v_1}{v_3}\operatorname{Im}m_{12}^2.
 \label{eq:imaginary_tadpoles}
\end{equation}
The remaining minimization conditions can be used to eliminate the diagonal mass parameters in favor of the VEVs,
\begin{align}
 m_{11}^2={}&\operatorname{Re}(m_{12}^2)\frac{v_2}{v_1}
 +\operatorname{Re}(m_{13}^2)\frac{v_3}{v_1}
 -\lambda_1v_1^2
 -\frac12\big[(\lambda_4+\lambda_7)v_2^2+(\lambda_5+\lambda_8)v_3^2\big],
 \label{eq:tad_m11}\\[2mm]
 m_{22}^2={}&\operatorname{Re}(m_{12}^2)\frac{v_1}{v_2}
 +\operatorname{Re}(m_{23}^2)\frac{v_3}{v_2}
 -\lambda_2v_2^2
 -\frac12\big[(\lambda_4+\lambda_7)v_1^2+(\lambda_6+\lambda_9)v_3^2\big],
 \label{eq:tad_m22}\\[2mm]
 m_{33}^2={}&\operatorname{Re}(m_{13}^2)\frac{v_1}{v_3}
 +\operatorname{Re}(m_{23}^2)\frac{v_2}{v_3}
 -\lambda_3v_3^2
 -\frac12\big[(\lambda_5+\lambda_8)v_1^2+(\lambda_6+\lambda_9)v_2^2\big].
 \label{eq:tad_m33}
\end{align}
In the following, we adopt the simplifying assumption that the tree-level Higgs potential is CP invariant and set
\begin{equation}
 \operatorname{Im}m_{12}^2=\operatorname{Im}m_{13}^2=
 \operatorname{Im}m_{23}^2=0.
 \label{eq:CPinv_assumption}
\end{equation}

\paragraph{Scalar mass matrices and diagonalization.}
In the CP-conserving limit, the physical spectrum after electroweak symmetry breaking consists of three neutral CP-even scalars $h_{1,2,3}$, two neutral CP-odd scalars $A_{1,2}$, and two charged-Higgs pairs $H_{1,2}^{\pm}$, in addition to the Goldstone modes $G^0$ and $G^\pm$.  The mass eigenstates are defined through
\begin{equation}
 \left(\begin{smallmatrix}h_1\\h_2\\h_3\end{smallmatrix}\right)
 =Z^h\left(\begin{smallmatrix}\rho_1\\\rho_2\\\rho_3\end{smallmatrix}\right),\qquad
 \left(\begin{smallmatrix}G^0\\A_1\\A_2\end{smallmatrix}\right)
 =Z^A\left(\begin{smallmatrix}\eta_1\\\eta_2\\\eta_3\end{smallmatrix}\right),\qquad
 \left(\begin{smallmatrix}G^+\\H_1^+\\H_2^+\end{smallmatrix}\right)
 =Z^\pm\left(\begin{smallmatrix}\phi_1^+\\\phi_2^+\\\phi_3^+\end{smallmatrix}\right).
 \label{eq:rotations}
\end{equation}
Following Ref.~\cite{Ma:2026G3HDMbs}, the three rotation matrices are parameterized directly after the definition of the mass eigenstates as
\begingroup
\small
\setlength{\arraycolsep}{3pt}
\begin{align}
 Z^A={}&
 \begin{pmatrix}0&0&1\\ \sin\gamma_A&\cos\gamma_A&0\\
 \cos\gamma_A&-\sin\gamma_A&0\end{pmatrix}
 \begin{pmatrix}\sin\beta'&0&-\cos\beta'\\0&1&0\\
 \cos\beta'&0&\sin\beta'\end{pmatrix}
 \begin{pmatrix}1&0&0\\0&\sin\beta&-\cos\beta\\
 0&\cos\beta&\sin\beta\end{pmatrix},
 \label{eq:ZA_param}\\[-1mm]
 Z^\pm={}&
 \begin{pmatrix}0&0&1\\ \sin\gamma_\pm&\cos\gamma_\pm&0\\
 \cos\gamma_\pm&-\sin\gamma_\pm&0\end{pmatrix}
 \begin{pmatrix}\sin\beta'&0&-\cos\beta'\\0&1&0\\
 \cos\beta'&0&\sin\beta'\end{pmatrix}
 \begin{pmatrix}1&0&0\\0&\sin\beta&-\cos\beta\\
 0&\cos\beta&\sin\beta\end{pmatrix},
 \label{eq:Zpm_param}\\[-1mm]
 Z^h={}&
 \begin{pmatrix}0&0&1\\ \sin\gamma_h&\cos\gamma_h&0\\
 \cos\gamma_h&-\sin\gamma_h&0\end{pmatrix}
 \begin{pmatrix}\cos\alpha'&0&\sin\alpha'\\0&1&0\\
 -\sin\alpha'&0&\cos\alpha'\end{pmatrix}
 \begin{pmatrix}1&0&0\\0&\cos\alpha&\sin\alpha\\
 0&-\sin\alpha&\cos\alpha\end{pmatrix}.
 \label{eq:Zh_param}
\end{align}
\endgroup
To match the notation used in the Feynman diagrams, we write $S_k\equiv h_k$ ($k=1,2,3$) for a generic neutral CP-even mass eigenstate; an unlabeled $S$ in Figs.~\ref{fig:top_feynman} and~\ref{fig:huc_feynman} denotes any of these states.  The observed SM-like Higgs is $S_1\equiv h_1\equiv h$, whereas $S_2\equiv h_2\equiv H$ and $S_3\equiv h_3\equiv H'$ are the two heavier CP-even states.  We similarly identify $(A_1,A_2)=(A,A')$.  When heavy masses cross, their labels continue to follow the corresponding generation-associated eigenvectors.  In the basis $(\eta_1,\eta_2,\eta_3)$, the CP-odd mass-squared matrix is
\begin{equation}
 {\cal M}_A^2=\begin{pmatrix}
 m_{12}^2v_2/v_1+m_{13}^2v_3/v_1&-m_{12}^2&-m_{13}^2\\
 -m_{12}^2&m_{12}^2v_1/v_2+m_{23}^2v_3/v_2&-m_{23}^2\\
 -m_{13}^2&-m_{23}^2&m_{13}^2v_1/v_3+m_{23}^2v_2/v_3
 \end{pmatrix},
 \label{eq:MA}
\end{equation}
The charged and CP-even mass-squared matrices can then be written compactly as
\begin{align}
 {\cal M}_{\pm}^2={}&{\cal M}_{A}^2+\frac12
 \begin{pmatrix}
 -v_2^2\lambda_7-v_3^2\lambda_8&v_1v_2\lambda_7&v_1v_3\lambda_8\\
 v_1v_2\lambda_7&-v_1^2\lambda_7-v_3^2\lambda_9&v_2v_3\lambda_9\\
 v_1v_3\lambda_8&v_2v_3\lambda_9&-v_1^2\lambda_8-v_2^2\lambda_9
 \end{pmatrix},
 \label{eq:Mcharged}\\
 {\cal M}_{h}^2={}&{\cal M}_{A}^2+
 \begin{pmatrix}
 2v_1^2\lambda_1&v_1v_2(\lambda_4+\lambda_7)&v_1v_3(\lambda_5+\lambda_8)\\
 v_1v_2(\lambda_4+\lambda_7)&2v_2^2\lambda_2&v_2v_3(\lambda_6+\lambda_9)\\
 v_1v_3(\lambda_5+\lambda_8)&v_2v_3(\lambda_6+\lambda_9)&2v_3^2\lambda_3
 \end{pmatrix}.
\label{eq:Mh}
\end{align}

The CP-odd, charged, and CP-even mass-squared matrices are diagonalized by the mixing matrices $Z^A$, $Z^\pm$, and $Z^h$, respectively,
\begin{align}
 &Z^A{\cal M}_A^2Z^{A\dagger}=
 \diag(0,M_{A_1}^2,M_{A_2}^2),
 \label{eq:diag_A}\\
 &Z^\pm{\cal M}_\pm^2Z^{\pm\dagger}=
 \diag(0,M_{H_1^\pm}^2,M_{H_2^\pm}^2),
 \label{eq:diag_Hpm}\\
 &Z^h{\cal M}_h^2Z^{h\dagger}=
 \diag(M_{h_1}^2,M_{h_2}^2,M_{h_3}^2).
 \label{eq:diag_h}
\end{align}
After the $\beta$ and $\beta'$ rotations, the CP-odd and charged matrices are partially diagonalized and the massless Goldstone bosons already appear as eigenstates.  The final rotations by $\gamma_A$ and $\gamma_\pm$ define the two massive pseudoscalars and the two charged-Higgs pairs.  These relations allow the soft parameters $m_{12}^2$, $m_{13}^2$, and $m_{23}^2$ to be traded for two physical pseudoscalar masses and the angle $\gamma_A$; the charged-Higgs masses and $\gamma_\pm$ are determined analogously.

\paragraph{Vacuum stability and perturbative unitarity.}
Following Ref.~\cite{Ma:2026G3HDMbs}, the necessary perturbative-unitarity preselection conditions used in the scan are
\begin{equation}
 |\lambda_1|,\,|\lambda_2|,\,|\lambda_3|\leq\frac{4\pi}{3},\qquad
 |\lambda_4+\lambda_7|,\,|\lambda_5+\lambda_8|,\,|\lambda_6+\lambda_9|\leq8\pi.
 \label{eq:unitarity_working}
\end{equation}
The conservative sufficient vacuum-stability conditions are
\begin{equation}
 \begin{aligned}
 \text{(a)}\quad&\lambda_1>0,\;\lambda_2>0,\;\lambda_3>0,\\[4pt]
 \text{(b)}\quad&\Lambda_{12}+2\sqrt{\lambda_1\lambda_2}\geq0,\;
 \Lambda_{13}+2\sqrt{\lambda_1\lambda_3}\geq0,\;
 \Lambda_{23}+2\sqrt{\lambda_2\lambda_3}\geq0,\\[6pt]
 \text{(c)}\quad&\sqrt{\lambda_1\lambda_2\lambda_3}
 +\frac12\Big(\Lambda_{12}\sqrt{\lambda_3}
 +\Lambda_{13}\sqrt{\lambda_2}
 +\Lambda_{23}\sqrt{\lambda_1}\Big)\\
 &+\frac12\sqrt{\Big(\Lambda_{12}+2\sqrt{\lambda_1\lambda_2}\Big)
 \Big(\Lambda_{13}+2\sqrt{\lambda_1\lambda_3}\Big)
 \Big(\Lambda_{23}+2\sqrt{\lambda_2\lambda_3}\Big)}\geq0.
 \end{aligned}
 \label{eq:BFB_conditions}
\end{equation}
\begin{equation}
 \Lambda_{12}=\lambda_4+\min(0,\lambda_7),\qquad
 \Lambda_{13}=\lambda_5+\min(0,\lambda_8),\qquad
 \Lambda_{23}=\lambda_6+\min(0,\lambda_9).
 \label{eq:LambdaBFB}
\end{equation}

\subsection{The fermion sector in the G3HDM}
\label{subsec:G3HDM_fermion}

The Yukawa interactions of the three Higgs doublets $\Phi_a$ ($a=1,2,3$) with the SM fermions are described by the most general renormalizable Yukawa Lagrangian
\begin{align}
 -\mathcal{L}^{\mathrm{Yuk}}_{3\mathrm{HDM}}={}&
 \sum_{a=1}^{3}\sum_{i,j=1}^{3}
 \Big(
 \lambda^{ua}_{ij}\,\bar q_{Li}\,\widetilde{\Phi}_a\,u_{Rj}
 +\lambda^{da}_{ij}\,\bar q_{Li}\,\Phi_a\,d_{Rj}
 +\lambda^{\ell a}_{ij}\,\bar \ell_{Li}\,\Phi_a\,e_{Rj}
 \Big)+\mathrm{h.c.},
 \label{eq:Yukawa_Lagrangian}
\end{align}
where $\widetilde{\Phi}_a\equiv i\sigma_2\Phi_a^{\ast}$ and $i,j$ are flavor indices.  Neutrino masses and mixing are neglected in the present setup.

To realize a \emph{generational} structure, we adopt the following schematic weak-basis Yukawa textures before the final fermion diagonalization~\cite{Altmannshofer:2016zrn,Altmannshofer:2017uvs,Altmannshofer:2018bch}:
\begin{subequations}\label{eq:Yukawa_textures}
\begin{align}
 \lambda_{u1}&\sim\frac{\sqrt2}{v_1}
 \begin{pmatrix}m_u&m_u&m_u\\m_u&m_u&m_u\\m_u&m_u&m_u\end{pmatrix},&
 \lambda_{u2}&\sim\frac{\sqrt2}{v_2}
 \begin{pmatrix}0&0&0\\0&m_c&m_c\\0&m_c&m_c\end{pmatrix},&
 \lambda_{u3}&\sim\frac{\sqrt2}{v_3}
 \begin{pmatrix}0&0&0\\0&0&0\\0&0&m_t\end{pmatrix},
 \label{eq:Yukawa_textures_u}\\[2mm]
 \lambda_{d1}&\sim\frac{\sqrt2}{v_1}
 \begin{pmatrix}m_d&m_s\lambda&m_b\lambda^3\\m_d&m_d&m_d\\m_d&m_d&m_d\end{pmatrix},&
 \lambda_{d2}&\sim\frac{\sqrt2}{v_2}
 \begin{pmatrix}0&0&0\\0&m_s&m_b\lambda^2\\0&m_s&m_s\end{pmatrix},&
 \lambda_{d3}&\sim\frac{\sqrt2}{v_3}
 \begin{pmatrix}0&0&0\\0&0&0\\0&0&m_b\end{pmatrix},
 \label{eq:Yukawa_textures_d}\\[2mm]
 \lambda_{\ell1}&\sim\frac{\sqrt2}{v_1}
 \begin{pmatrix}m_e&m_e&m_e\\m_e&m_e&m_e\\m_e&m_e&m_e\end{pmatrix},&
 \lambda_{\ell2}&\sim\frac{\sqrt2}{v_2}
 \begin{pmatrix}0&0&0\\0&m_\mu&m_\mu\\0&m_\mu&m_\mu\end{pmatrix},&
 \lambda_{\ell3}&\sim\frac{\sqrt2}{v_3}
 \begin{pmatrix}0&0&0\\0&0&0\\0&0&m_\tau\end{pmatrix}.
 \label{eq:Yukawa_textures_l}
\end{align}
\end{subequations}
Here ``$\sim$'' specifies only the parametric order of the entries, and $\lambda\simeq|V_{us}|$ denotes the Wolfenstein parameter.  The exact $\mathcal O(1)$ coefficients are correlated so that every Yukawa matrix remains rank one; they cannot be varied independently entry by entry without violating the defining G3HDM assumption.  Consequently, each Higgs doublet couples to one linear combination of the three fermion generations.  This rank-one statement, rather than the schematic zeros alone, is the core of the generational construction.

We choose the flavor-basis convention in which the CKM matrix originates entirely from the left-handed down-quark rotation,
\begin{equation*}
 U_L^u=I,\qquad U_L^d=V_{\rm CKM},\qquad
 U_R^u=U_R^d=U_L^\ell=U_R^\ell=I,
\end{equation*}
where $I$ denotes the $3\times3$ identity matrix.  Thus all rotation matrices other than $U_L^d$ are taken to be identity matrices.  In this convention, we define in the fermion mass-eigenstate basis the following mass parameters
\begin{align}
 m^{f_1}_{ff'}&=\frac{v_1}{\sqrt2}\langle f_L|\lambda_{f_1}|f'_R\rangle,\qquad
 m^{f_2}_{ff'}=\frac{v_2}{\sqrt2}\langle f_L|\lambda_{f_2}|f'_R\rangle,\qquad
 m^{f_3}_{ff'}=\frac{v_3}{\sqrt2}\langle f_L|\lambda_{f_3}|f'_R\rangle,
 \label{eq:mf_def}
\end{align}
which satisfy
\begin{equation}
 m^{f_3}_{ff'}+m^{f_2}_{ff'}+m^{f_1}_{ff'}=m_f\delta_{ff'},
 \label{eq:mf_sumrule}
\end{equation}
with $m_f$ denoting the physical fermion masses.

The couplings to physical scalar mass eigenstates can be written directly in terms of these mass-basis matrices.  For $k=1,2,3$ and $r=1,2$, define
\begin{align*}
 ({\cal Y}^{f,h_k})_{ij}&=\sum_{a=1}^3 Z^h_{ka}\frac{m^{f_a}_{ij}}{v_a},&
 ({\cal Y}^{f,A_r})_{ij}&=\sum_{a=1}^3 Z^A_{r+1,a}\frac{m^{f_a}_{ij}}{v_a},&
 ({\cal Y}^{f,H_r^+})_{ij}&=\sum_{a=1}^3 Z^\pm_{r+1,a}\frac{m^{f_a}_{ij}}{v_a}.
\end{align*}
The row $r+1$ in the CP-odd and charged sectors omits the Goldstone mode.  In this notation, the neutral interactions are
\begin{align*}
 -\Lag_Y^{\rm neutral}={}&
 \sum_{f=u,d,\ell}\sum_{k=1}^3\sum_{i,j}h_k\,\bar f_i
 \left[({\cal Y}^{f,h_k})_{ij}P_R+({\cal Y}^{f,h_k})_{ji}^*P_L\right]f_j\\
 &+i\sum_{f=u,d,\ell}\xi_f\sum_{r=1}^2\sum_{i,j}A_r\,\bar f_i
 \left[({\cal Y}^{f,A_r})_{ij}P_R-({\cal Y}^{f,A_r})_{ji}^*P_L\right]f_j.
\end{align*}
Here $\xi_{d,\ell}=1$ and $\xi_u=-1$; the relative up-type sign follows from the neutral component of $\widetilde\Phi_a$.  The charged-scalar interactions are
\begin{align*}
 -\Lag_Y^{H^\pm}={}&\sqrt2\sum_{r=1}^2\sum_{i,j}H_r^+\Big\{
 \bar u_i\Big[(V_{\rm CKM}{\cal Y}^{d,H_r^+})_{ij}P_R\\[-1mm]
 &\hspace{39mm}-\big[({\cal Y}^{u,H_r^+})^\dagger V_{\rm CKM}\big]_{ij}P_L\Big]d_j
 +\bar\nu_i({\cal Y}^{\ell,H_r^+})_{ij}P_R\ell_j\Big\}+\hc.
\end{align*}
These expressions fix the complex conjugations and CKM matrix ordering used in the loop amplitudes.

Expanding to leading order in the ratios of first-to-second and second-to-third generation masses, one obtains~\cite{AltmannshoferToner:2025}
\begin{align}
 \frac{m^{u_1}_{qq'}}{m_u}&\simeq
 \begin{pmatrix}
 1&x_{uc}&x_{ut}\\
 x_{cu}&x_{cu}x_{uc}&x_{cu}x_{ut}\\
 x_{tu}&x_{tu}x_{uc}&x_{tu}x_{ut}
 \end{pmatrix},
 &
 \frac{m^{u_2}_{qq'}}{m_c}&\simeq
 \begin{pmatrix}
 \dfrac{m_u^2}{m_c^2}x_{uc}x_{cu}&-\dfrac{m_u}{m_c}x_{uc}&-\dfrac{m_u}{m_c}x_{uc}y_{ct}\\
 -\dfrac{m_u}{m_c}x_{cu}&1&y_{ct}\\
 -\dfrac{m_u}{m_c}y_{tc}x_{cu}&y_{tc}&y_{tc}y_{ct}
 \end{pmatrix},
 \label{eq:mu12}
\end{align}
\begin{equation}
\resizebox{0.98\textwidth}{!}{$\displaystyle
 \frac{m^{u_3}_{qq'}}{m_t}\simeq
 \begin{pmatrix}
 \dfrac{m_u^2}{m_t^2}(x_{ut}-x_{uc}y_{ct})(x_{tu}-y_{tc}x_{cu})&
 \dfrac{m_um_c}{m_t^2}(x_{ut}-x_{uc}y_{ct})y_{tc}&
 -\dfrac{m_u}{m_t}(x_{ut}-x_{uc}y_{ct})\\
 \dfrac{m_um_c}{m_t^2}y_{ct}(x_{tu}-y_{tc}x_{cu})&
 \dfrac{m_c^2}{m_t^2}y_{ct}y_{tc}&-\dfrac{m_c}{m_t}y_{ct}\\
 -\dfrac{m_u}{m_t}(x_{tu}-y_{tc}x_{cu})&-\dfrac{m_c}{m_t}y_{tc}&1
 \end{pmatrix}.$}
 \label{eq:mu3}
\end{equation}
\begin{equation}
 \frac{m^{d_1}_{qq'}}{m_d}\simeq
 \begin{pmatrix}
 1&\dfrac{m_s}{m_d}V_{ud}^*V_{us}&\dfrac{m_b}{m_d}V_{ud}^*V_{ub}\\
 x_{sd}&x_{sd}\dfrac{m_s}{m_d}V_{ud}^*V_{us}&x_{sd}\dfrac{m_b}{m_d}V_{ud}^*V_{ub}\\
 x_{bd}&x_{bd}\dfrac{m_s}{m_d}V_{ud}^*V_{us}&x_{bd}\dfrac{m_b}{m_d}V_{ud}^*V_{ub}
 \end{pmatrix}.
 \label{eq:md1}
\end{equation}
\begin{equation}
 \frac{m^{d_2}_{qq'}}{m_s}\simeq
 \begin{pmatrix}
 -\dfrac{m_d}{m_s}V_{cd}^*V_{cs}x_{sd}&V_{cd}^*V_{cs}&\dfrac{m_b}{m_s}V_{cd}^*V_{cb}\\
 -\dfrac{m_d}{m_s}x_{sd}&1&\dfrac{m_b}{m_s}V_{cs}^*V_{cb}\\
 -\dfrac{m_d}{m_s}y_{bs}x_{sd}&y_{bs}&y_{bs}\dfrac{m_b}{m_s}V_{cs}^*V_{cb}
 \end{pmatrix}.
 \label{eq:md12}
\end{equation}
\begin{equation}
 \frac{m^{d_3}_{qq'}}{m_b}\simeq
 \begin{pmatrix}
 -\dfrac{m_d}{m_b}V_{td}^*V_{tb}(x_{bd}-y_{bs}x_{sd})&-\dfrac{m_s}{m_b}V_{td}^*V_{tb}y_{bs}&V_{td}^*V_{tb}\\
 -\dfrac{m_d}{m_b}V_{ts}^*V_{tb}(x_{bd}-y_{bs}x_{sd})&-\dfrac{m_s}{m_b}V_{ts}^*V_{tb}y_{bs}&V_{ts}^*V_{tb}\\
 -\dfrac{m_d}{m_b}(x_{bd}-y_{bs}x_{sd})&-\dfrac{m_s}{m_b}y_{bs}&1
 \end{pmatrix}.
 \label{eq:md3}
\end{equation}
The charged-lepton mass parameters follow from Eqs.~\eqref{eq:mu12} and~\eqref{eq:mu3} through the replacements $(u,c,t)\to(e,\mu,\tau)$ and $(x^u,y^u)\to(x^\ell,y^\ell)$.  In the above expressions, $x_{ij}$ and $y_{ij}$ are free, in general complex, $\mathcal O(1)$ parameters that encode additional sources of flavor and CP violation beyond the SM.  The displayed correlations follow both from reproducing the fermion masses and the CKM matrix and from the assumed rank-one Yukawa matrices~\cite{AltmannshoferToner:2025}.

\section{Observables and numerical procedure}

\subsection{Flavor observables}

We parameterize the flavor-changing $h_1$ interactions as
\begin{equation}
 -\Lag\supset h_1\,\overline f_i
 \left(g^L_{ijh_1}P_L+g^R_{ijh_1}P_R\right)f_j+\hc.
 \label{eq:effective_h}
\end{equation}
For the CP-even mass eigenstates, the corresponding Yukawa matrix is
\begin{equation}
 ({\cal Y}^{f,h_k})_{ij}=\sum_{a=1}^3\frac{Z^h_{ka}}{v_a}m^{f_a}_{ij},
 \qquad
 g^R_{ijh_1}=({\cal Y}^{f,h_1})_{ij},\qquad
 g^L_{ijh_1}=({\cal Y}^{f,h_1})_{ji}^*.
 \label{eq:physical_h_Yukawa}
\end{equation}
Using Eqs.~\eqref{eq:mu12} and~\eqref{eq:mu3}, and the analogous charged-lepton matrices, the relevant independent entries are directly denoted by $y_{ct}$, $y_{tc}$, $y_{\mu\tau}$, and $y_{\tau\mu}$.  To leading order in the fermion-mass hierarchy,
\begin{align}
 g^R_{cth_1}&\simeq m_cy_{ct}D_{23}, &
 g^L_{cth_1}&\simeq m_cy_{tc}^*D_{23},
 \nonumber\\
 g^R_{\mu\tau h_1}&\simeq m_\mu y_{\mu\tau}D_{23}, &
 g^L_{\mu\tau h_1}&\simeq m_\mu y_{\tau\mu}^*D_{23},
 \label{eq:dominant_couplings}
\end{align}
where
\begin{equation}
 D_{23}=\frac{Z^h_{12}}{v_2}-\frac{Z^h_{13}}{v_3},
 \qquad \Delta_{23}=|D_{23}|.
 \label{eq:D23}
\end{equation}
All texture phases are set to zero in the numerical sample, so the shorthand $y_{ij}^2$ used below equals $|y_{ij}|^2$.
The tree-level width for $t\to qh_1$ is
\begin{align}
 \Gamma(t\to qh_1)={}&\frac{\lambda^{1/2}(m_t^2,m_q^2,m_{h_1}^2)}{32\pi m_t^3}
 \nonumber\\[-1mm]
 &\times\Big[(|g^L_{qt}|^2+|g^R_{qt}|^2)
 (m_t^2+m_q^2-m_{h_1}^2)
 \nonumber\\
 &\hspace{30mm}+4m_tm_q\operatorname{Re}(g^L_{qt}g^{R*}_{qt})\Big],
 \label{eq:top_width}
\end{align}
where $\lambda(x,y,z)=x^2+y^2+z^2-2xy-2xz-2yz$.  In the limit $m_c\ll m_t$, Eq.~\eqref{eq:dominant_couplings} yields
\begin{equation}
 \Gamma(t\to ch_1)_{\rm tree}\simeq
 \frac{m_t}{32\pi}\left(1-\frac{m_{h_1}^2}{m_t^2}\right)^2
 m_c^2(y_{ct}^2+y_{tc}^2)\Delta_{23}^2.
 \label{eq:top_driver}
\end{equation}

For $i\ne j$, the sum of the two charge-conjugate Higgs decay modes is
\begin{align}
 \Gamma(h_1\to f_i\overline f_j+\overline f_i f_j)
 \simeq\frac{N_c m_{h_1}}{8\pi}
 (|g^L_{ij}|^2+|g^R_{ij}|^2),
 \label{eq:h_width}
\end{align}
when the final-state masses are neglected in phase space.  Thus
\begin{equation}
 \Gamma(h_1\to\mu\tau)_{\rm tree}\propto
 m_\mu^2(y_{\mu\tau}^2+y_{\tau\mu}^2)\Delta_{23}^2.
 \label{eq:mutau_driver}
\end{equation}
We calculate the quoted $t\to qh_1$ and $h_1\to f_i\bar f_j+\bar f_i f_j$ branching ratios with the complete tree plus one-loop amplitudes and with the total widths evaluated at the same parameter point.  The tree-only results are retained separately to quantify the loop corrections.

Radiative charged-lepton decays are described by the dipole operators
\begin{equation}
 \Lag_{\rm dip}=\frac{em_{\ell_i}}{2}\,
 \overline\ell_j\sigma^{\mu\nu}
 (A_L^{ji}P_L+A_R^{ji}P_R)\ell_iF_{\mu\nu}+\hc,
 \label{eq:dipole}
\end{equation}
for which
\begin{equation}
 \Br(\ell_i\to\ell_j\gamma)=
 \frac{48\pi^3\alpha}{G_F^2}
 (|A_L^{ji}|^2+|A_R^{ji}|^2)
 \Br(\ell_i\to\ell_j\nu_i\overline\nu_j).
 \label{eq:lrad}
\end{equation}
The coefficients include neutral- and charged-scalar one-loop contributions; the broader relation between Higgs-sector CP/flavor structure, electric dipole moments, and $\mu\to e\gamma$ has been analyzed in Refs.~\cite{Altmannshofer:2020shb,Altmannshofer:2024edmmeg}.  Electric dipole moments in the CP-violating G3HDM have also been investigated in Ref.~\cite{Ma:2026G3HDMEDM}.  For $\ell_i\to3\ell_j$ we include photon and $Z$ penguins, scalar penguins, box diagrams, their interference, and tree-level neutral-scalar exchange whenever present.  Coherent conversion is normalized as
\begin{equation}
 \mathrm{CR}(\mu N\to eN)=\frac{\Gamma(\mu N\to eN)}{\Gamma_{\rm capt}(N)},
 \label{eq:conversion}
\end{equation}
using target-specific nuclear overlap and capture inputs for both aluminum and gold.  Retaining non-dipole operators is important because it weakens the correlation between $\mu\to e\gamma$ and conversion.

\subsection{Computational Implementation and Representative Diagrams}

The G3HDM was implemented with \textsc{SARAH}~4.15.3~\cite{Staub:2013tta}.  The generated numerical code was evaluated with \textsc{SPheno}~4.0.5~\cite{Porod:2003um,Porod:2011nf}, which provides the spectrum and the charged-lepton-flavor-violating observables, including $h_1\to\ell_i^\pm\ell_j^\mp$.  The complete top-FCNC amplitudes and $h_1\to u\bar c+\bar u c$ amplitudes were evaluated using \textsc{LoopTools}~2.16~\cite{Hahn:1998yk} for the one-loop integral functions.

\begin{figure}[htbp]
 \centering

 \begin{minipage}{0.32\textwidth}
  \centering
  \includegraphics[width=0.65\linewidth,keepaspectratio]{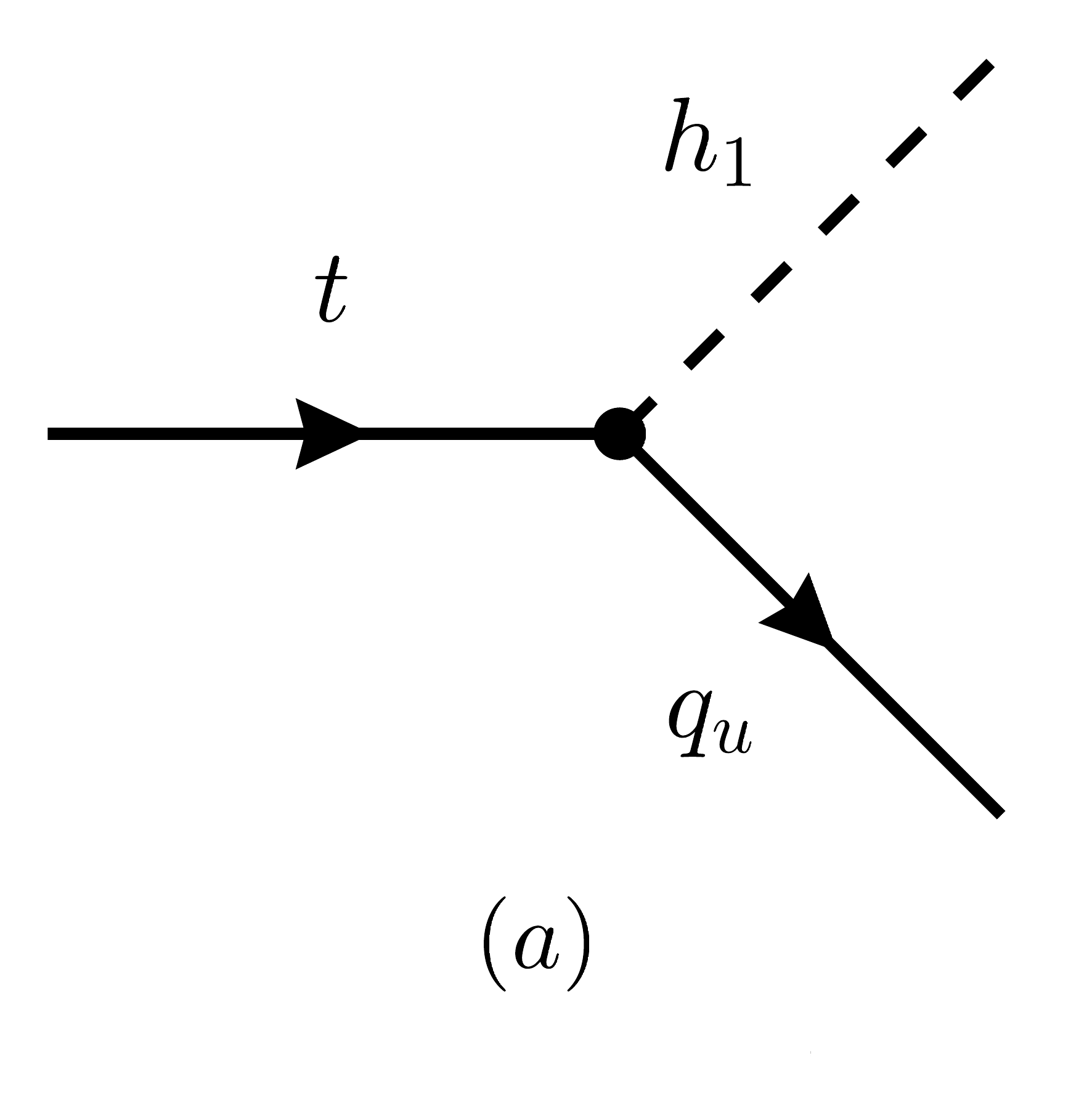}
 \end{minipage}

 \par\vspace{1mm}

 \begin{minipage}{0.32\textwidth}
  \centering
  \includegraphics[width=0.65\linewidth,keepaspectratio]{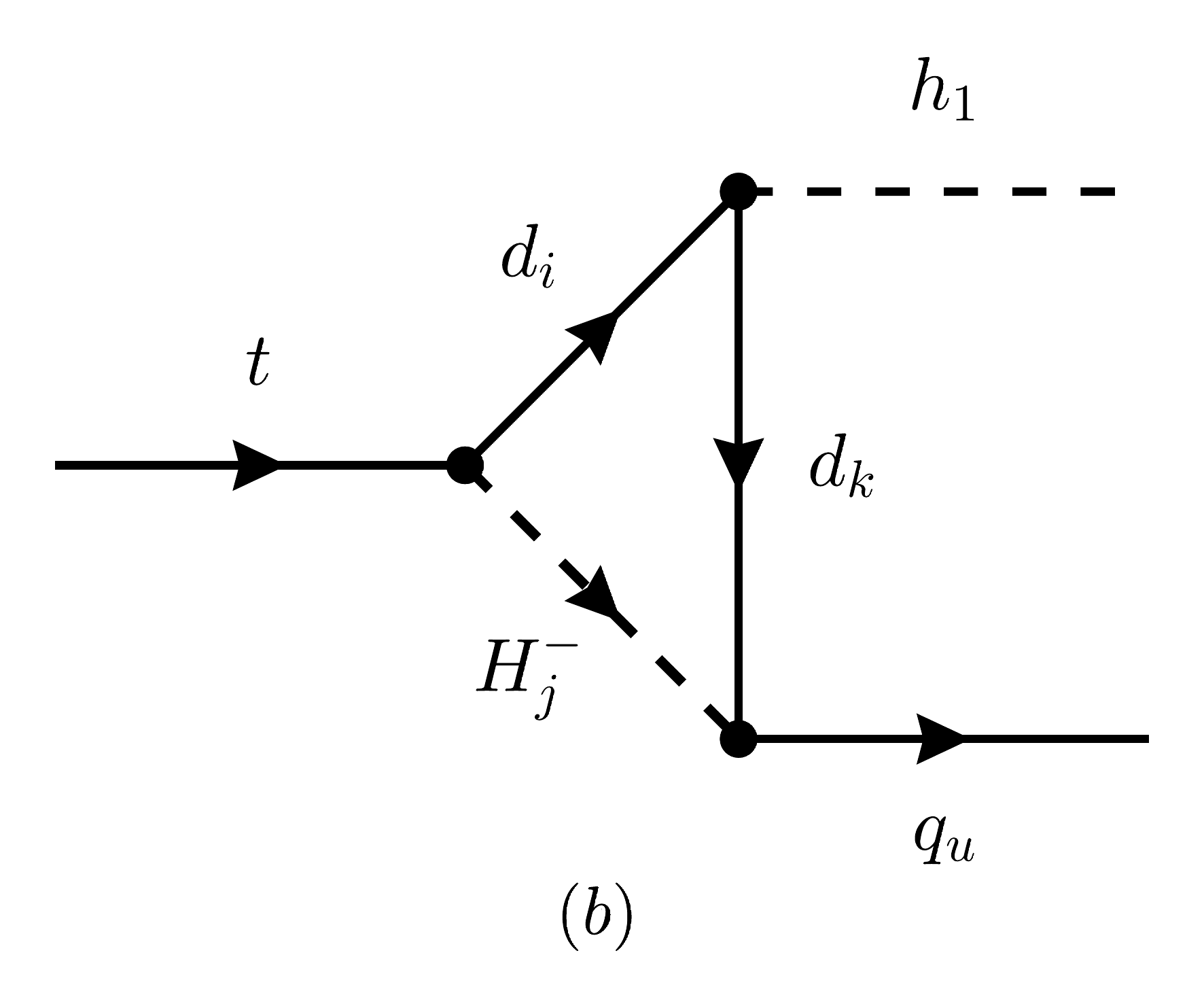}
 \end{minipage}\hfill
 \begin{minipage}{0.32\textwidth}
  \centering
  \includegraphics[width=0.65\linewidth,keepaspectratio]{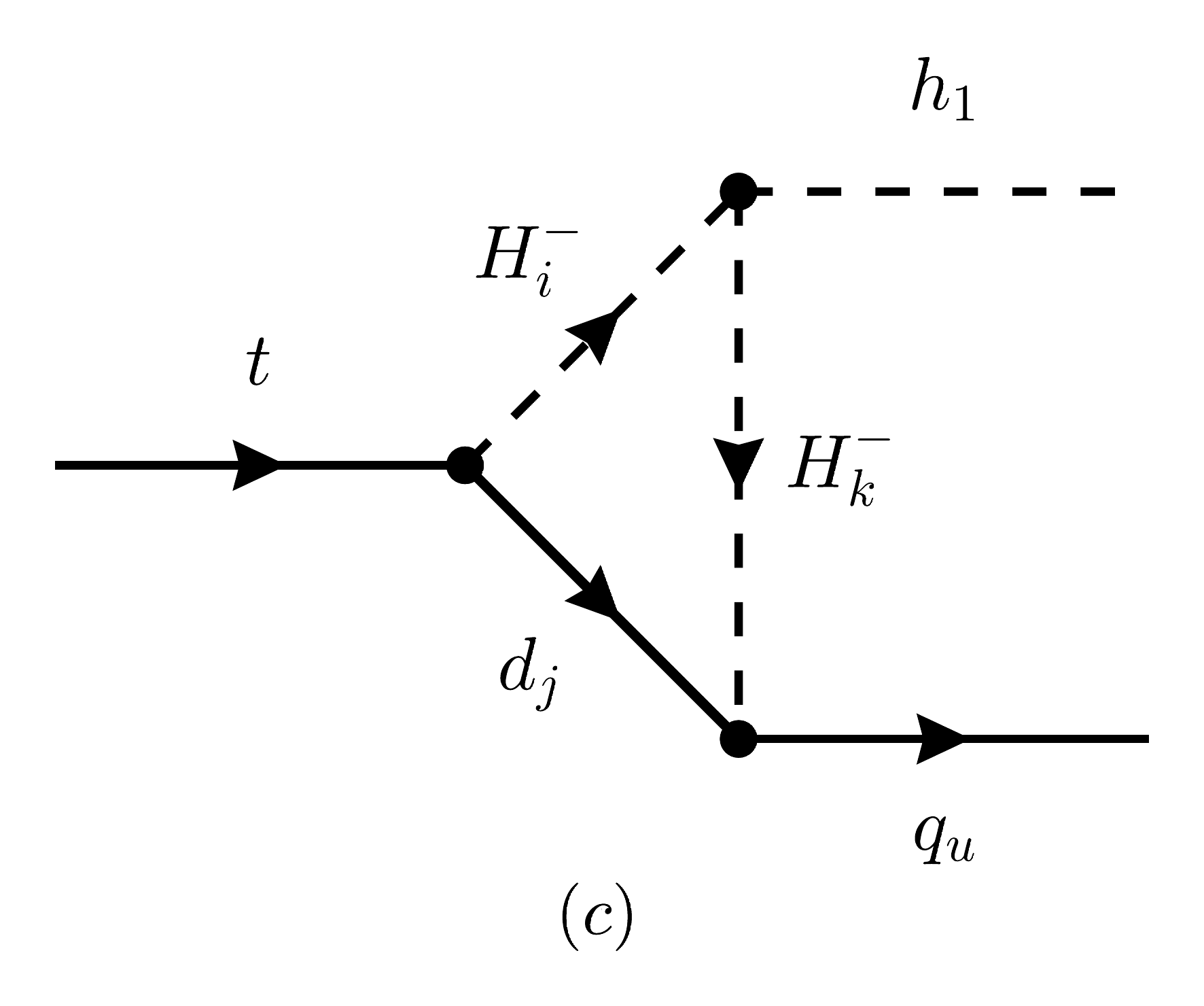}
 \end{minipage}\hfill
 \begin{minipage}{0.32\textwidth}
  \centering
  \includegraphics[width=0.65\linewidth,keepaspectratio]{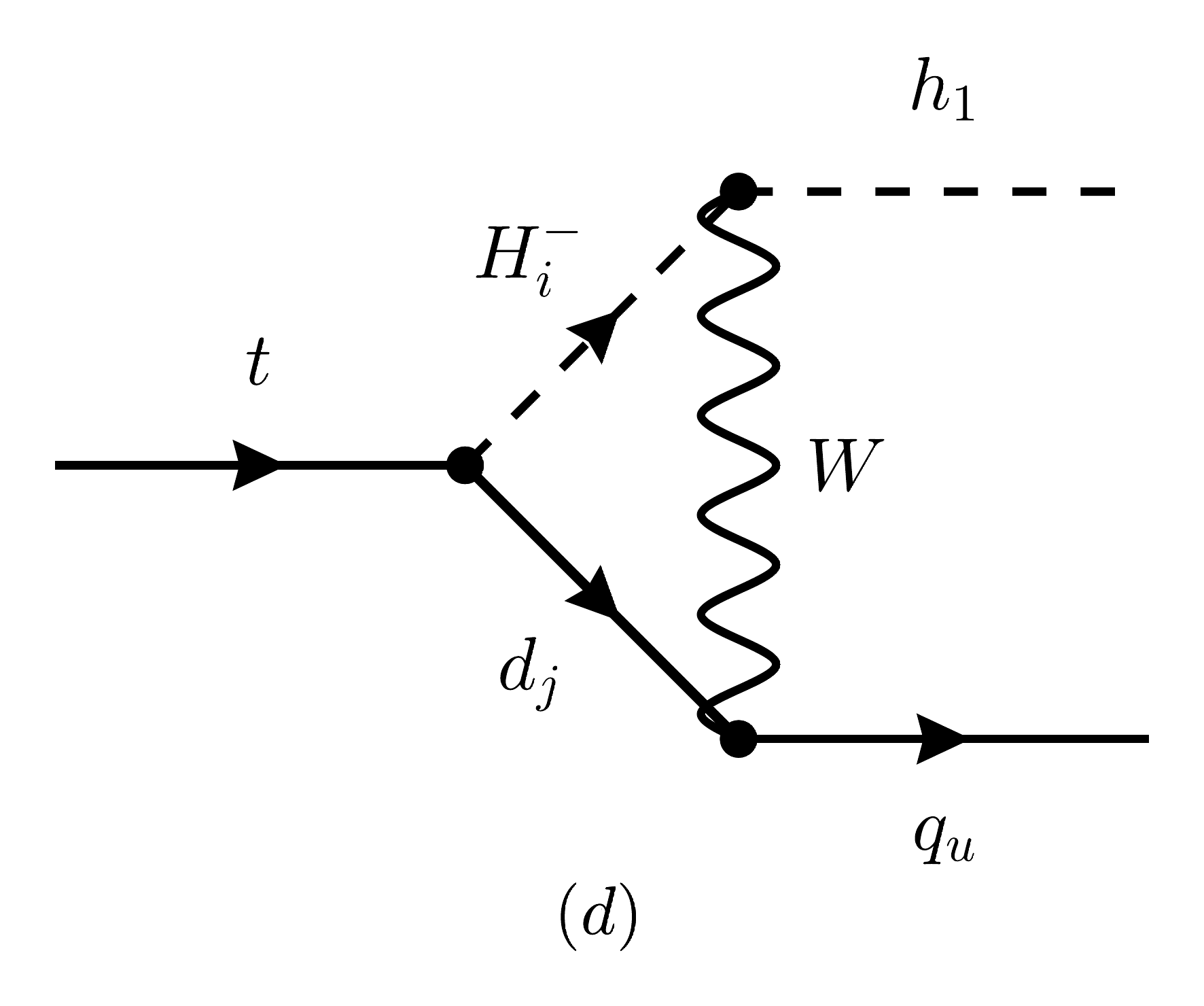}
 \end{minipage}

 \par\vspace{1mm}

 \begin{minipage}{0.32\textwidth}
  \centering
  \includegraphics[width=0.65\linewidth,keepaspectratio]{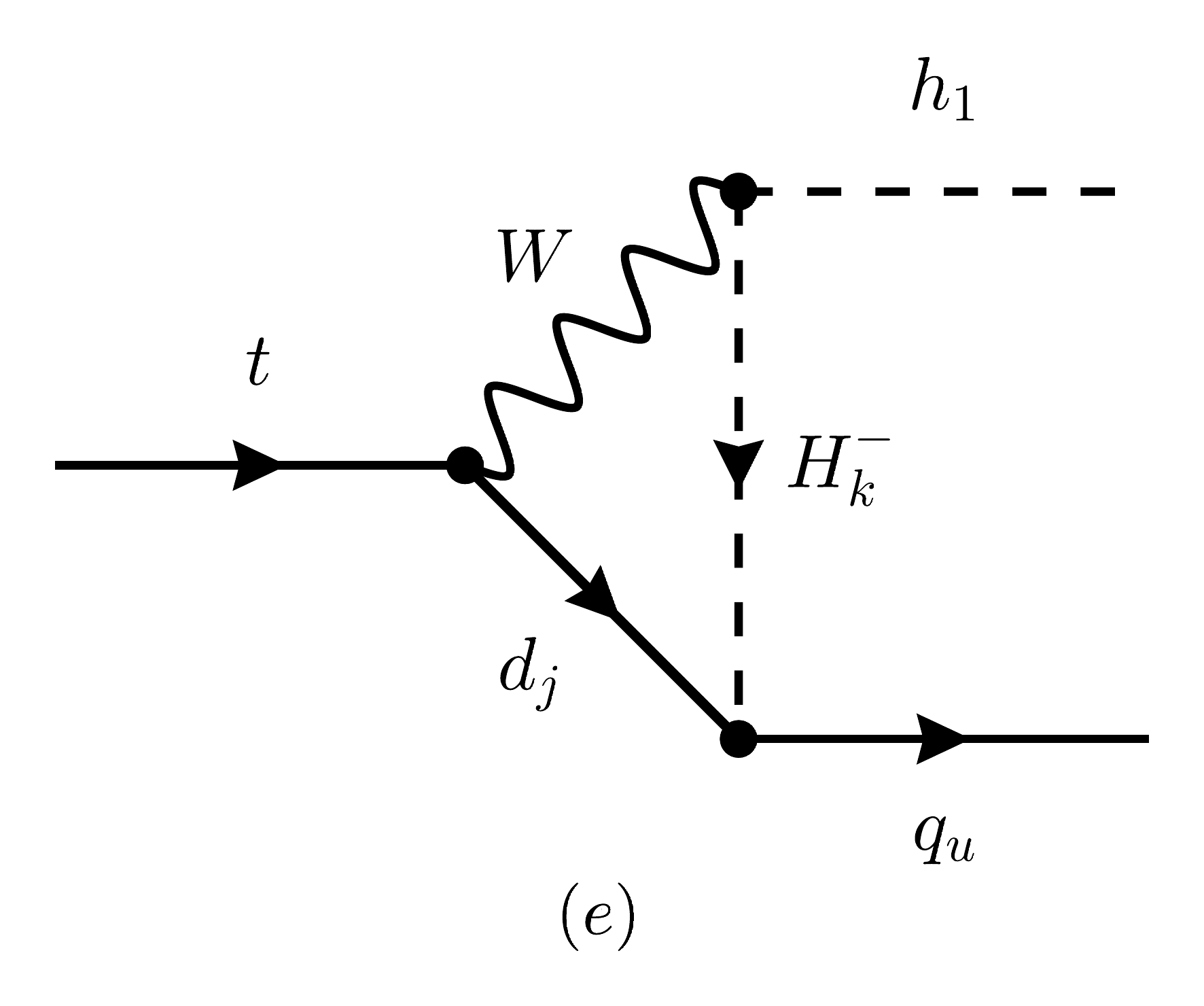}
 \end{minipage}\hfill
 \begin{minipage}{0.32\textwidth}
  \centering
  \includegraphics[width=0.65\linewidth,keepaspectratio]{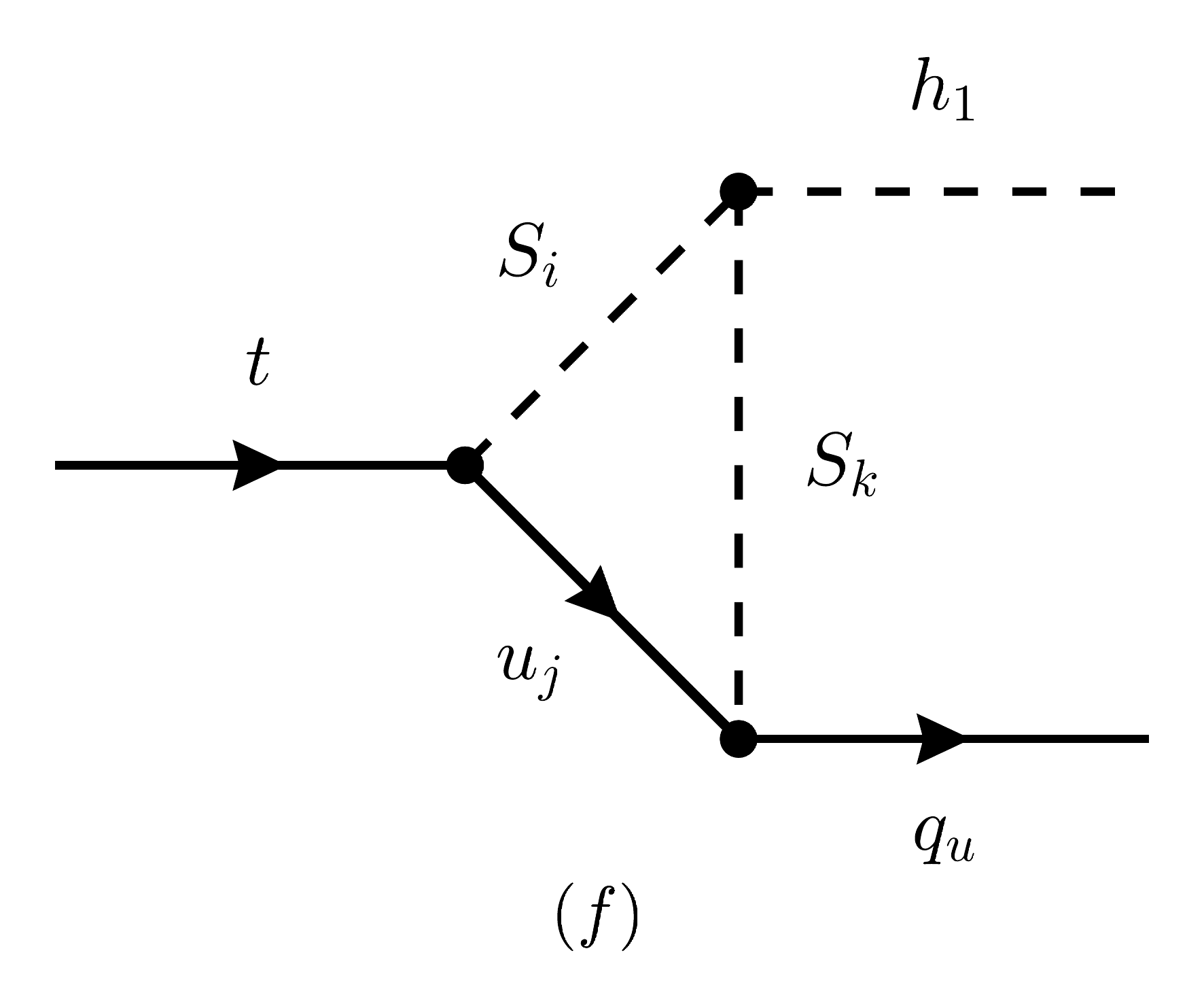}
 \end{minipage}\hfill
 \begin{minipage}{0.32\textwidth}
  \centering
  \includegraphics[width=0.65\linewidth,keepaspectratio]{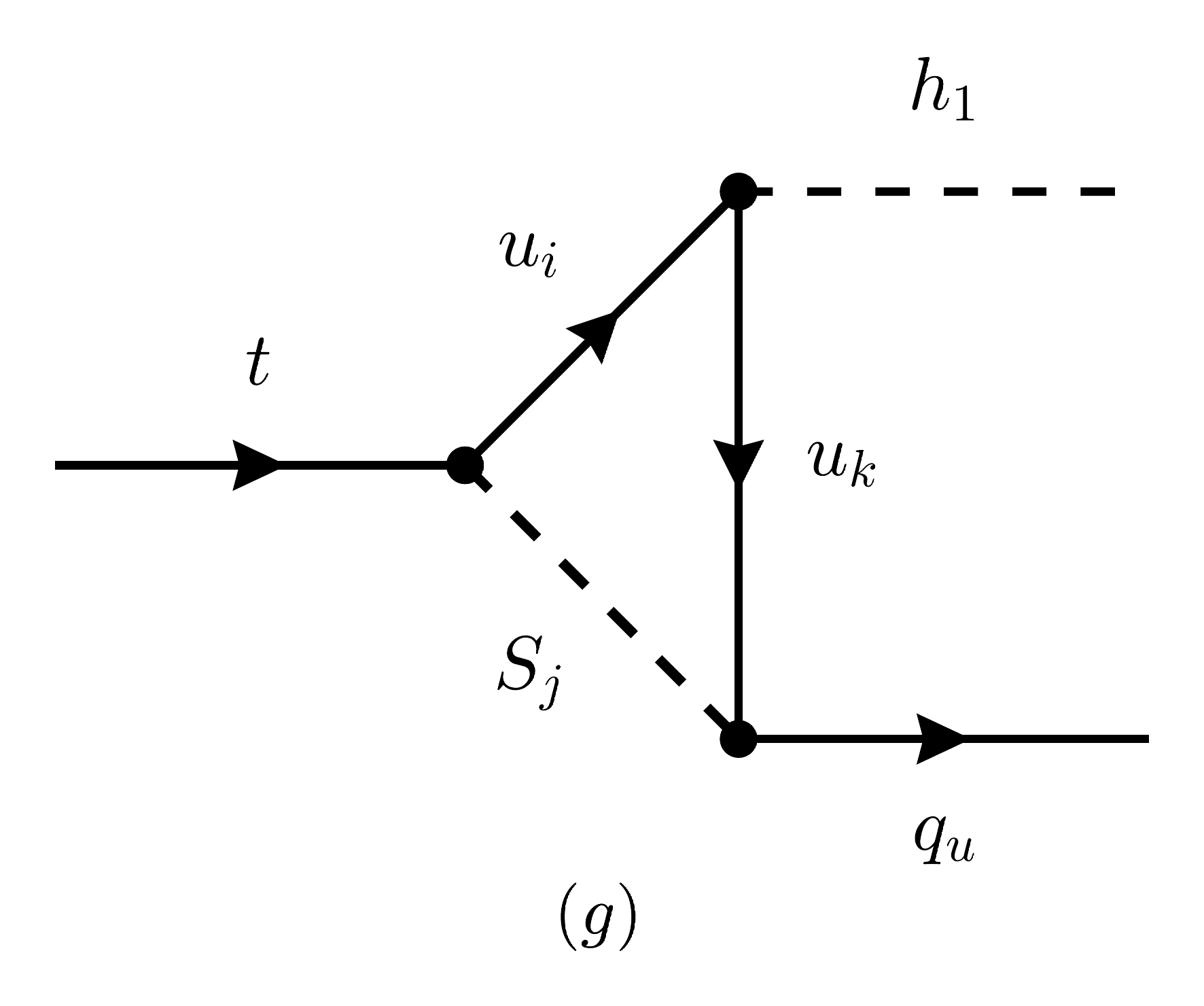}
 \end{minipage}

 \caption{Representative tree-level and one-loop diagrams for $t\to qh_1$, where $q=u,c$. Panel (a) is the tree-level FCNC contribution, while panels (b)--(g) show representative one-loop topologies. Here $S_i$ denotes a neutral scalar and $H_i^\pm$ a charged scalar; repeated internal-particle indices are summed.}
 \label{fig:top_feynman}
\end{figure}

\begin{figure}[htbp]
 \centering

 \begin{minipage}{0.32\textwidth}
  \centering
  \includegraphics[width=0.65\linewidth,keepaspectratio]{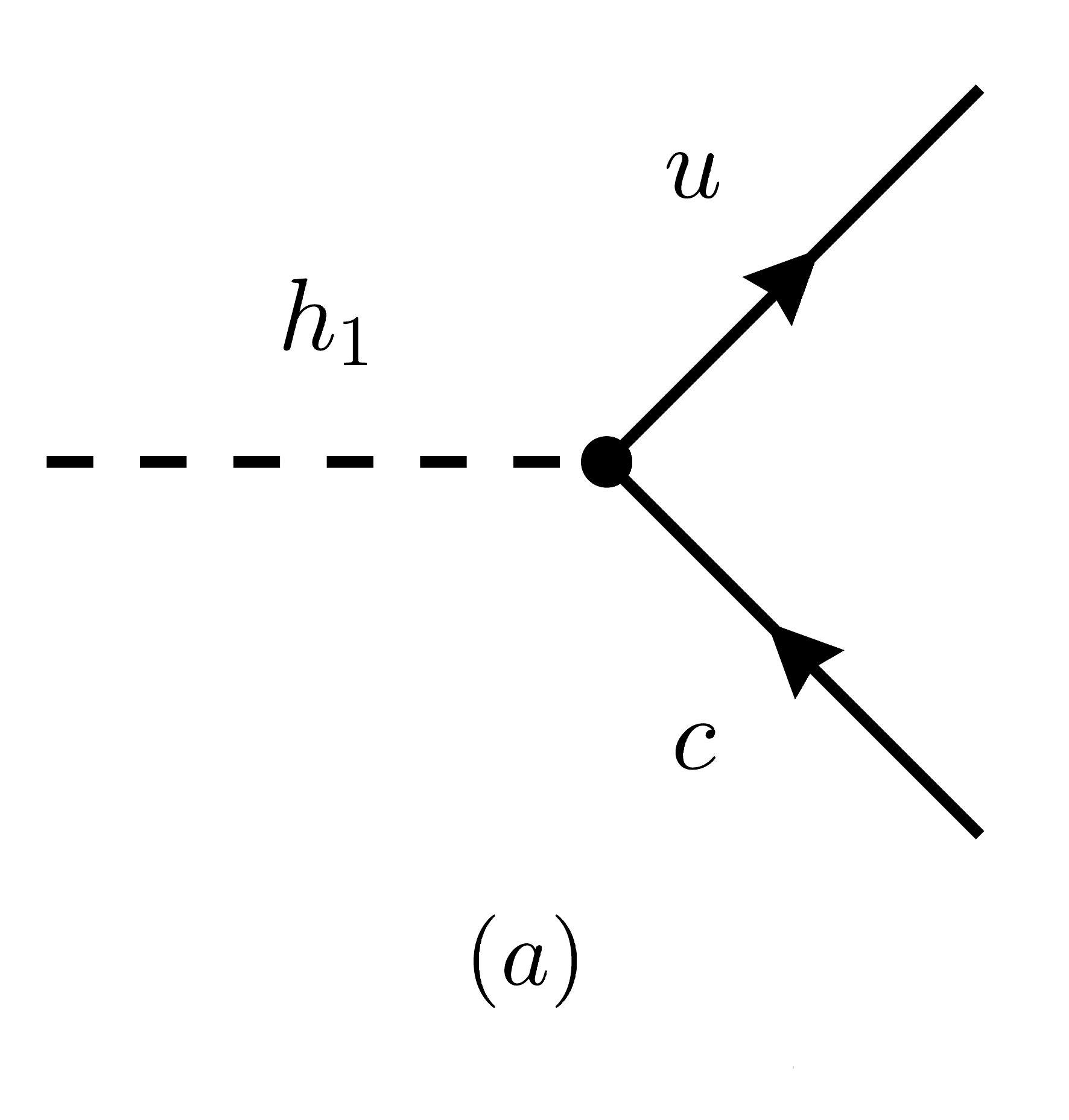}
 \end{minipage}

 \par\vspace{1mm}

 \begin{minipage}{0.32\textwidth}
  \centering
  \includegraphics[width=0.65\linewidth,keepaspectratio]{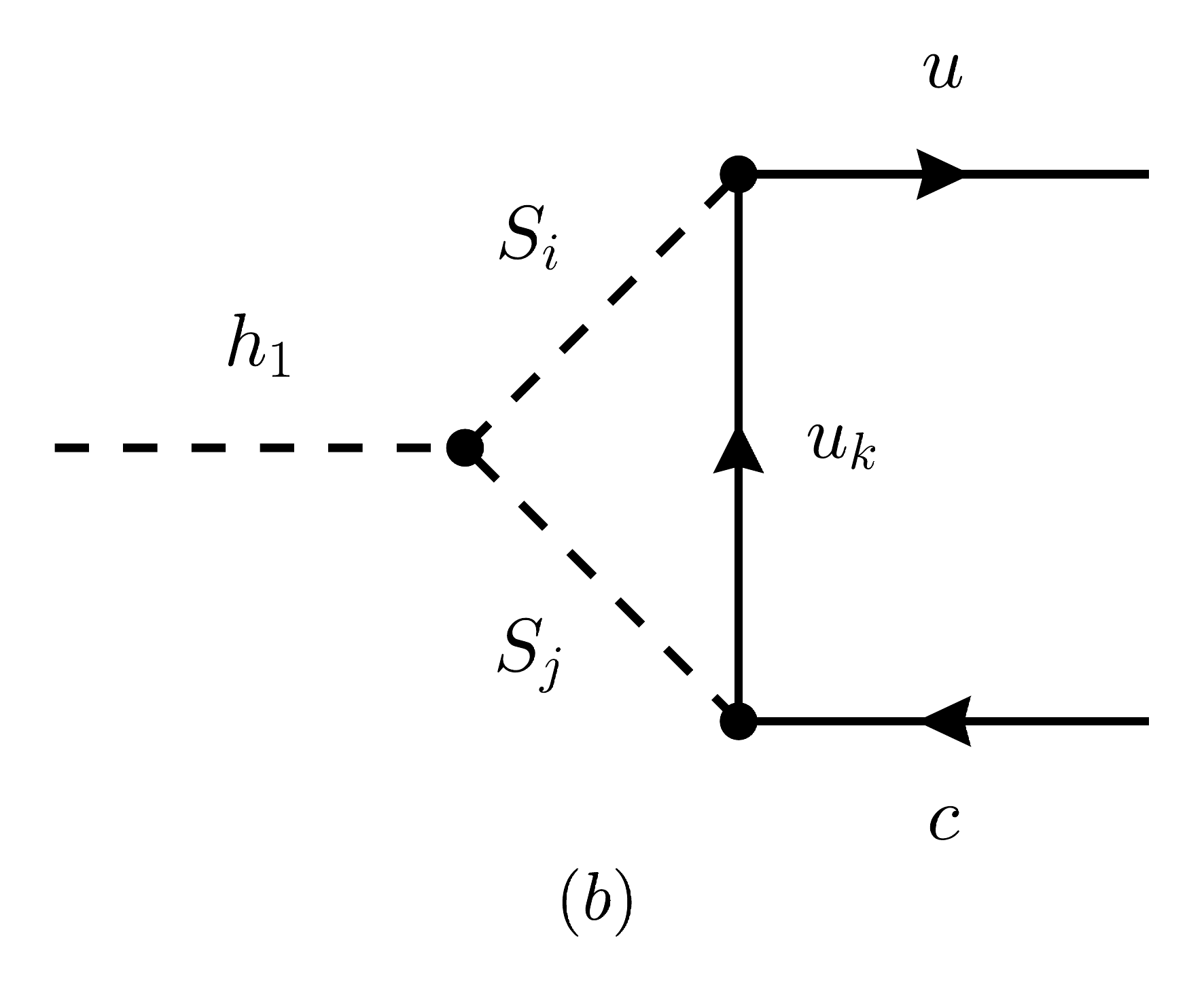}
 \end{minipage}\hfill
 \begin{minipage}{0.32\textwidth}
  \centering
  \includegraphics[width=0.65\linewidth,keepaspectratio]{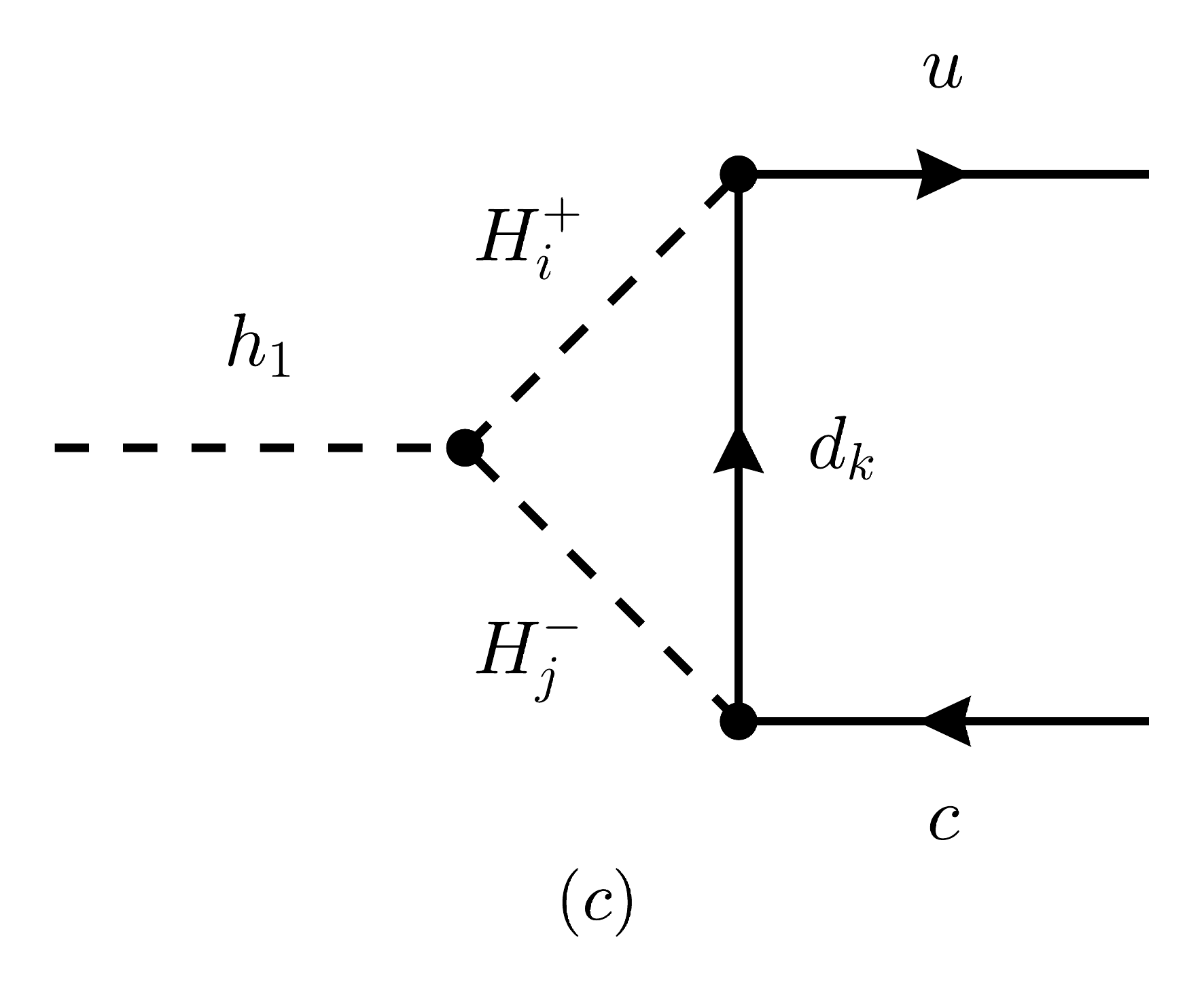}
 \end{minipage}\hfill
 \begin{minipage}{0.32\textwidth}
  \centering
  \includegraphics[width=0.65\linewidth,keepaspectratio]{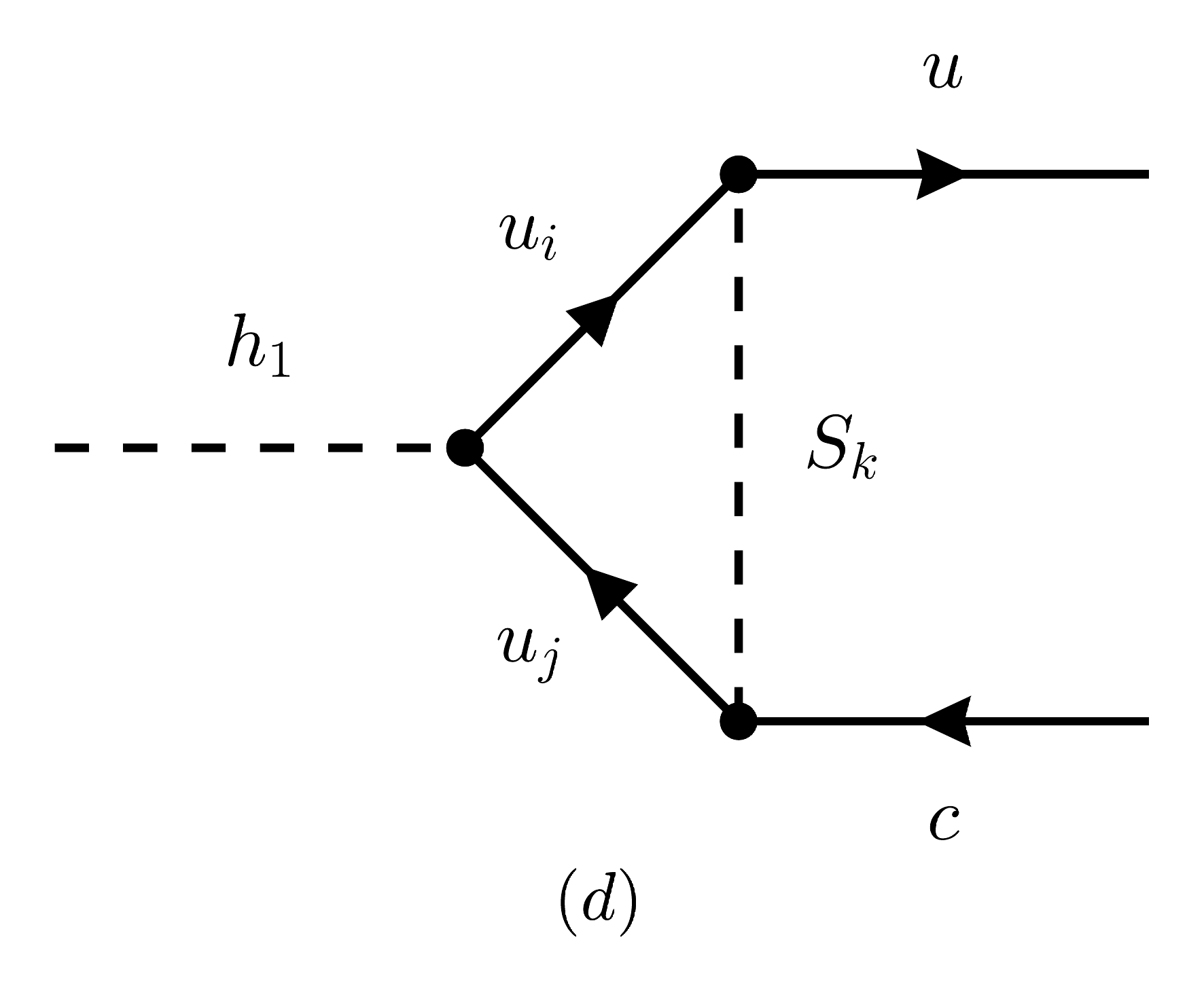}
 \end{minipage}

 \par\vspace{1mm}

 \begin{minipage}{0.32\textwidth}
  \centering
  \includegraphics[width=0.65\linewidth,keepaspectratio]{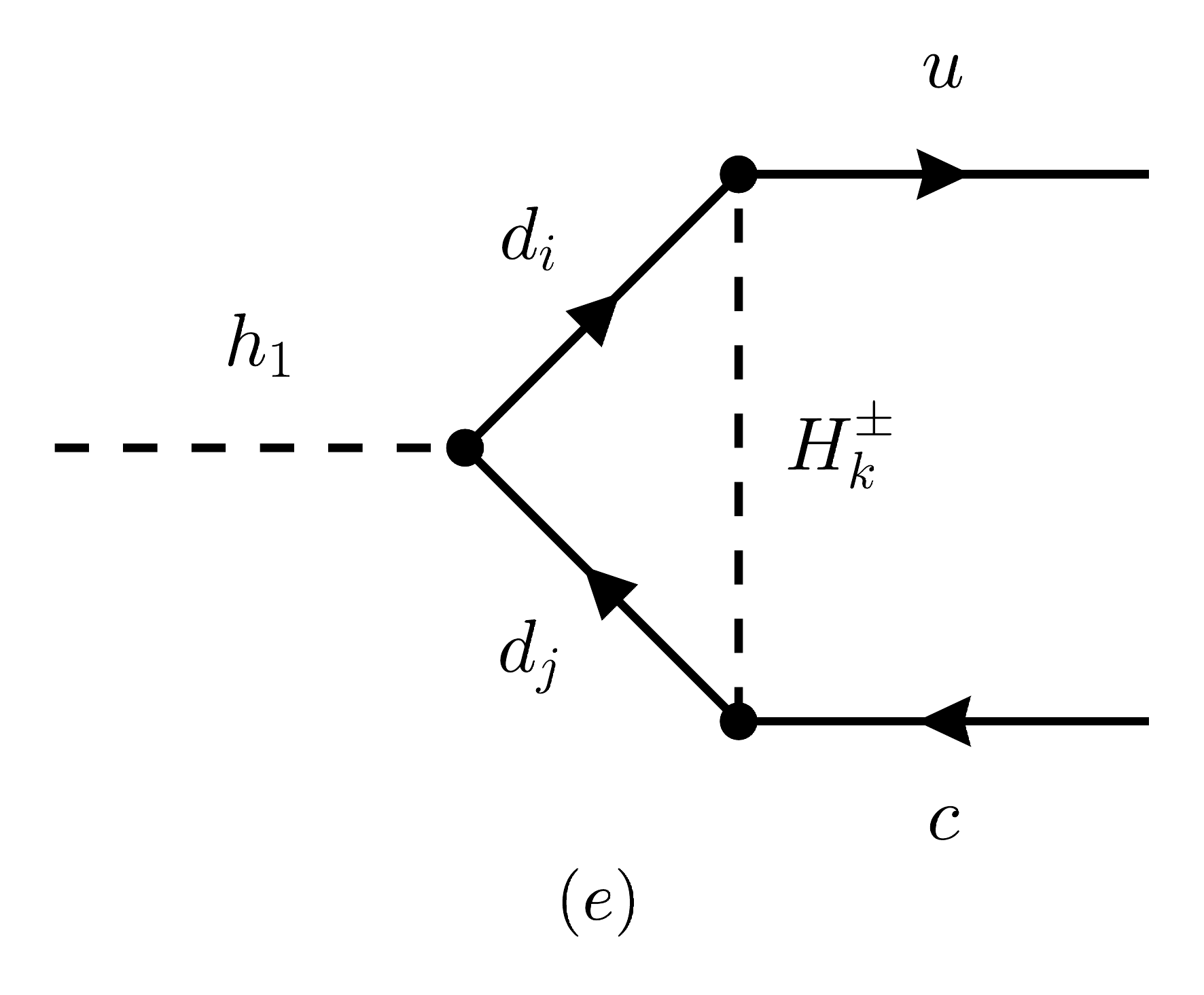}
 \end{minipage}\hfill
 \begin{minipage}{0.32\textwidth}
  \centering
  \includegraphics[width=0.65\linewidth,keepaspectratio]{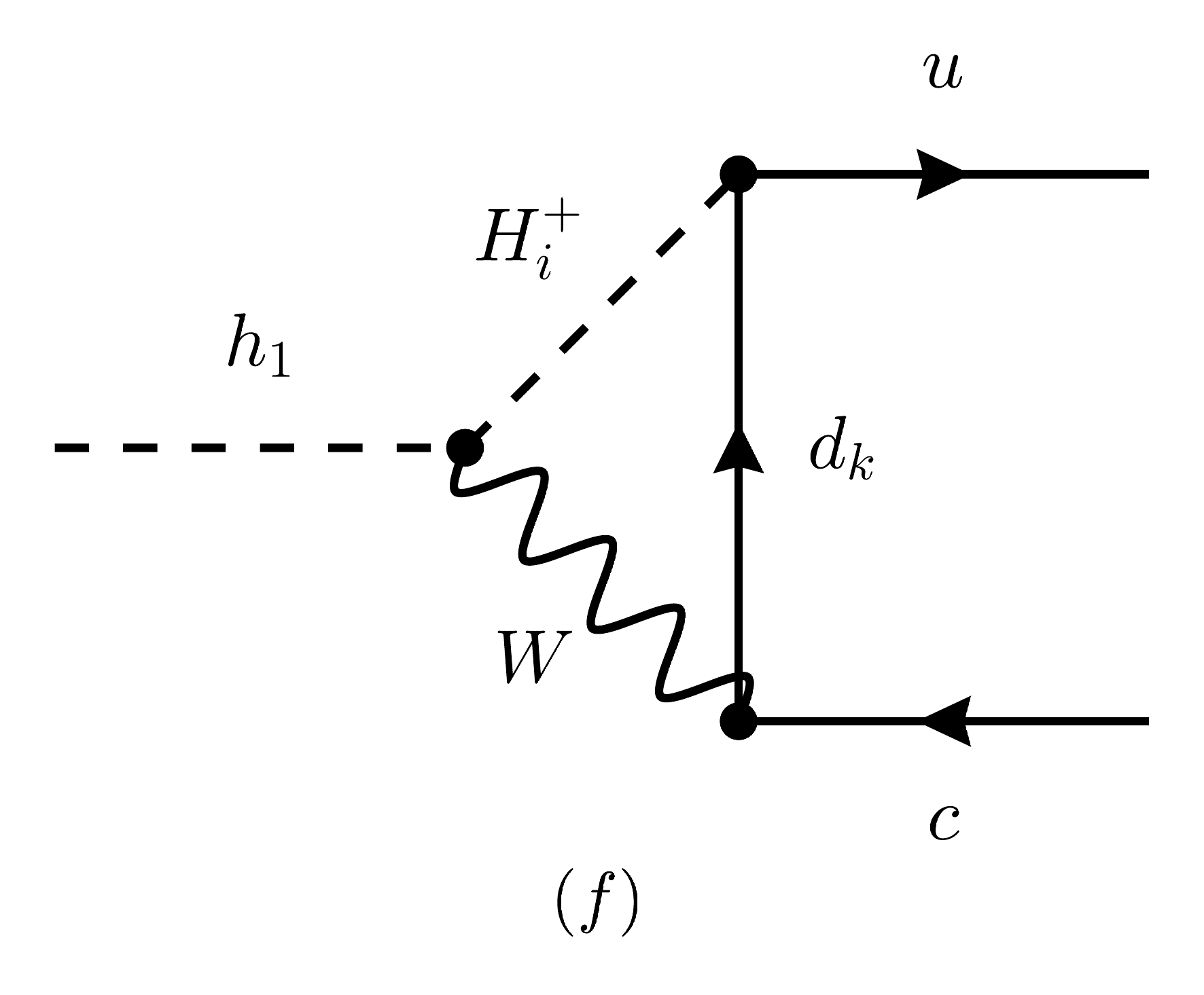}
 \end{minipage}\hfill
 \begin{minipage}{0.32\textwidth}
  \centering
  \includegraphics[width=0.65\linewidth,keepaspectratio]{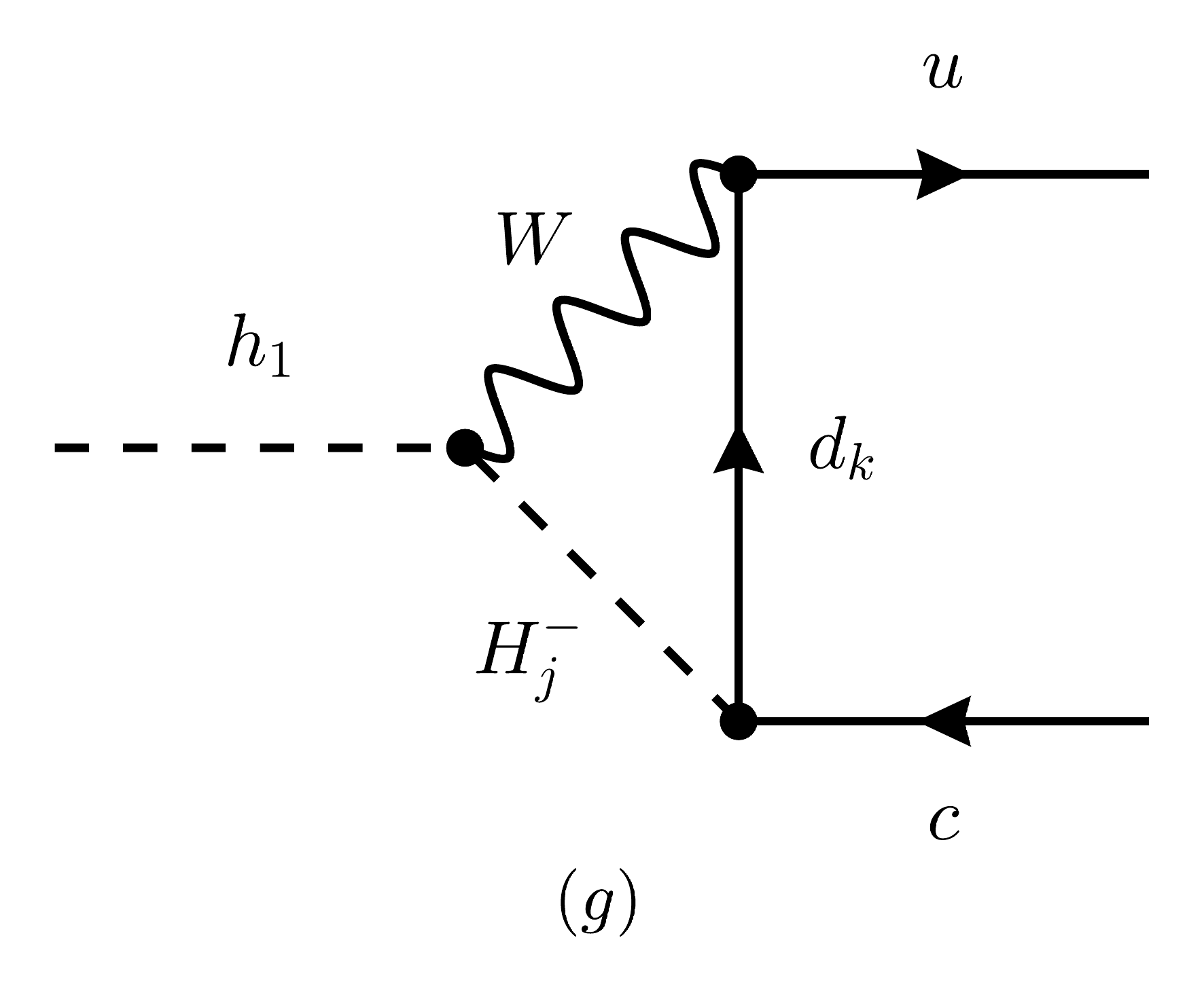}
 \end{minipage}

 \caption{Representative tree-level and one-loop diagrams for $h_1\to u\bar c+\bar u c$. Panel (a) is the tree-level FCNC contribution, and panels (b)--(g) are representative one-loop topologies. The notation for the internal neutral and charged scalars is the same as in Fig.~\ref{fig:top_feynman}.}
 \label{fig:huc_feynman}
\end{figure}

Both quark-flavor-changing processes already occur at tree level because the rank-one Yukawa matrices are misaligned in the fermion mass basis.  In Figs.~\ref{fig:top_feynman} and~\ref{fig:huc_feynman}, panel (a) displays this tree-level neutral-Higgs vertex.  The remaining panels illustrate the neutral-scalar/up-quark and charged-scalar/down-quark loop classes, including diagrams involving a $W$ boson.  They are representative rather than an exhaustive listing; all relevant diagrams and interference terms are retained in the numerical amplitudes.

\FloatBarrier
We next specify the scanned parameter ranges and the theoretical and experimental selections imposed on the numerical sample.

\subsection{Input Parameters, Scan, and Constraints}

Instead of scanning the three soft masses directly, we use the two physical charged-Higgs masses and the charged-sector angle $\gamma_\pm$.  Diagonalizing Eq.~\eqref{eq:Mcharged} gives
\begin{align}
 m_{12}^2={}&\frac12\lambda_7v_1v_2
 +\frac{v_1v_2}{v_1^2+v_2^2}
 (c_\gamma^2M_{H_{\max}^\pm}^2+s_\gamma^2M_{H_{\min}^\pm}^2)
 \nonumber\\
 &+\frac{v_3(v_1^2-v_2^2)}{v(v_1^2+v_2^2)}c_\gamma s_\gamma
 (M_{H_{\max}^\pm}^2-M_{H_{\min}^\pm}^2)
 \nonumber\\
 &-\frac{v_1v_2v_3^2}{v^2(v_1^2+v_2^2)}
 (c_\gamma^2M_{H_{\min}^\pm}^2+s_\gamma^2M_{H_{\max}^\pm}^2),
 \label{eq:m12_input}
\end{align}

\begin{align}
 m_{13}^2={}&\frac12\lambda_8v_1v_3
 +\frac{v_1v_3}{v^2}
 (c_\gamma^2M_{H_{\min}^\pm}^2+s_\gamma^2M_{H_{\max}^\pm}^2)
 \nonumber\\
 &+\frac{v_2}{v}c_\gamma s_\gamma
 (M_{H_{\max}^\pm}^2-M_{H_{\min}^\pm}^2),
 \label{eq:m13_input}\\
 m_{23}^2={}&\frac12\lambda_9v_2v_3
 +\frac{v_2v_3}{v^2}
 (c_\gamma^2M_{H_{\min}^\pm}^2+s_\gamma^2M_{H_{\max}^\pm}^2)
 \nonumber\\
 &-\frac{v_1}{v}c_\gamma s_\gamma
 (M_{H_{\max}^\pm}^2-M_{H_{\min}^\pm}^2),
 \label{eq:m23_input}
\end{align}
where $c_\gamma=\cos\gamma_\pm$ and $s_\gamma=\sin\gamma_\pm$.

The scan ranges are shown in Table~\ref{tab:scan}.  The $x_{ij}$ and $y_{ij}$ denote the real fermion-texture inputs in the corresponding sector; their phases $\theta_f$ are set to zero.  The down-sector range is narrower because neutral-meson observables strongly constrain generic down-quark flavor violation.  The scan is CP conserving throughout.

\begin{table}[h!]
\caption{Parameter ranges used in the numerical scan.  Dimensionful inputs are in GeV.  Endpoints that fail the strict BFB inequalities are discarded.}
\label{tab:scan}
\begin{tabular}{lccccccc}
\toprule
 & $\tan\beta$ & $\tan\beta'$ & $\lambda_1$ & $\lambda_2$ & $\lambda_3$ & $\lambda_{i\ne1,2,3}$\\
\midrule
Minimum & 3 & 15 & 0 & 0 & 0 & $0$\\
Maximum & 10 & 25 & 4 & 4 & 1 & 4\\
\midrule
 & $M_{H_{\min}^{\pm}}$ & $M_{H_{\max}^{\pm}}$ & $\gamma_\pm/\pi$ & $x^u_{ij},y^u_{ij}$ & $x^d_{ij},y^d_{ij}$ & $x^\ell_{ij},y^\ell_{ij}$ & $\theta_f/\pi$\\
\midrule
Minimum & 300 & 4000 & $-1$ & $-1.5$ & $-0.1$ & $-1.5$ & 0\\
Maximum & 3000 & 6000 & 1 & 1.5 & 0.1 & 1.5 & 0\\
\bottomrule
\end{tabular}
\end{table}

The scalar-sector selection applies the conservative sufficient BFB conditions in Eq.~\eqref{eq:BFB_conditions} and the necessary perturbative-unitarity preselection in Eq.~\eqref{eq:unitarity_working}, following Ref.~\cite{Ma:2026G3HDMbs}.  The measured $h_1$ mass and the Higgs signal strengths reported by ATLAS and CMS are also required~\cite{ATLAS:2022vkf,CMS:2022dwd}.  The signal strengths used in the selection are
\begin{align}
 &\mu_{\gamma\gamma}=1.10\pm0.06,
 &&\mu_{WW^*}=1.00\pm0.08,
 \nonumber\\
 &\mu_{ZZ^*}=1.02\pm0.08,
 &&\mu_{b\bar b}=0.99\pm0.12,
 \nonumber\\
 &\mu_{\tau^+\tau^-}=0.91\pm0.09.
 \label{eq:signal_strengths}
\end{align}
We also impose the measured rare-$B$-decay constraints~\cite{Altmannshofer:2012az,Altmannshofer:2017wqy,Aebischer:2019mlg,Altmannshofer:2021qrr,CMS:2022mgd,LHCb:2021moh,LHCb:2017rmj},
\begin{align}
 \Br(\overline B\to X_s\gamma)&=(3.49\pm0.19)\times10^{-4},
 \nonumber\\
 \Br(B_s^0\to\mu^+\mu^-)&=(3.01\pm0.35)\times10^{-9},
 \label{eq:B_constraints}
\end{align}
For neutral-meson mixing, we define $C_{B_q}=\Delta M_{B_q}/\Delta M_{B_q}^{\rm SM}$, $C_{\Delta M_K}=\Delta M_K/\Delta M_K^{\rm SM}$, and $C_{\epsilon_K}=\epsilon_K/\epsilon_K^{\rm SM}$.  The Summer 2025 UTfit new-physics fit gives $C_{B_d}=1.06\pm0.08$, $C_{B_s}=1.10\pm0.06$, and $C_{\epsilon_K}=1.05\pm0.10$~\cite{UTfit:2025NP}.  We therefore impose their $2\sigma$ intervals,
\begin{align}
 0.90<&\frac{\Delta M_{B_d}}{\Delta M_{B_d}^{\rm SM}}<1.22,&
 0.98<&\frac{\Delta M_{B_s}}{\Delta M_{B_s}^{\rm SM}}<1.22,
 \nonumber\\
 0<&\frac{\Delta M_K}{\Delta M_K^{\rm SM}}<2.0,&
 0.85<&\frac{\epsilon_K}{\epsilon_K^{\rm SM}}<1.25.
 \label{eq:mixing_windows}
\end{align}
The $B_d$, $B_s$, and $\epsilon_K$ windows are the UTfit 2025 $2\sigma$ ranges.  For $C_{\Delta M_K}$, for which the same fit table quotes no corresponding interval, we adopt the conservative range $0<C_{\Delta M_K}<2$.  We follow the standard weak-effective-Hamiltonian treatment~\cite{Buchalla:1995vs,Buras:2010mh,Buras:2012fs,Buras:2013rqa,Aebischer:2020dsw,Lenz:2020awd}.

\Needspace{9\baselineskip}
The electroweak oblique parameters are required to agree with the three-parameter electroweak fit quoted by the Particle Data Group~\cite{ParticleDataGroup:2024cfk},
\begin{equation}
 S=-0.04\pm0.10,\qquad T=0.01\pm0.12,\qquad
 U=-0.01\pm0.11.
 \label{eq:STU}
\end{equation}
The quoted one-dimensional $S$, $T$, and $U$ ranges are applied independently at the $1\sigma$ level.  Multi-doublet contributions are evaluated with the standard oblique-parameter expressions of Ref.~\cite{Grimus:2007if}.  Other experimental averages, lattice inputs, and SM parameters are taken consistently from Refs.~\cite{HeavyFlavorAveragingGroup:2022wzx,FlavourLatticeAveragingGroupFLAG:2021npn,FermilabLattice:2016ipl,ETM:2013jap}.  After all selections, 4839 parameter points remain.  Correlations are quantified using the Spearman rank coefficient $\rho_S$, which is sensitive to monotonic relations without assuming linearity.  For log-log comparisons we also quote the Pearson coefficient $r_{\log}$.

\section{Numerical Analysis}

\subsection{Global pattern of flavor-violating decays}

\begin{figure}[htbp]
 \centering
 \includegraphics[width=0.88\textwidth]{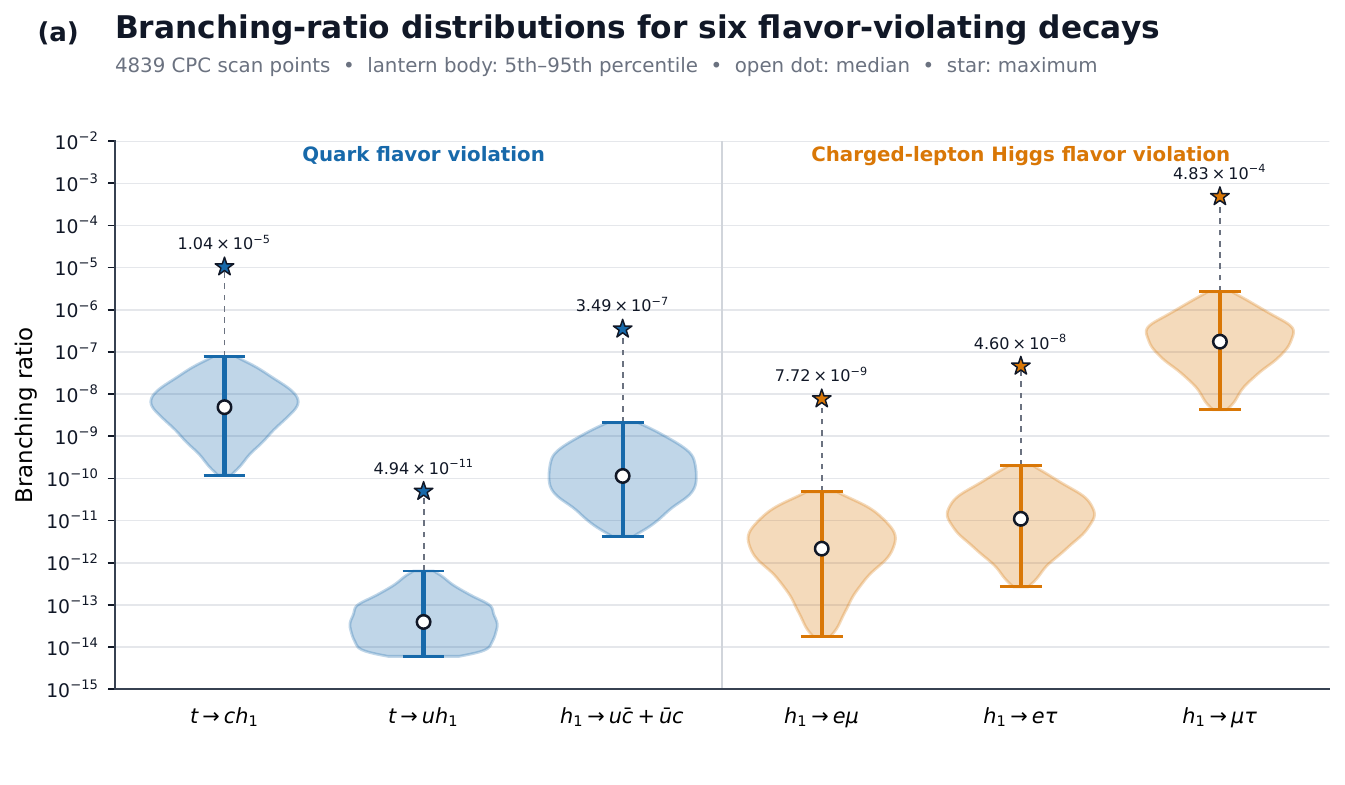}\\[4mm]
 \includegraphics[width=0.88\textwidth]{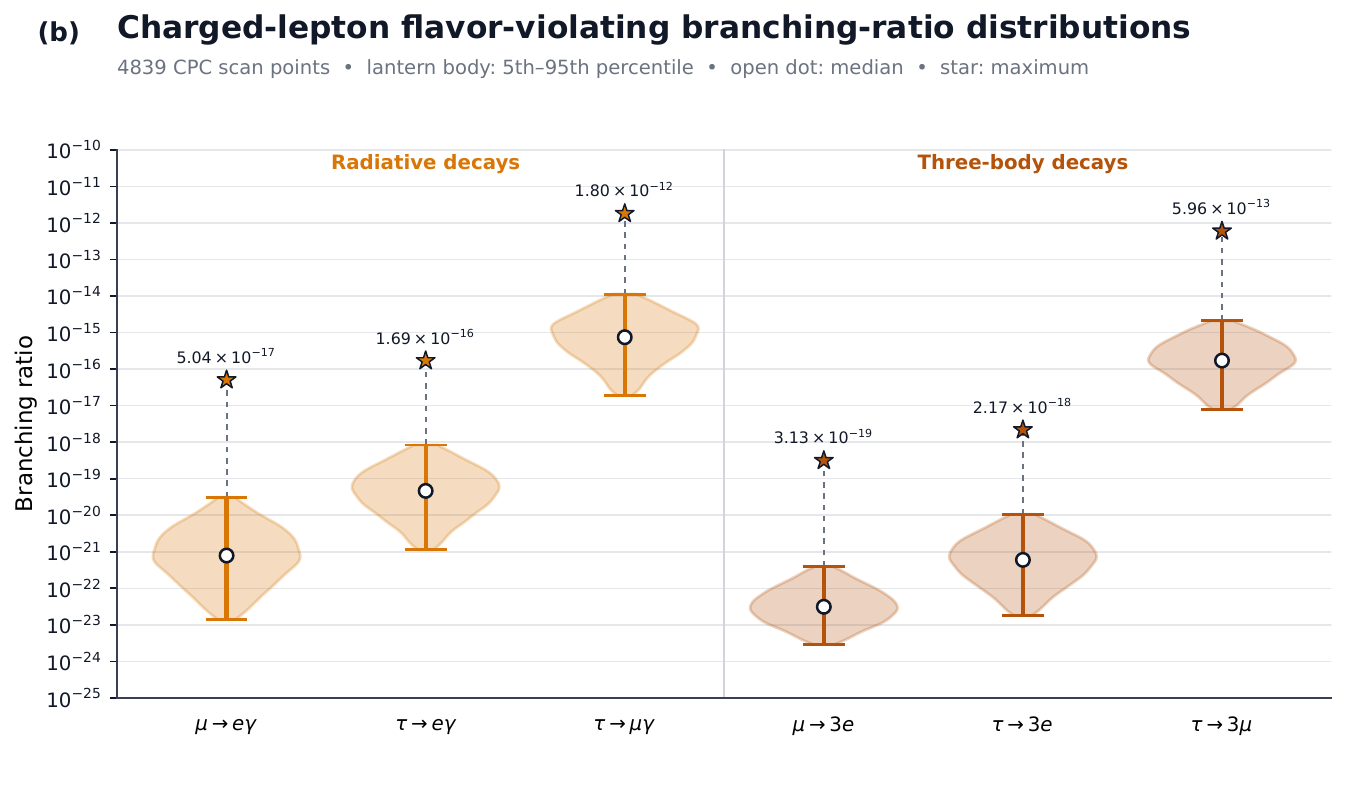}
 \caption{Branching-ratio distributions for (a) flavor-violating top and $h_1$ decays and (b) radiative and three-body charged-lepton flavor-violating decays.  The lantern body contains the 5th--95th percentile range, the open circle is the median, and the star is the maximum among the 4839 accepted CP-conserving points.}
 \label{fig:lanterns}
\end{figure}

Figure~\ref{fig:lanterns} summarizes the branching-ratio distributions.  Each lantern shows the 5th--95th percentile interval, the open circle is the median, and the star is the maximum of the accepted sample.  Table~\ref{tab:maxima_limits} compares the scan maxima with representative direct experimental upper limits.

\begin{table}[t]
\centering
\caption{Maximal flavor-violating rates in the accepted G3HDM sample and representative experimental bounds or projected benchmarks.  Collider limits are quoted at 95\% C.L. and charged-lepton limits at 90\% C.L.  The top limits are the ATLAS combination, while the Higgs LFV limits are from ATLAS and CMS~\cite{ATLAS:2024mih,ATLAS:2023mvd,CMS:2023pte}.  Low-energy decay limits are from MEG~II and the PDG/HFLAV summaries~\cite{MEGII:2025gzr,ParticleDataGroup:2024cfk,HeavyFlavorAveragingGroup:2022wzx}.  The Au entry is the observed SINDRUM-II bound, whereas the Al entry is the projected Mu2e Run-I 90\% C.L. upper limit~\cite{SINDRUMII:2006,Mu2e:2023RunI}.}
\label{tab:maxima_limits}
\begin{tabular}{lcc}
\toprule
Process & G3HDM maximum & Experimental benchmark\\
\midrule
$t\to uh_1$ & $4.94\times10^{-11}$ & $2.6\times10^{-4}$\\
$t\to ch_1$ & $1.04\times10^{-5}$ & $3.4\times10^{-4}$\\
$h_1\to u\bar c+\bar u c$ & $3.49\times10^{-7}$ & ---\\
$h_1\to e\mu$ & $7.72\times10^{-9}$ & $4.4\times10^{-5}$\\
$h_1\to e\tau$ & $4.60\times10^{-8}$ & $2.0\times10^{-3}$\\
$h_1\to\mu\tau$ & $4.83\times10^{-4}$ & $1.8\times10^{-3}$\\
\midrule
$\mu\to e\gamma$ & $5.04\times10^{-17}$ & $1.5\times10^{-13}$\\
$\tau\to e\gamma$ & $1.69\times10^{-16}$ & $3.3\times10^{-8}$\\
$\tau\to\mu\gamma$ & $1.80\times10^{-12}$ & $4.4\times10^{-8}$\\
$\mu\to3e$ & $3.13\times10^{-19}$ & $1.0\times10^{-12}$\\
$\tau\to3e$ & $2.17\times10^{-18}$ & $2.7\times10^{-8}$\\
$\tau\to3\mu$ & $5.96\times10^{-13}$ & $2.1\times10^{-8}$\\
\midrule
$\mathrm{CR}(\mu\mathrm{Al}\to e\mathrm{Al})$ & $4.9503\times10^{-15}$ & $6.2\times10^{-16}$ (projected)\\
$\mathrm{CR}(\mu\mathrm{Au}\to e\mathrm{Au})$ & $1.0479\times10^{-14}$ & $7.0\times10^{-13}$\\
\bottomrule
\end{tabular}
\end{table}

The table reports no experimental benchmark for $h_1\to u\bar c+\bar u c$ because no model-independent direct upper limit is currently available.  For conversion, the Au maximum is about a factor of 67 below the current SINDRUM-II limit $7\times10^{-13}$~\cite{SINDRUMII:2006}.  The Al maximum is about eight times the projected Mu2e Run-I 90\% C.L. upper-limit sensitivity $6.2\times10^{-16}$~\cite{Mu2e:2023RunI}.  It also lies between the COMET Phase-I single-event sensitivity $3.1\times10^{-15}$ and its projected 90\% C.L. upper-limit benchmark $7\times10^{-15}$~\cite{COMET:2020PhaseI}.  Thus the upper tail of the accepted Al distribution is directly testable by the next generation of conversion experiments.

The hierarchy among the predicted collider modes follows directly from the mass-basis matrices in Eqs.~\eqref{eq:mu12} and~\eqref{eq:mu3}.  In particular,
\begin{align}
 &({\cal Y}^{u,h_1})_{ct}\simeq m_c y_{ct}D_{23},\qquad
 ({\cal Y}^{u,h_1})_{tc}\simeq m_c y_{tc}D_{23},
 \nonumber\\
 &({\cal Y}^{u,h_1})_{ut},\;({\cal Y}^{u,h_1})_{tu},\;
 ({\cal Y}^{u,h_1})_{uc},\;({\cal Y}^{u,h_1})_{cu}={\cal O}(m_u),
 \label{eq:quark_hierarchy}
\end{align}
up to texture parameters and the appropriate scalar-mixing combinations.  Thus the first-generation up-quark amplitudes are suppressed relative to $t\to ch_1$ by approximately $m_u/m_c$ at amplitude level.  The different parent widths and phase spaces explain why the numerical ratios of branching fractions do not equal the simple squared mass ratio exactly.  The charged-lepton replacement rule gives
\begin{equation}
 ({\cal Y}^{\ell,h_1})_{\mu\tau},({\cal Y}^{\ell,h_1})_{\tau\mu}={\cal O}(m_\mu),
 \qquad
 ({\cal Y}^{\ell,h_1})_{e\tau},({\cal Y}^{\ell,h_1})_{\tau e},
 ({\cal Y}^{\ell,h_1})_{e\mu},({\cal Y}^{\ell,h_1})_{\mu e}={\cal O}(m_e).
 \label{eq:lepton_hierarchy}
\end{equation}
This $m_e/m_\mu$ amplitude suppression, together with the independent first-generation texture factors, explains why $h_1\to\mu\tau$ is much larger than $h_1\to e\tau$ and $h_1\to e\mu$ throughout the accepted sample.

As Table~\ref{tab:maxima_limits} makes explicit, the radiative and three-body charged-lepton modes remain several orders of magnitude below their present bounds.  The upper tail of $h_1\to\mu\tau$ is much closer to collider reach than the first-generation modes and is therefore a particularly useful test of the model.  The top decay is strongly enhanced relative to the minute SM expectation, although its accepted values remain below current direct sensitivity.

\subsection{Parameter dependence of the dominant channels}

\begin{figure}[htbp]
 \centering
 \begin{minipage}{0.48\textwidth}
  \centering
  \includegraphics[width=\linewidth]{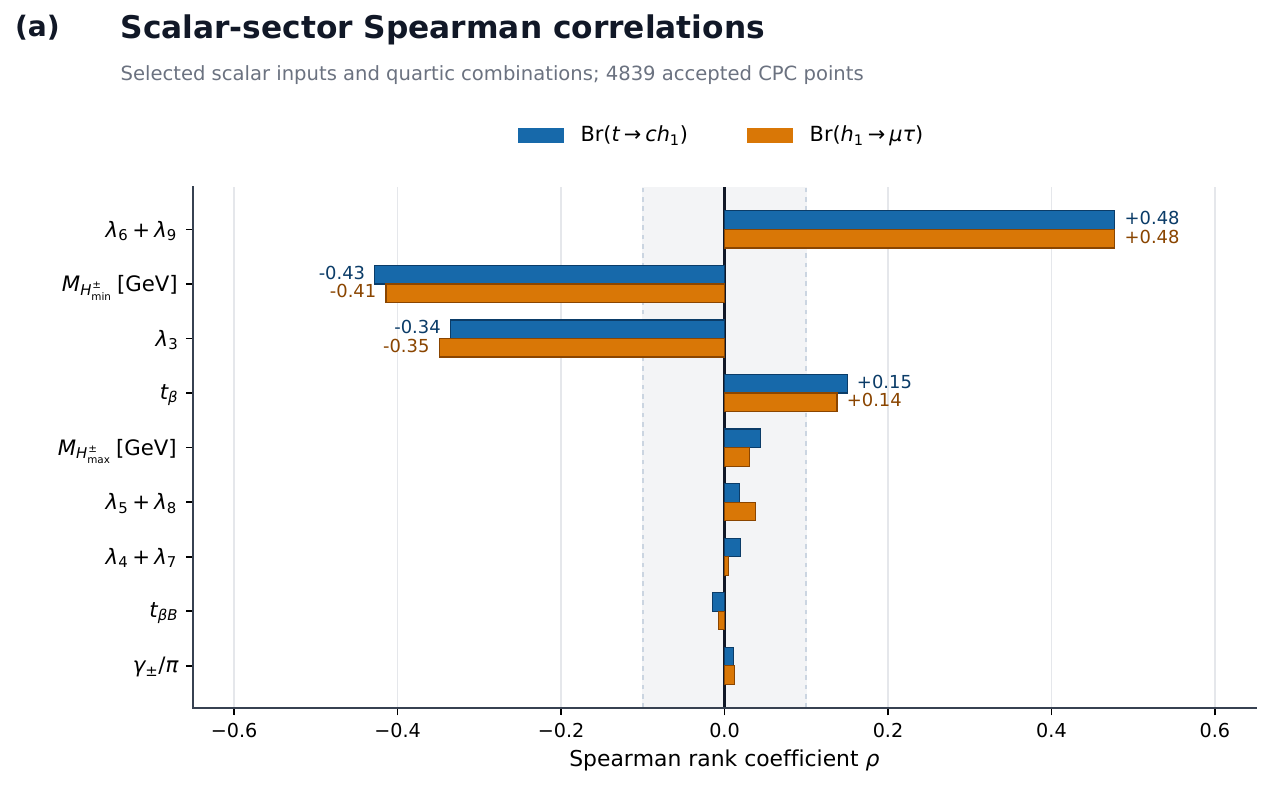}
 \end{minipage}\hfill
 \begin{minipage}{0.48\textwidth}
  \centering
  \includegraphics[width=\linewidth]{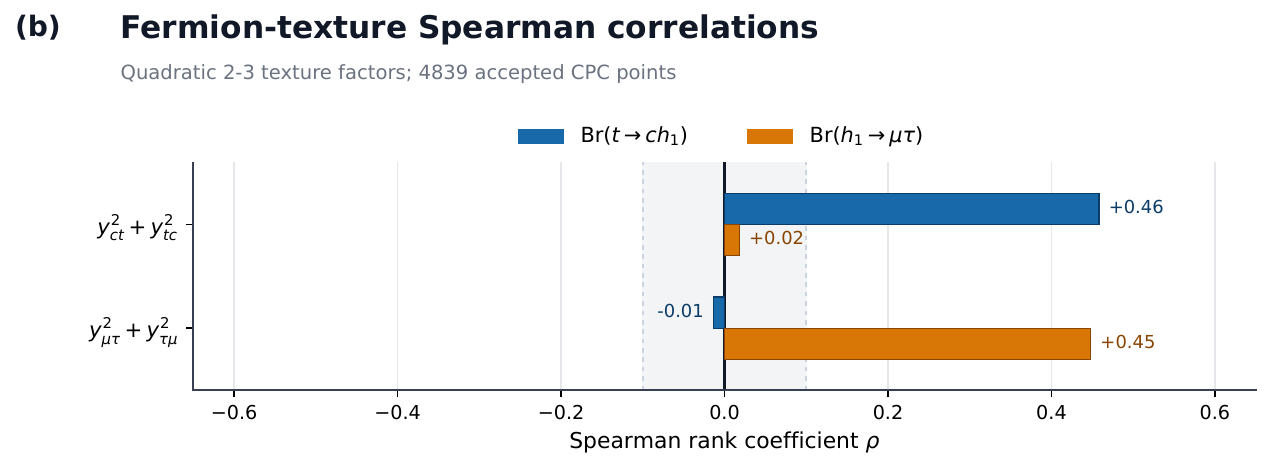}
 \end{minipage}
 \caption{Spearman correlations of the two dominant branching ratios with (a) selected scalar inputs and quartic combinations and (b) the quadratic $2$--$3$ fermion-texture factors. The coefficients are marginal correlations over the accepted sample.}
 \label{fig:spearman}
\end{figure}

The marginal Spearman coefficients are displayed in Fig.~\ref{fig:spearman}.  The light charged-Higgs mass is the most important scanned scalar input, with
\begin{align}
 \rho_S[M_{H_{\min}^{\pm}},\Br(t\to ch_1)]&=-0.428,
 \nonumber\\
 \rho_S[M_{H_{\min}^{\pm}},\Br(h_1\to\mu\tau)]&=-0.414.
 \label{eq:mass_correlations}
\end{align}
For interpreting this numerical trend, we use the leading scalar-misalignment proxy
\begin{equation}
 \Delta_{23}^{(0)}=\frac{v\,|2\lambda_3-\lambda_6-\lambda_9|}
 {M_{H_{\min}^{\pm}}^2}.
 \label{eq:delta_proxy}
\end{equation}
The quartic combination $\lambda_6+\lambda_9$ is positively correlated with both rates, with $\rho_S=0.478$, whereas $\lambda_3$ gives $\rho_S=-0.335$ for $t\to ch_1$ and $-0.349$ for $h_1\to\mu\tau$.  All other displayed scalar inputs have small marginal correlations.  These observations are consistent with the leading numerator $2\lambda_3-\lambda_6-\lambda_9$ and the inverse-mass-squared dependence in Eq.~\eqref{eq:delta_proxy}.

The fermion sector shows the complementary half of the factorized result.  The quark texture $y_{ct}^2+y_{tc}^2$ has $\rho_S=0.458$ with $\Br(t\to ch_1)$ and no appreciable correlation with $h_1\to\mu\tau$.  Conversely, $y_{\mu\tau}^2+y_{\tau\mu}^2$ has $\rho_S=0.448$ with $\Br(h_1\to\mu\tau)$ and essentially none with the top decay.  Quadratic combinations are used because the rates depend on coupling moduli and because signed linear inputs would suffer artificial cancellations.

\begin{figure}[p]
 \centering

 \includegraphics[width=0.4\textwidth]{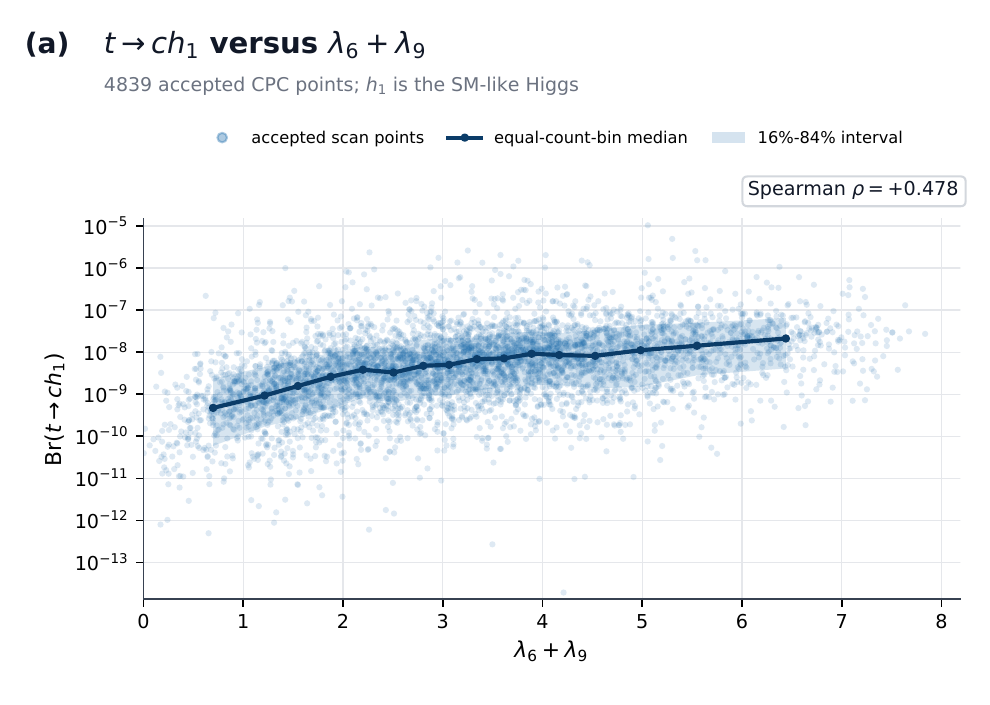} \hspace{5mm}
 \includegraphics[width=0.4\textwidth]{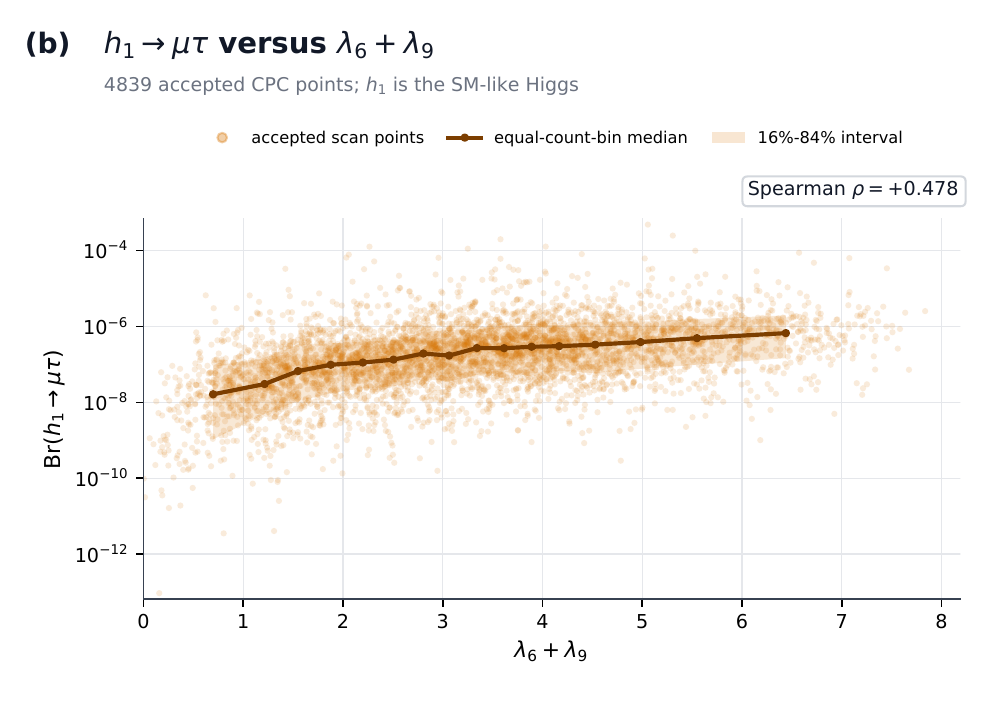}\\[3mm]

 \includegraphics[width=0.4\textwidth]{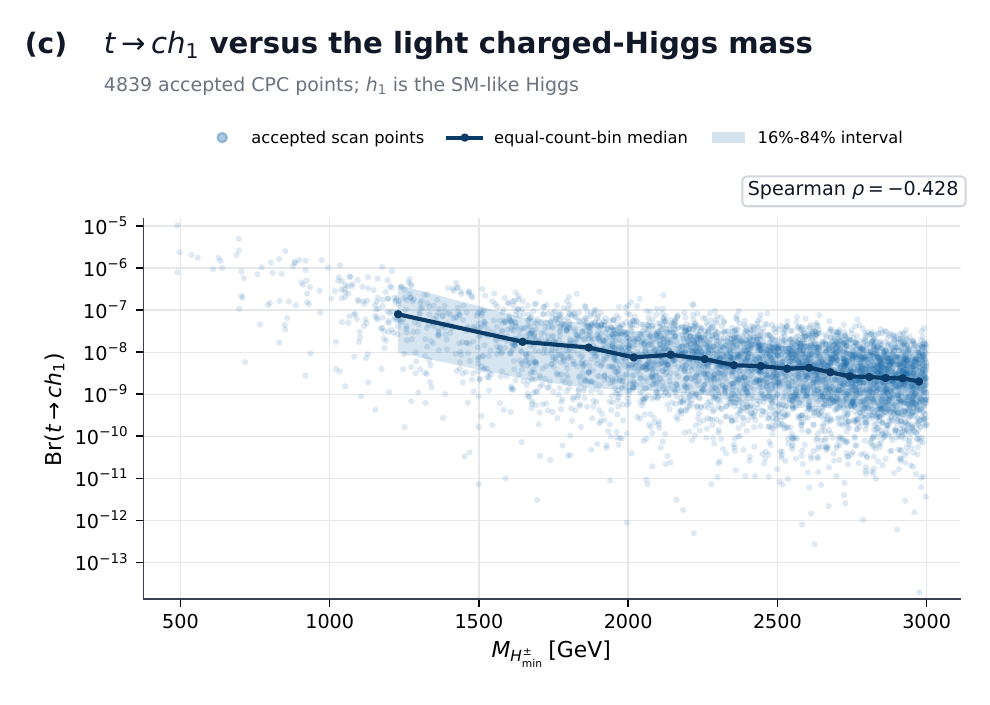} \hspace{5mm}
 \includegraphics[width=0.4\textwidth]{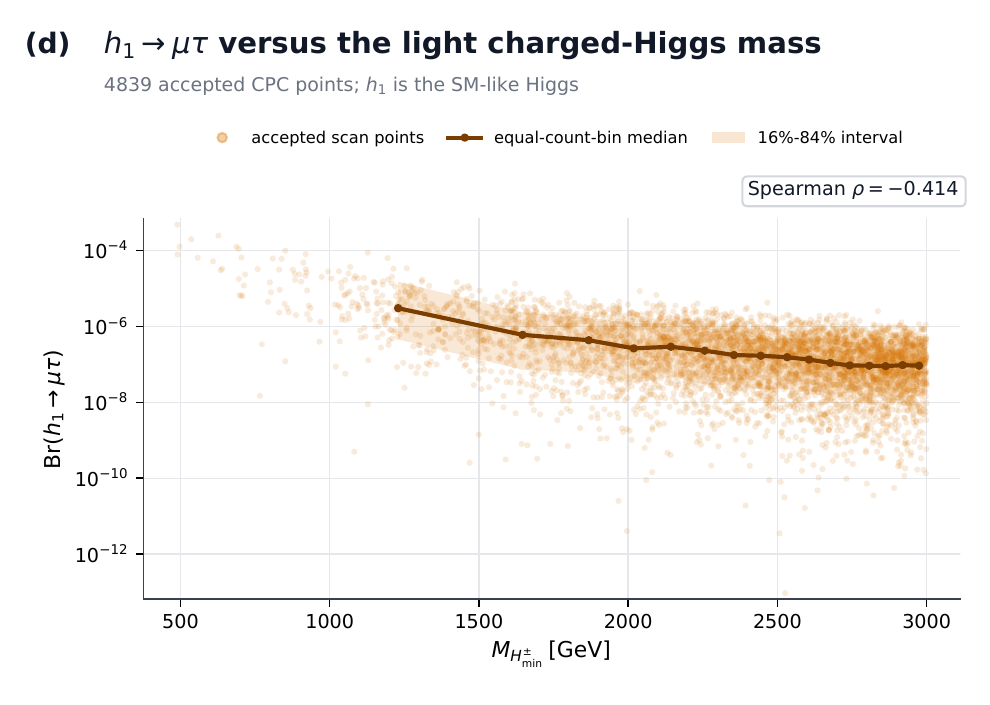}\\[3mm]

 \includegraphics[width=0.4\textwidth]{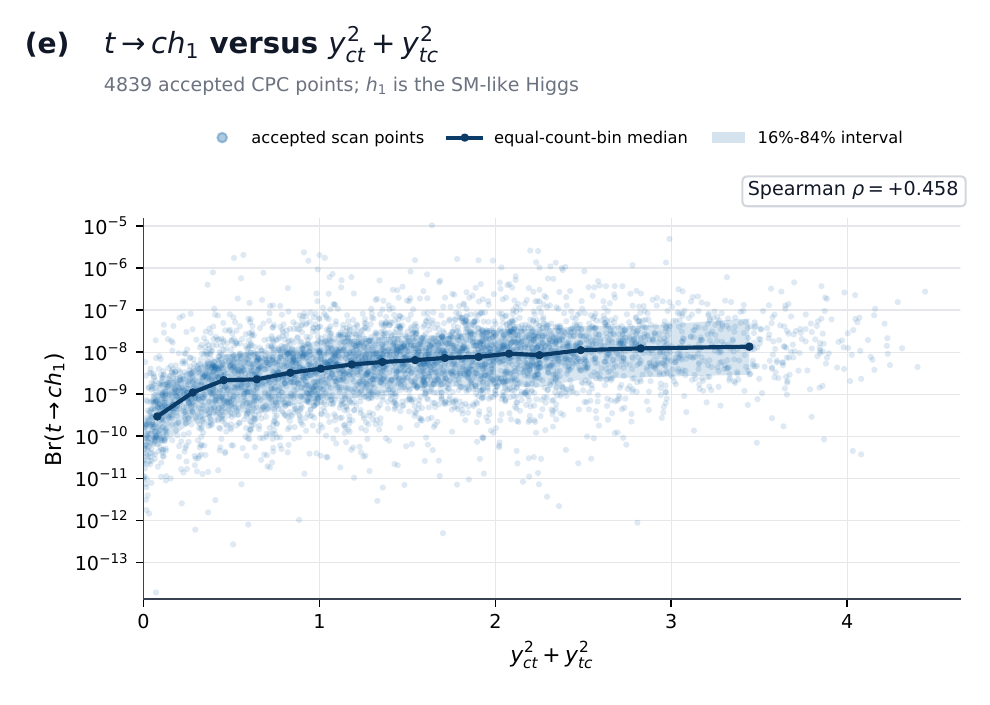} \hspace{5mm}
 \includegraphics[width=0.4\textwidth]{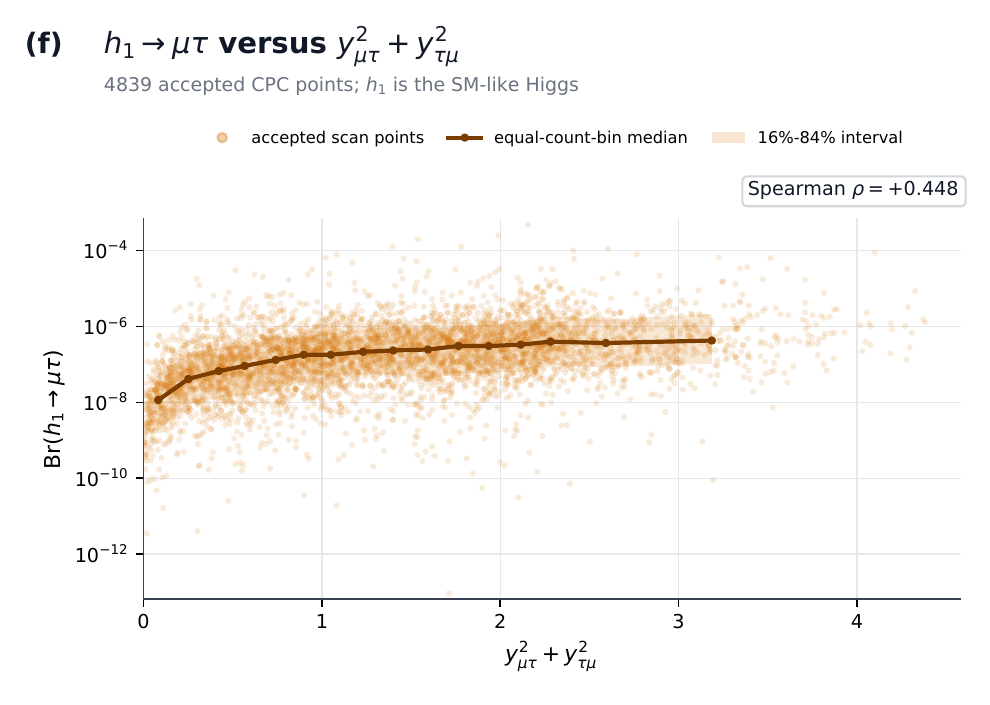}\\[3mm]

 \includegraphics[width=0.4\textwidth]{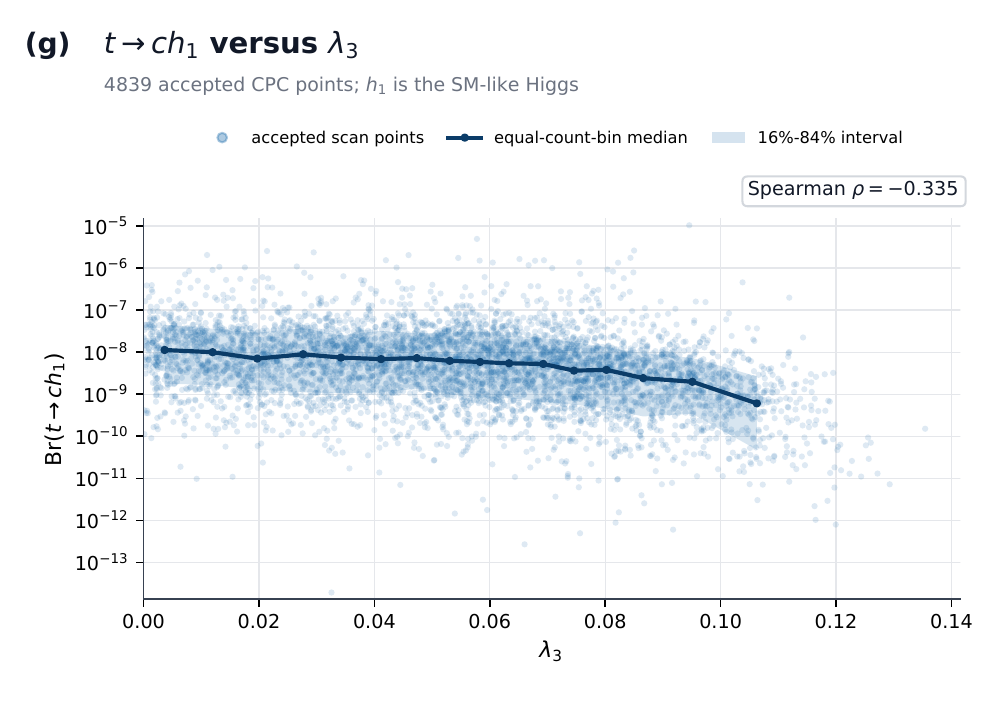} \hspace{5mm}
 \includegraphics[width=0.4\textwidth]{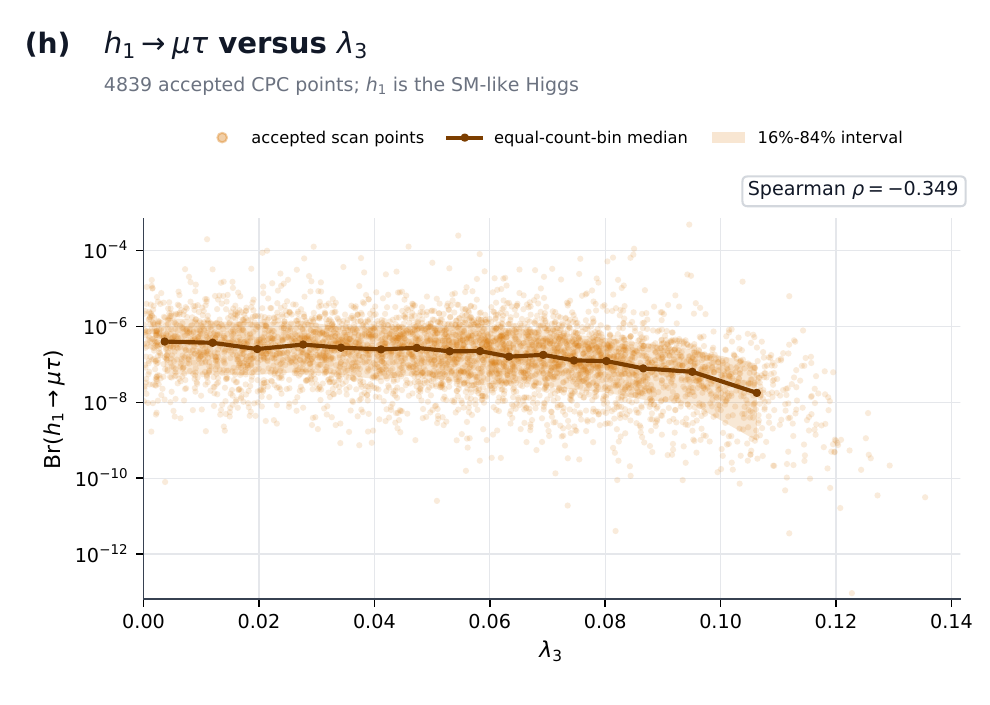}

 \caption{Dependence of $\Br(t\to ch_1)$ and $\Br(h_1\to\mu\tau)$ on the scalar-sector and fermion-texture parameters. Panels (a)--(d) show their dependence on $\lambda_6+\lambda_9$ and the light charged-Higgs mass, while panels (e)--(h) show their dependence on the corresponding quadratic fermion-texture factors and on $\lambda_3$. Solid curves show the medians in equal-count bins, and shaded bands indicate the 16th--84th percentile intervals.}
 \label{fig:scatter_dependence}
\end{figure}

Figures~\ref{fig:scatter_dependence} show the corresponding point distributions.  Both rates rise with $\lambda_6+\lambda_9$ and decrease as either $M_{H_{\min}^{\pm}}$ or $\lambda_3$ increases.  They also grow with their own quadratic texture factor.  The broad vertical spread in every two-parameter projection is expected: a marginal plot displays only one part of the product in Eqs.~\eqref{eq:top_driver} and~\eqref{eq:mutau_driver}.  The binned median nevertheless traces the monotonic trend.

\subsection{Tree-level origin and factorized drivers}

\begin{figure}[htbp]
 \centering
 \includegraphics[width=0.4\textwidth]{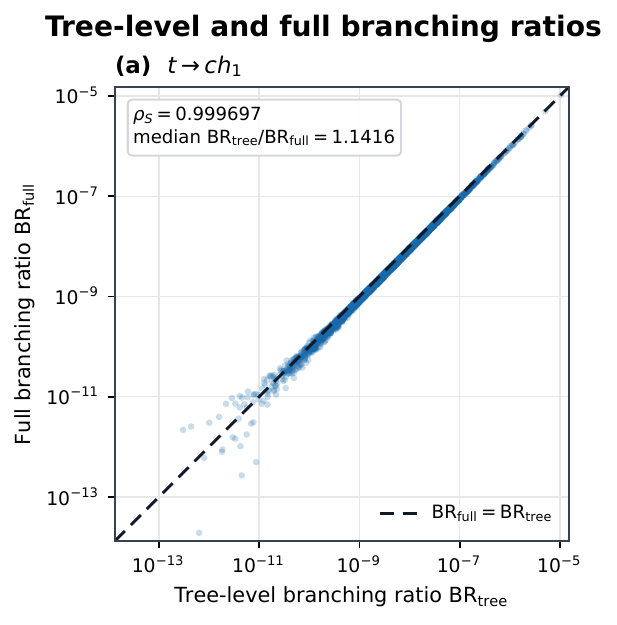} \hspace{5mm}
 \includegraphics[width=0.4\textwidth]{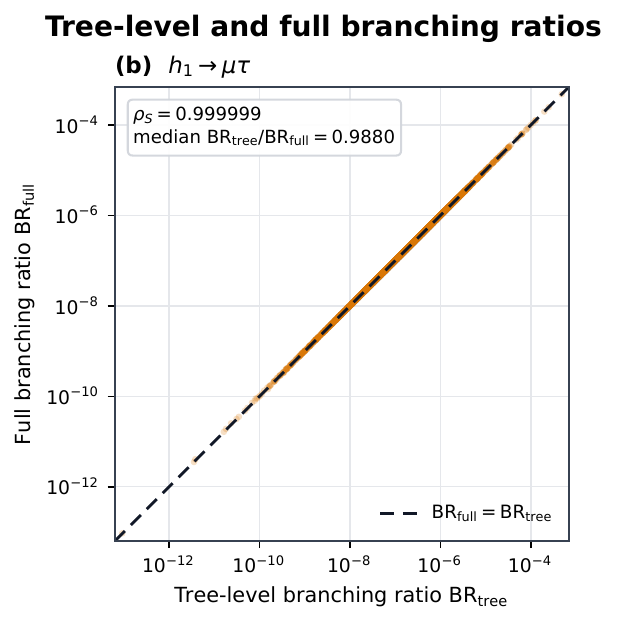}
 \caption{Tree-level branching ratios versus the complete tree plus one-loop predictions for (a) $t\to ch_1$ and (b) $h_1\to\mu\tau$.  The diagonal line denotes equality.}
 \label{fig:treefull}
\end{figure}

Figure~\ref{fig:treefull} compares the tree-only and complete branching ratios.  The Spearman coefficients are $0.999697$ for $t\to ch_1$ and $0.999999$ for $h_1\to\mu\tau$.  The median ratios are
\begin{align}
 \operatorname{median}\frac{\Br_{\rm tree}}{\Br_{\rm full}}
 &=1.1416\quad(t\to ch_1),
 \nonumber\\
 &=0.9880\quad(h_1\to\mu\tau).
 \label{eq:treefull_ratios}
\end{align}

The enhanced rates therefore originate from the tree-level neutral-Higgs Yukawa coupling, not from a loop enhancement.  Loop contributions remain part of the full prediction: they change the median $h_1\to\mu\tau$ result at the percent level and give a modest, visible correction to $t\to ch_1$.

\begin{figure}[htbp]
 \centering
 \includegraphics[width=0.4\textwidth]{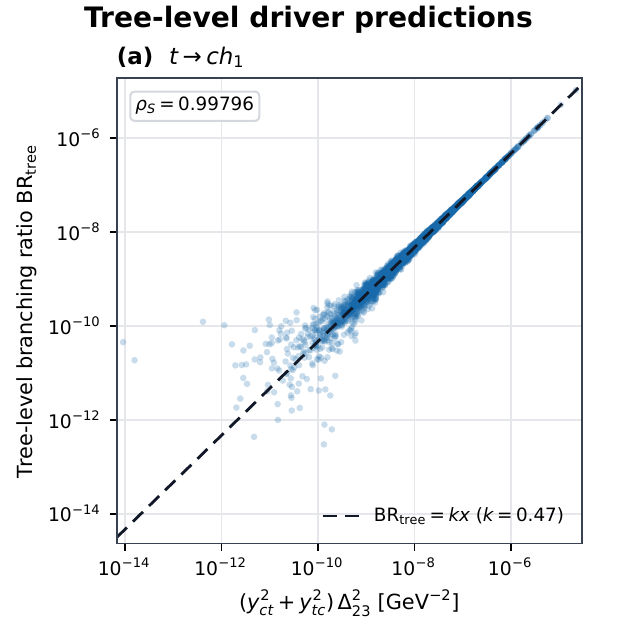} \hspace{5mm}
 \includegraphics[width=0.4\textwidth]{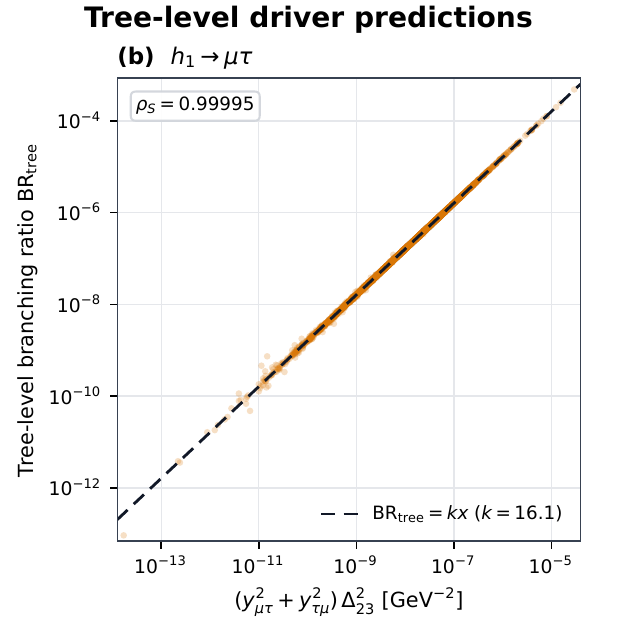}\\[3mm]
 \includegraphics[width=0.4\textwidth]{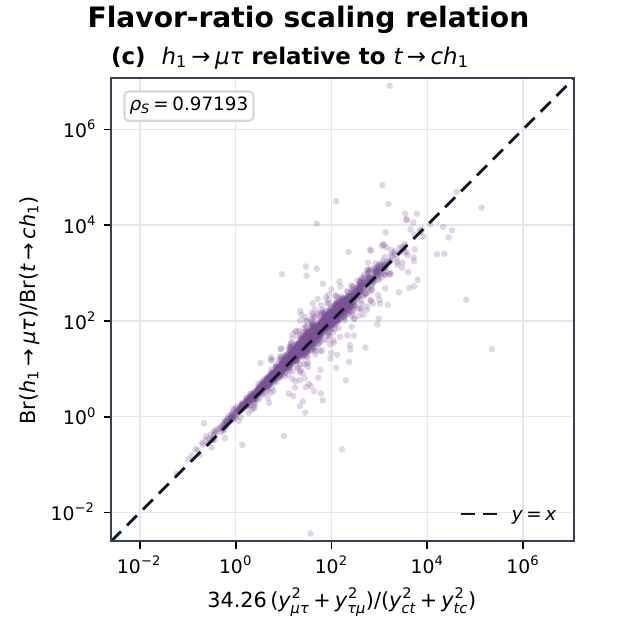}
 \caption{Tests of the factorized analytic drivers.  Panels (a) and (b) show the tree-level branching ratios as functions of $(y_{ct}^2+y_{tc}^2)\Delta_{23}^2$ and $(y_{\mu\tau}^2+y_{\tau\mu}^2)\Delta_{23}^2$, respectively.  Panel (c) tests the flavor-ratio scaling relation after the common scalar factor cancels.  The nearly perfect rank correlations validate Eqs.~\eqref{eq:top_driver} and~\eqref{eq:mutau_driver}.}
 \label{fig:drivers}
\end{figure}

The factorized analytic drivers provide a still sharper test.  As shown in Fig.~\ref{fig:drivers}, the tree-level point-by-point dependence is accurately captured by
\begin{align}
 \Br(t\to ch_1)_{\rm tree}&\simeq
 0.47~\GeV^2\,(y_{ct}^2+y_{tc}^2)\Delta_{23}^2,
 \nonumber\\
 \Br(h_1\to\mu\tau)_{\rm tree}&\simeq
 16.1~\GeV^2\,(y_{\mu\tau}^2+y_{\tau\mu}^2)\Delta_{23}^2.
 \label{eq:numeric_drivers}
\end{align}
The respective Spearman coefficients are $0.99796$ and $0.99995$.  The different numerical coefficients mainly reflect the different parent masses, total widths, phase space, and the common $m_c^2$ or $m_\mu^2$ Yukawa normalization.  The same $\Delta_{23}$ controls both processes.

Dividing the two relations in Eq.~\eqref{eq:numeric_drivers} cancels the common scalar factor and gives
\begin{equation}
 \frac{\Br(h_1\to\mu\tau)}{\Br(t\to ch_1)}\simeq
 34.26\,\frac{y_{\mu\tau}^2+y_{\tau\mu}^2}{y_{ct}^2+y_{tc}^2}.
 \label{eq:flavor_ratio}
\end{equation}
Panel (c) of Fig.~\ref{fig:drivers} follows this diagonal scaling with $\rho_S=0.97193$.  The remaining spread is consistent with the modest loop correction in the top channel and with subleading terms beyond the leading factorized approximation.  This ratio is especially useful because it tests the generation-dependent Yukawa textures with greatly reduced sensitivity to the scalar-sector parameters.

\begin{figure}[htbp]
 \centering
 \includegraphics[width=0.49\textwidth]{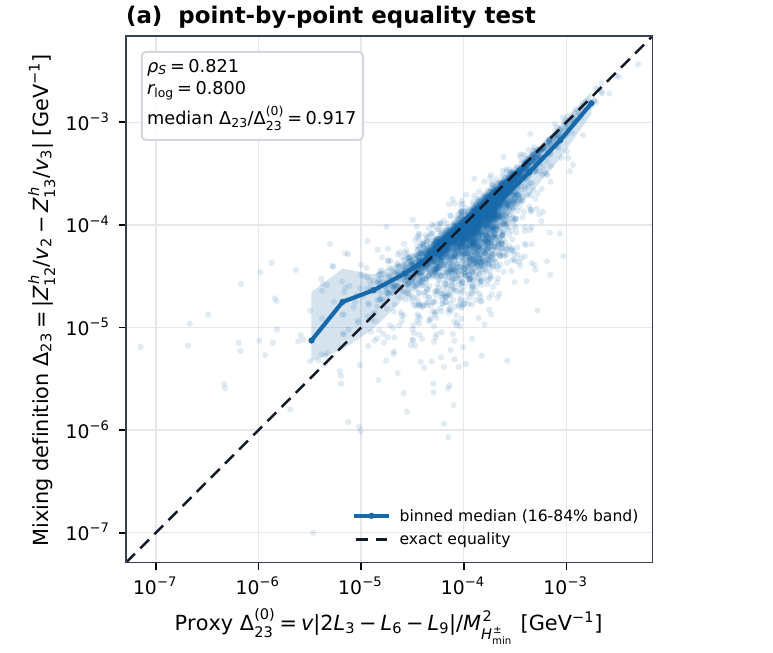}\hfill
 \includegraphics[width=0.49\textwidth]{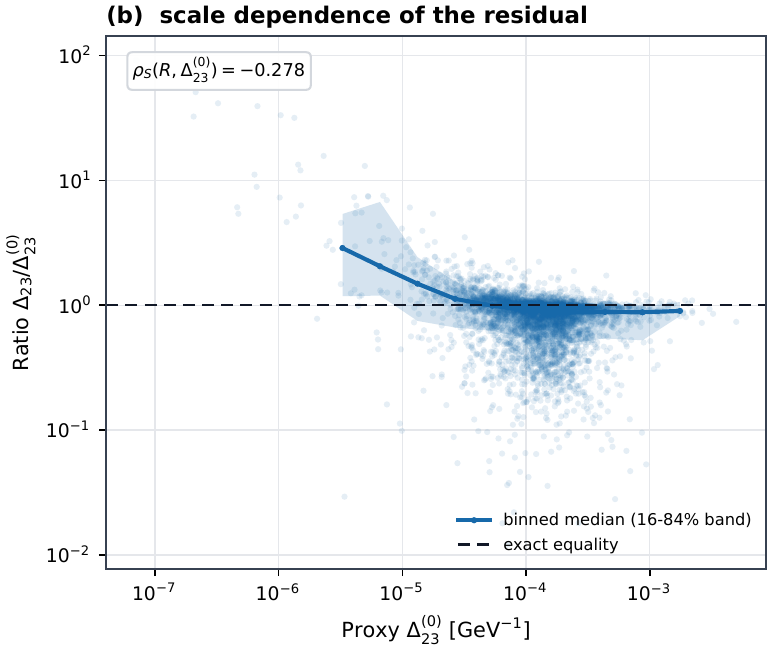}\\[1mm]
 \includegraphics[width=0.49\textwidth]{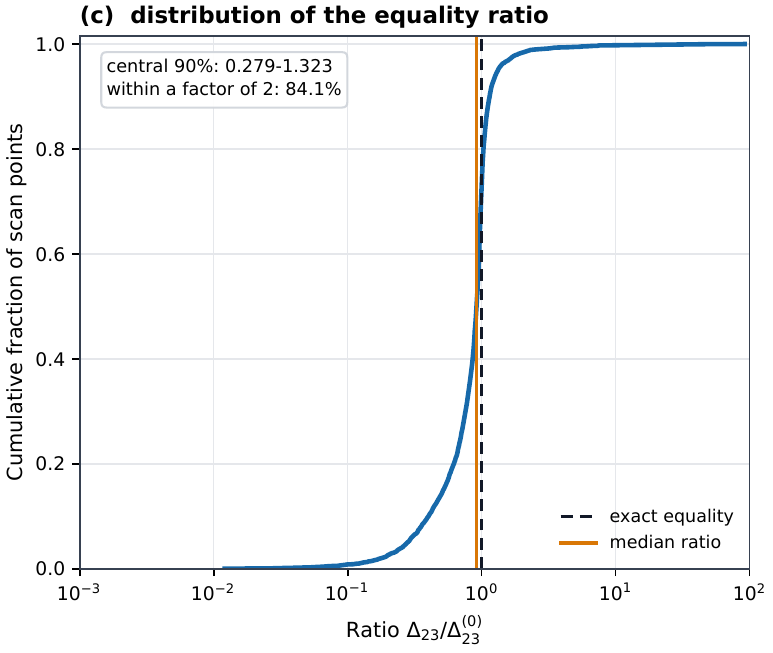}\hfill
 \includegraphics[width=0.49\textwidth]{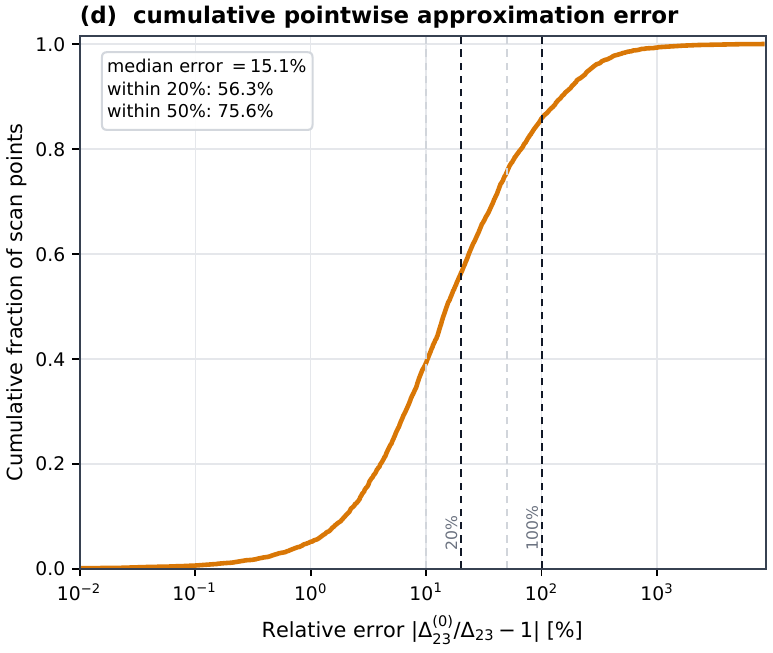}
 \caption{Test of the approximate scalar-misalignment proxy $\Delta_{23}^{(0)}=v|2\lambda_3-\lambda_6-\lambda_9|/M_{H_{\min}^{\pm}}^2$: (a) exact versus approximate values, (b) scale dependence of their ratio, (c) cumulative distribution of the ratio, and (d) cumulative pointwise relative error.}
 \label{fig:delta_test}
\end{figure}

Finally, Fig.~\ref{fig:delta_test} compares the exact mixing-matrix definition of $\Delta_{23}$ with the charged-Higgs proxy in Eq.~\eqref{eq:delta_proxy}.  The point-by-point comparison gives $\rho_S=0.821$, $r_{\log}=0.800$, and a median ratio $\Delta_{23}/\Delta_{23}^{(0)}=0.917$.  The central 90\% interval of this ratio is $0.279$--$1.323$, and 84.1\% of the accepted points lie within a factor of two.  The median pointwise relative error is 15.1\%; 56.3\% (75.6\%) of the points are reproduced within 20\% (50\%).

The approximation is therefore reliable for interpreting the dominant scale and parameter trends, but it is not a replacement for exact diagonalization in precision predictions.  Residual deviations arise from higher orders in $v^2/M_A^2$, nonzero charged- and neutral-sector angle differences, and the remaining quartic combinations.  The result also clarifies why the marginal correlation with $\lambda_3$ alone is modest: $\lambda_3$ is scanned over the comparatively narrow interval $[0,1]$, while the observable depends on a combination of three quartics and on the charged-Higgs mass.

\subsection{Relations among low-energy lepton-flavor observables}

\begin{figure}[p]
 \centering
 \includegraphics[width=0.48\textwidth]{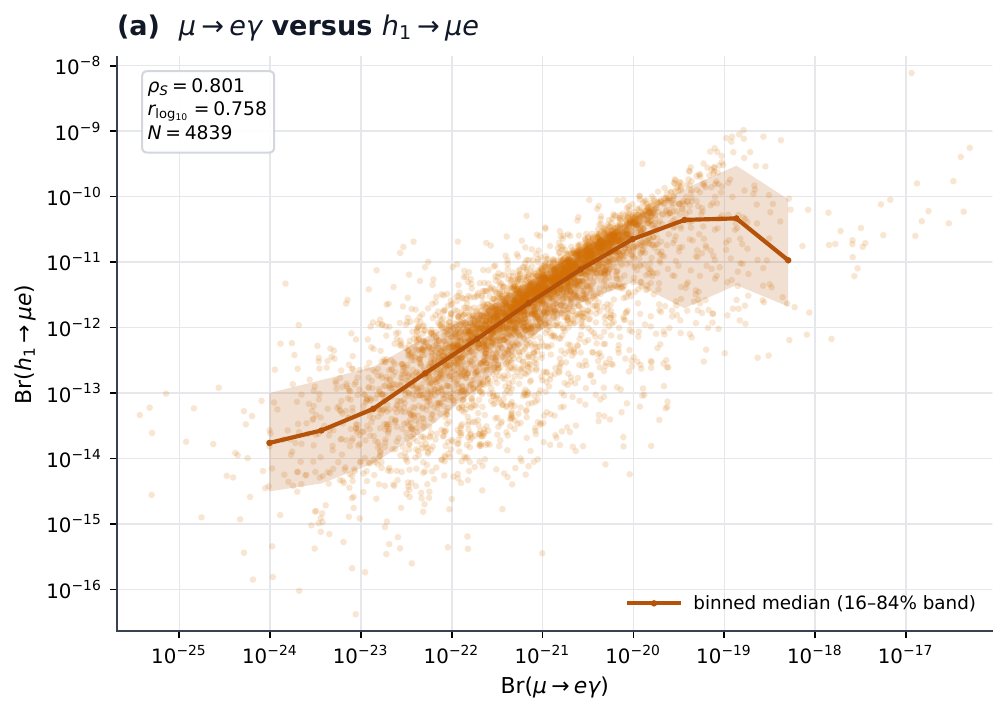}\hfill
 \includegraphics[width=0.48\textwidth]{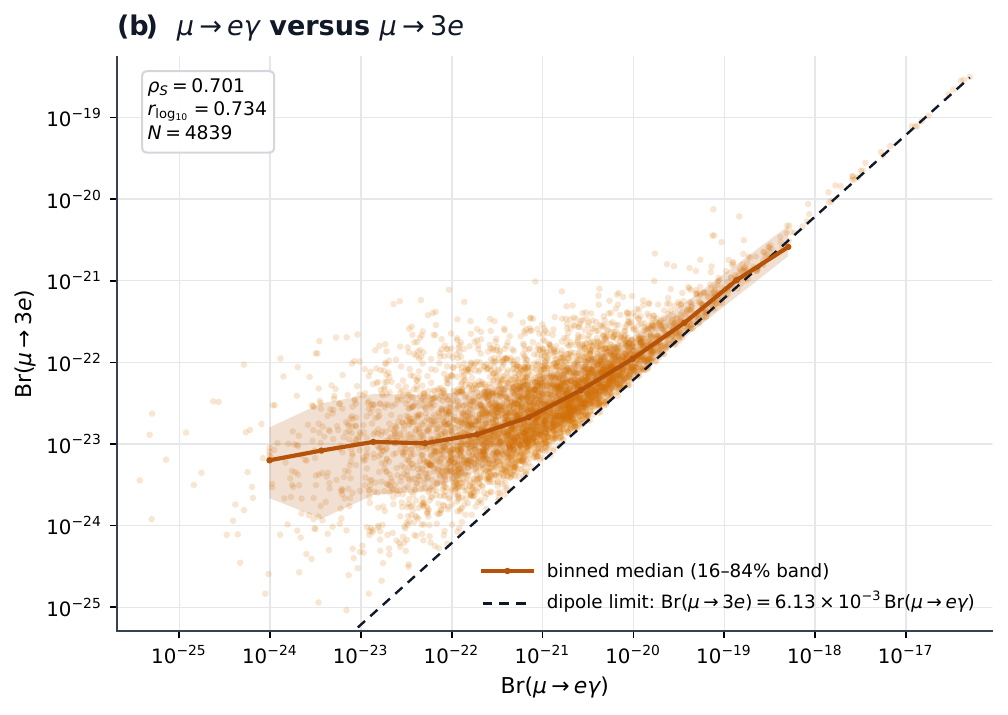}\\[3mm]
 \includegraphics[width=0.48\textwidth]{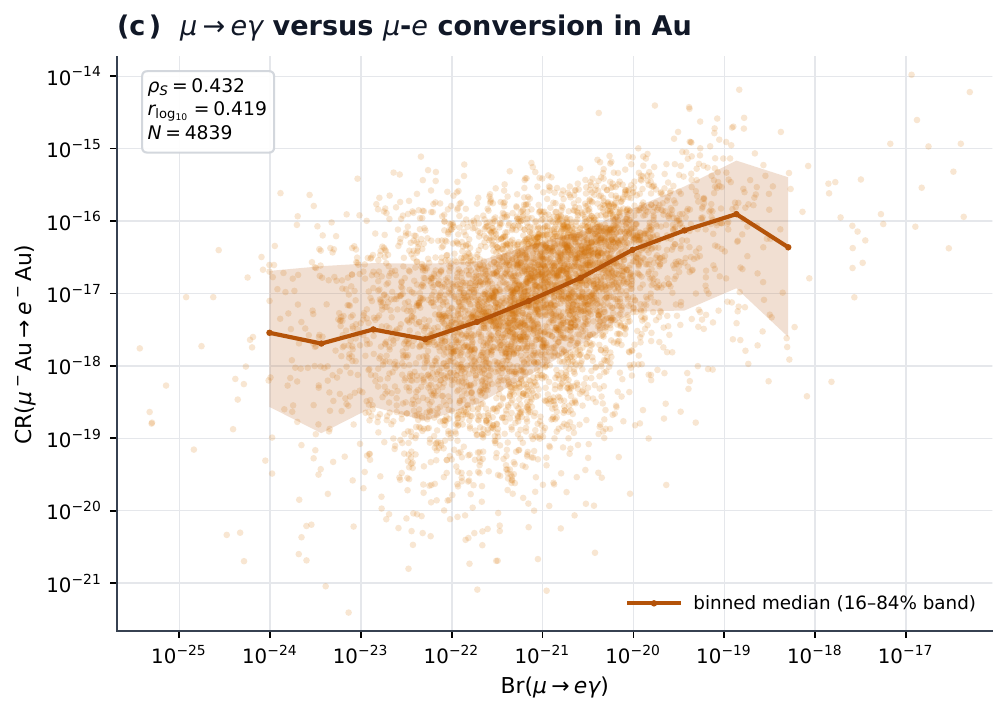}\hfill
 \includegraphics[width=0.48\textwidth]{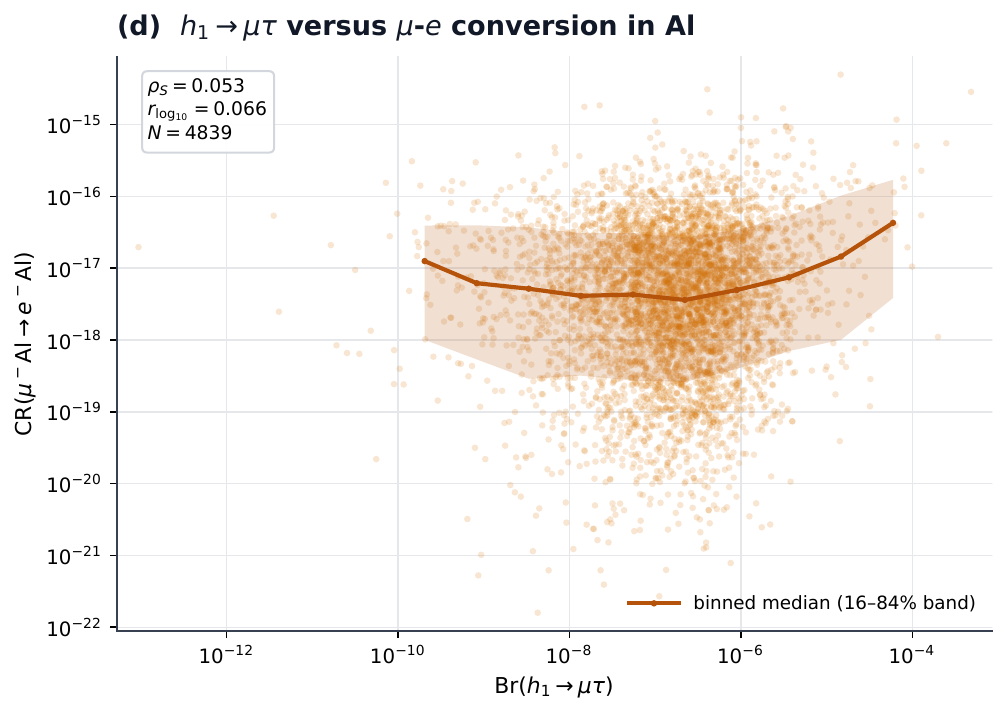}\\[3mm]
 \includegraphics[width=0.48\textwidth]{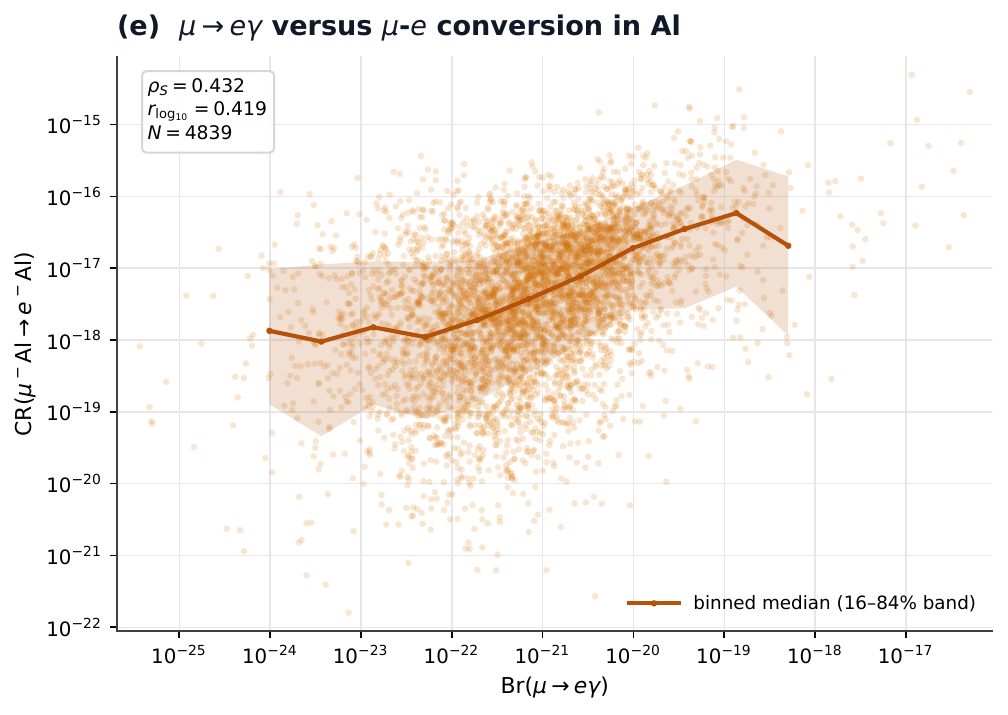}
 \caption{Correlations among low-energy LFV and flavor-changing Higgs observables: (a) $\Br(h_1\to\mu e)$ versus $\Br(\mu\to e\gamma)$, (b) $\Br(\mu\to3e)$ versus $\Br(\mu\to e\gamma)$, (c) coherent $\mu$-$e$ conversion in Au versus $\Br(\mu\to e\gamma)$, (d) conversion in Al versus $\Br(h_1\to\mu\tau)$, and (e) conversion in Al versus $\Br(\mu\to e\gamma)$.  Solid curves and bands denote binned medians and 16th--84th percentile intervals.  The dashed line in panel (b) is the dipole-dominance relation $\Br(\mu\to3e)=6.13\times10^{-3}\Br(\mu\to e\gamma)$.}
 \label{fig:lfv_correlations}
\end{figure}

The correlations involving the electron-muon sector are shown in Fig.~\ref{fig:lfv_correlations}.  The $\mu\to e\gamma$ rate is strongly correlated with the flavor-changing Higgs decay $h_1\to\mu e$, with $\rho_S=0.801$ and $r_{\log}=0.758$.  The relationship is not exact because the Higgs decay probes a tree-level $h_1$ coupling, whereas $\mu\to e\gamma$ is a loop observable that samples the full neutral- and charged-scalar spectrum.

The $\mu\to3e$ channel has a stronger correlation with $\mu\to e\gamma$, $\rho_S=0.701$ and $r_{\log}=0.734$.  Many points approach the familiar dipole-dominance estimate
\begin{equation}
 \Br(\mu\to3e)\simeq6.13\times10^{-3}\Br(\mu\to e\gamma),
 \label{eq:dipole_relation}
\end{equation}
with visible departures caused by non-dipole penguins, boxes, and scalar exchange.  Conversion in both Au and Al has the more moderate correlation $\rho_S=0.432$ and $r_{\log}=0.419$ with $\mu\to e\gamma$.  The two targets probe the same underlying dipole, scalar, and vector coefficients but weight them with different nuclear overlap integrals, so the overall ordering of points is similar while the absolute rates differ.  By contrast, Al conversion and $h_1\to\mu\tau$ are essentially uncorrelated, with $\rho_S=0.053$ and $r_{\log}=0.066$.  This is expected because conversion probes first--second-generation lepton textures, whereas $h_1\to\mu\tau$ is controlled primarily by independent second--third-generation textures.  These complementary correlations would help identify the operator structure if LFV were observed in more than one channel.

\FloatBarrier
\section{Conclusions}

We have investigated top-quark FCNC and charged-lepton flavor violation in the CP-conserving generational three-Higgs-doublet model.  The numerical analysis contains 4839 points satisfying the scalar-sector conditions, the Higgs-mass and signal-strength requirements, electroweak precision data, the UTfit 2025 neutral-meson-mixing selection, $\overline B\to X_s\gamma$, and $B_s\to\mu^+\mu^-$.

The dominant collider predictions are $\Br(t\to ch_1)\leq1.04\times10^{-5}$ and $\Br(h_1\to\mu\tau)\leq4.83\times10^{-4}$ in the scanned domain.  First-generation modes, radiative LFV decays, and three-body LFV decays are much smaller.  The maximal conversion rates are $4.9503\times10^{-15}$ in Al and $1.0479\times10^{-14}$ in Au.  The Au maximum is safely below the present SINDRUM-II bound, while the Al upper tail is within the projected Mu2e Run-I and COMET Phase-I reach.  Direct comparison of tree-only and complete results demonstrates that the two leading collider branching ratios are generated predominantly by tree-level neutral-Higgs flavor-changing couplings, with a modest loop correction in the top channel.

The central physical result is the factorization
\begin{align}
 \Br(t\to ch_1)&\propto(y_{ct}^2+y_{tc}^2)\Delta_{23}^2,
 \nonumber\\
 \Br(h_1\to\mu\tau)&\propto(y_{\mu\tau}^2+y_{\tau\mu}^2)\Delta_{23}^2,
 \nonumber\\
 \Delta_{23}&=\left|\frac{Z^h_{12}}{v_2}-\frac{Z^h_{13}}{v_3}\right|.
 \label{eq:conclusion_factorization}
\end{align}
Obviously, $\Br(t\to ch_1),\;\Br(h_1\to\mu\tau)$ can be enhanced by a quadratic $2$--$3$ fermion-texture factor $\Delta_{23}$, and both of them vanish in the alignment limit.  Over most of the accepted region, its scale is well captured by $v|2\lambda_3-\lambda_6-\lambda_9|/M_{H_{\min}^{\pm}}^2$, explaining the positive correlations with $\lambda_6+\lambda_9$ and the negative correlations with the light charged-Higgs mass.


\clearpage
\begin{acknowledgments}
This work was supported by the National Natural Science Foundation of China (NNSFC) under Grants No.~12075074, 12235008, 11535002, and 11705045; the Natural Science Foundation for Distinguished Young Scholars of Hebei Province under Grant No.~A2022201017; the Youth Top-notch Talent Support Program of Hebei Province; and the Midwest Universities Comprehensive Strength Promotion Project.
\end{acknowledgments}

\bibliographystyle{apsrev4-2}
\bibliography{references}

@article{Crivellin:2013wna,
    author = "Crivellin, Andreas and Kokulu, Ahmet and Greub, Christoph",
    title = "{Flavor-phenomenology of two-Higgs-doublet models with generic Yukawa structure}",
    eprint = "1303.5877",
    archivePrefix = "arXiv",
    primaryClass = "hep-ph",
    doi = "10.1103/PhysRevD.87.094031",
    journal = "Phys. Rev. D",
    volume = "87",
    number = "9",
    pages = "094031",
    year = "2013"
}

@article{Glashow:1976nt,
    author = "Glashow, Sheldon L. and Weinberg, Steven",
    title = "{Natural Conservation Laws for Neutral Currents}",
    reportNumber = "HUTP-76-A158",
    doi = "10.1103/PhysRevD.15.1958",
    journal = "Phys. Rev. D",
    volume = "15",
    pages = "1958",
    year = "1977"
}

@article{Altmannshofer:2015esa,
    author = "Altmannshofer, Wolfgang and Gori, Stefania and Kagan, Alexander L. and Silvestrini, Luca and Zupan, Jure",
    title = "{Uncovering Mass Generation Through Higgs Flavor Violation}",
    eprint = "1507.07927",
    archivePrefix = "arXiv",
    primaryClass = "hep-ph",
    doi = "10.1103/PhysRevD.93.031301",
    journal = "Phys. Rev. D",
    volume = "93",
    number = "3",
    pages = "031301",
    year = "2016"
}

@article{Altmannshofer:2024jyv,
    author = "Altmannshofer, Wolfgang and Greljo, Admir",
    title = "{Recent Progress in Flavor Model Building}",
    eprint = "2412.04549",
    archivePrefix = "arXiv",
    primaryClass = "hep-ph",
    doi = "10.1146/annurev-nucl-121423-100950",
    journal = "Ann. Rev. Nucl. Part. Sci.",
    volume = "75",
    pages = "201--322",
    year = "2025"
}

@article{Egana-Ugrinovic:2019dqu,
    author = "Egana-Ugrinovic, Daniel and Homiller, Samuel and Meade, Patrick Roddy",
    title = "{Higgs bosons with large couplings to light quarks}",
    eprint = "1908.11376",
    archivePrefix = "arXiv",
    primaryClass = "hep-ph",
    reportNumber = "YITP-SB-19-25",
    doi = "10.1103/PhysRevD.100.115041",
    journal = "Phys. Rev. D",
    volume = "100",
    number = "11",
    pages = "115041",
    year = "2019"
}

@article{Altmannshofer:2019ogm,
    author = "Altmannshofer, Wolfgang and Maddock, Brian and Tuckler, Douglas",
    title = "{Rare Top Decays as Probes of Flavorful Higgs Bosons}",
    eprint = "1904.10956",
    archivePrefix = "arXiv",
    primaryClass = "hep-ph",
    doi = "10.1103/PhysRevD.100.015003",
    journal = "Phys. Rev. D",
    volume = "100",
    number = "1",
    pages = "015003",
    year = "2019"
}

@article{Blechman:2010cs,
    author = "Blechman, Andrew E. and Petrov, Alexey A. and Yeghiyan, Gagik",
    title = "{The Flavor puzzle in multi-Higgs models}",
    eprint = "1009.1612",
    archivePrefix = "arXiv",
    primaryClass = "hep-ph",
    reportNumber = "WSU-HEP-1003",
    doi = "10.1007/JHEP11(2010)075",
    journal = "JHEP",
    volume = "2010",
    number = "11",
    pages = "075",
    year = "2010"
}

@article{Das:1995df,
    author = "Das, Ashok K. and Kao, Chung",
    title = "{A Two Higgs doublet model for the top quark}",
    eprint = "hep-ph/9511329",
    archivePrefix = "arXiv",
    reportNumber = "UR-1446",
    doi = "10.1016/0370-2693(96)00031-7",
    journal = "Phys. Lett. B",
    volume = "372",
    pages = "106--112",
    year = "1996"
}

@article{Botella:2016krk,
    author = "Botella, F. J. and Branco, G. C. and Rebelo, M. N. and Silva-Marcos, J. I.",
    title = "{What if the masses of the first two quark families are not generated by the standard model Higgs boson?}",
    eprint = "1602.08011",
    archivePrefix = "arXiv",
    primaryClass = "hep-ph",
    reportNumber = "CERN-TH-2016-039",
    doi = "10.1103/PhysRevD.94.115031",
    journal = "Phys. Rev. D",
    volume = "94",
    number = "11",
    pages = "115031",
    year = "2016"
}

@article{Ghosh:2015gpa,
    author = "Ghosh, Diptimoy and Gupta, Rick Sandeepan and Perez, Gilad",
    title = "{Is the Higgs Mechanism of Fermion Mass Generation a Fact? A Yukawa-less First-Two-Generation Model}",
    eprint = "1508.01501",
    archivePrefix = "arXiv",
    primaryClass = "hep-ph",
    doi = "10.1016/j.physletb.2016.02.059",
    journal = "Phys. Lett. B",
    volume = "755",
    pages = "504--508",
    year = "2016"
}

@article{CMS:2022dwd,
    author = "Tumasyan, Armen and others",
    collaboration = "CMS",
    title = "{A portrait of the Higgs boson by the CMS experiment ten years after the discovery.}",
    eprint = "2207.00043",
    archivePrefix = "arXiv",
    primaryClass = "hep-ex",
    reportNumber = "CMS-HIG-22-001, CERN-EP-2022-039",
    doi = "10.1038/s41586-022-04892-x",
    journal = "Nature",
    volume = "607",
    number = "7917",
    pages = "60--68",
    year = "2022",
    note = "[Erratum: Nature 623, (2023)]"
}

@article{ATLAS:2022vkf,
    author = "Aad, Georges and others",
    collaboration = "ATLAS",
    title = "{A detailed map of Higgs boson interactions by the ATLAS experiment ten years after the discovery}",
    eprint = "2207.00092",
    archivePrefix = "arXiv",
    primaryClass = "hep-ex",
    reportNumber = "CERN-EP-2022-057",
    doi = "10.1038/s41586-022-04893-w",
    journal = "Nature",
    volume = "607",
    number = "7917",
    pages = "52--59",
    year = "2022",
    note = "[Erratum: Nature 612, E24 (2022)]"
}

@article{Das:2019yad,
    author = "Das, Dipankar and Saha, Ipsita",
    title = "{Alignment limit in three Higgs-doublet models}",
    eprint = "1904.03970",
    archivePrefix = "arXiv",
    primaryClass = "hep-ph",
    doi = "10.1103/PhysRevD.100.035021",
    journal = "Phys. Rev. D",
    volume = "100",
    number = "3",
    pages = "035021",
    year = "2019"
}

@article{Pilaftsis:2016erj,
    author = "Pilaftsis, Apostolos",
    title = "{Symmetries for standard model alignment in multi-Higgs doublet models}",
    eprint = "1602.02017",
    archivePrefix = "arXiv",
    primaryClass = "hep-ph",
    reportNumber = "MAN-HEP-2016-03, CERN-TH-2016-057",
    doi = "10.1103/PhysRevD.93.075012",
    journal = "Phys. Rev. D",
    volume = "93",
    number = "7",
    pages = "075012",
    year = "2016"
}

@article{Boto:2023nyi,
    author = "Boto, Rafael and Das, Dipankar and Lourenco, Luis and Romao, Jorge C. and Silva, Joao P.",
    title = "{Fingerprinting the type-Z three-Higgs-doublet models}",
    eprint = "2304.13494",
    archivePrefix = "arXiv",
    primaryClass = "hep-ph",
    doi = "10.1103/PhysRevD.108.015020",
    journal = "Phys. Rev. D",
    volume = "108",
    number = "1",
    pages = "015020",
    year = "2023"
}

@article{Das:2022gbm,
    author = "Das, Dipankar and Levy, Miguel and Pal, Palash B. and Prasad, Anugrah M. and Saha, Ipsita and Srivastava, Ayushi",
    title = "{Democratic three-Higgs-doublet models: The custodial limit and wrong-sign Yukawa coupling}",
    eprint = "2301.00231",
    archivePrefix = "arXiv",
    primaryClass = "hep-ph",
    doi = "10.1103/PhysRevD.107.055035",
    journal = "Phys. Rev. D",
    volume = "107",
    number = "5",
    pages = "055035",
    year = "2023"
}

@article{Boto:2021qgu,
    author = "Boto, Rafael and Rom\~ao, Jorge C. and Silva, Jo\~ao P.",
    title = "{Current bounds on the type-Z Z3 three-Higgs-doublet model}",
    eprint = "2106.11977",
    archivePrefix = "arXiv",
    primaryClass = "hep-ph",
    reportNumber = "CFTP/21-010",
    doi = "10.1103/PhysRevD.104.095006",
    journal = "Phys. Rev. D",
    volume = "104",
    number = "9",
    pages = "095006",
    year = "2021"
}

@article{Keus:2014jha,
    author = "Keus, Venus and King, Stephen F. and Moretti, Stefano and Sokolowska, Dorota",
    title = "{Dark Matter with Two Inert Doublets plus One Higgs Doublet}",
    eprint = "1407.7859",
    archivePrefix = "arXiv",
    primaryClass = "hep-ph",
    doi = "10.1007/JHEP11(2014)016",
    journal = "JHEP",
    volume = "2014",
    number = "11",
    pages = "016",
    year = "2014"
}

@article{Cordero:2017owj,
    author = "Cordero, A. and Hernandez-Sanchez, J. and Keus, V. and King, S. F. and Moretti, S. and Rojas, D. and Sokolowska, D.",
    title = "{Dark Matter Signals at the LHC from a 3HDM}",
    eprint = "1712.09598",
    archivePrefix = "arXiv",
    primaryClass = "hep-ph",
    doi = "10.1007/JHEP05(2018)030",
    journal = "JHEP",
    volume = "2018",
    number = "05",
    pages = "030",
    year = "2018"
}

@article{Aranda:2019vda,
    author = "Aranda, A. and Hern\'andez-Otero, D. and Hern\'andez-Sanchez, J. and Keus, V. and Moretti, S. and Rojas-Ciofalo, D. and Shindou, T.",
    title = "{Z$_3$ symmetric inert ( 2+1 )-Higgs-doublet model}",
    eprint = "1907.12470",
    archivePrefix = "arXiv",
    primaryClass = "hep-ph",
    doi = "10.1103/PhysRevD.103.015023",
    journal = "Phys. Rev. D",
    volume = "103",
    number = "1",
    pages = "015023",
    year = "2021"
}

@article{Khater:2021wcx,
    author = "Khater, W. and Kun\v{c}inas, A. and Ogreid, O. M. and Osland, P. and Rebelo, M. N.",
    title = "{Dark matter in three-Higgs-doublet models with S$_{3}$ symmetry}",
    eprint = "2108.07026",
    archivePrefix = "arXiv",
    primaryClass = "hep-ph",
    doi = "10.1007/JHEP01(2022)120",
    journal = "JHEP",
    volume = "2022",
    number = "01",
    pages = "120",
    year = "2022"
}

@article{Kuncinas:2022whn,
    author = "Kun\v{c}inas, A. and Ogreid, O. M. and Osland, P. and Rebelo, M. N.",
    title = "{Dark matter in a CP-violating three-Higgs-doublet model with S3 symmetry}",
    eprint = "2204.05684",
    archivePrefix = "arXiv",
    primaryClass = "hep-ph",
    doi = "10.1103/PhysRevD.106.075002",
    journal = "Phys. Rev. D",
    volume = "106",
    number = "7",
    pages = "075002",
    year = "2022"
}

@article{Grzadkowski:2009bt,
    author = "Grzadkowski, B. and Ogreid, O. M. and Osland, P.",
    title = "{Natural Multi-Higgs Model with Dark Matter and CP Violation}",
    eprint = "0904.2173",
    archivePrefix = "arXiv",
    primaryClass = "hep-ph",
    reportNumber = "NORDITA-2009-24, IFT-09-05",
    doi = "10.1103/PhysRevD.80.055013",
    journal = "Phys. Rev. D",
    volume = "80",
    pages = "055013",
    year = "2009"
}

@article{Ahriche:2015mea,
    author = "Ahriche, Amine and Faisel, Gaber and Ho, Shu-Yu and Nasri, Salah and Tandean, Jusak",
    title = "{Effects of two inert scalar doublets on Higgs boson interactions and the electroweak phase transition}",
    eprint = "1501.06605",
    archivePrefix = "arXiv",
    primaryClass = "hep-ph",
    doi = "10.1103/PhysRevD.92.035020",
    journal = "Phys. Rev. D",
    volume = "92",
    number = "3",
    pages = "035020",
    year = "2015"
}

@article{Hartmann:2014ppa,
    author = "Hartmann, Florian and Kilian, Wolfgang",
    title = "{Flavour Models with Three Higgs Generations}",
    eprint = "1405.1901",
    archivePrefix = "arXiv",
    primaryClass = "hep-ph",
    reportNumber = "SI-HEP-2014-09",
    doi = "10.1140/epjc/s10052-014-3055-4",
    journal = "Eur. Phys. J. C",
    volume = "74",
    pages = "3055",
    year = "2014"
}

@article{Keus:2013hya,
    author = "Keus, Venus and King, Stephen F. and Moretti, Stefano",
    title = "{Three-Higgs-doublet models: symmetries, potentials and Higgs boson masses}",
    eprint = "1310.8253",
    archivePrefix = "arXiv",
    primaryClass = "hep-ph",
    doi = "10.1007/JHEP01(2014)052",
    journal = "JHEP",
    volume = "2014",
    number = "01",
    pages = "052",
    year = "2014"
}

@article{Ivanov:2010ww,
    author = "Ivanov, I. P. and Nishi, C. C.",
    title = "{Properties of the general NHDM. I. The Orbit space}",
    eprint = "1004.1799",
    archivePrefix = "arXiv",
    primaryClass = "hep-th",
    doi = "10.1103/PhysRevD.82.015014",
    journal = "Phys. Rev. D",
    volume = "82",
    pages = "015014",
    year = "2010"
}

@article{Ivanov:2010wz,
    author = "Ivanov, I. P.",
    title = "{Properties of the general NHDM. II. Higgs potential and its symmetries}",
    eprint = "1004.1802",
    archivePrefix = "arXiv",
    primaryClass = "hep-th",
    doi = "10.1007/JHEP07(2010)020",
    journal = "JHEP",
    volume = "2010",
    number = "07",
    pages = "020",
    year = "2010"
}

@article{Ivanov:2012ry,
    author = "Ivanov, I. P. and Vdovin, Evgeny",
    title = "{Discrete symmetries in the three-Higgs-doublet model}",
    eprint = "1206.7108",
    archivePrefix = "arXiv",
    primaryClass = "hep-ph",
    doi = "10.1103/PhysRevD.86.095030",
    journal = "Phys. Rev. D",
    volume = "86",
    pages = "095030",
    year = "2012"
}

@article{Darvishi:2019dbh,
    author = "Darvishi, Neda and Pilaftsis, Apostolos",
    title = "{Classifying Accidental Symmetries in Multi-Higgs Doublet Models}",
    eprint = "1912.00887",
    archivePrefix = "arXiv",
    primaryClass = "hep-ph",
    reportNumber = "MAN/HEP/2019/009",
    doi = "10.1103/PhysRevD.101.095008",
    journal = "Phys. Rev. D",
    volume = "101",
    number = "9",
    pages = "095008",
    year = "2020"
}

@article{Darvishi:2021txa,
    author = "Darvishi, Neda and Masouminia, M. R. and Pilaftsis, Apostolos",
    title = "{Maximally symmetric three-Higgs-doublet model}",
    eprint = "2106.03159",
    archivePrefix = "arXiv",
    primaryClass = "hep-ph",
    reportNumber = "IPPP/20/110",
    doi = "10.1103/PhysRevD.104.115017",
    journal = "Phys. Rev. D",
    volume = "104",
    number = "11",
    pages = "115017",
    year = "2021"
}

@article{Boto:2022uwv,
    author = "Boto, Rafael and Rom\~ao, Jorge C. and Silva, Jo\~ao P.",
    title = "{Bounded from below conditions on a class of symmetry constrained 3HDM}",
    eprint = "2208.01068",
    archivePrefix = "arXiv",
    primaryClass = "hep-ph",
    reportNumber = "CFTP/22-004",
    doi = "10.1103/PhysRevD.106.115010",
    journal = "Phys. Rev. D",
    volume = "106",
    number = "11",
    pages = "115010",
    year = "2022"
}

@article{Penuelas:2017ikk,
    author = "Pe\~nuelas, Ana and Pich, Antonio",
    title = "{Flavour alignment in multi-Higgs-doublet models}",
    eprint = "1710.02040",
    archivePrefix = "arXiv",
    primaryClass = "hep-ph",
    reportNumber = "IFIC-17-32, FTUV-17-1005",
    doi = "10.1007/JHEP12(2017)084",
    journal = "JHEP",
    volume = "2017",
    number = "12",
    pages = "084",
    year = "2017"
}

@article{Grimus:2007if,
    author = "Grimus, W. and Lavoura, L. and Ogreid, O. M. and Osland, P.",
    title = "{A Precision constraint on multi-Higgs-doublet models}",
    eprint = "0711.4022",
    archivePrefix = "arXiv",
    primaryClass = "hep-ph",
    reportNumber = "UWTHPH-2007-28",
    doi = "10.1088/0954-3899/35/7/075001",
    journal = "J. Phys. G",
    volume = "35",
    pages = "075001",
    year = "2008"
}

@article{CMS:2022mgd,
    author = "Tumasyan, Armen and others",
    collaboration = "CMS",
    title = "{Measurement of the $B^0_s\to \mu^+\mu^-$ decay properties and search for the $B^0\to\mu^+\mu^-$ decay in proton-proton collisions at $\sqrt{s}$ = 13 TeV}",
    eprint = "2212.10311",
    archivePrefix = "arXiv",
    primaryClass = "hep-ex",
    reportNumber = "CMS-BPH-21-006, CERN-EP-2022-270",
    doi = "10.1016/j.physletb.2023.137955",
    journal = "Phys. Lett. B",
    volume = "842",
    pages = "137955",
    year = "2023"
}

@article{LHCb:2021moh,
    author = "Aaij, R. and others",
    collaboration = "LHCb",
    title = "{Precise determination of the $B_{\mathrm{s}}^0$\textendash{}$\overline B_{\mathrm{s}}^0$ oscillation frequency}",
    eprint = "2104.04421",
    archivePrefix = "arXiv",
    primaryClass = "hep-ex",
    reportNumber = "LHCb-PAPER-2021-005, CERN-EP-2021-047",
    doi = "10.1038/s41567-021-01394-x",
    journal = "Nature Phys.",
    volume = "18",
    number = "1",
    pages = "1--5",
    year = "2022"
}

@article{ParticleDataGroup:2024cfk,
    author = "Navas, S. and others",
    collaboration = "Particle Data Group",
    title = "{Review of particle physics}",
    doi = "10.1103/PhysRevD.110.030001",
    journal = "Phys. Rev. D",
    volume = "110",
    number = "3",
    pages = "030001",
    year = "2024"
}

@article{Nishi:2007nh,
    author = "Nishi, Celso C.",
    title = "{The Structure of potentials with N Higgs doublets}",
    eprint = "0706.2685",
    archivePrefix = "arXiv",
    primaryClass = "hep-ph",
    doi = "10.1103/PhysRevD.76.055013",
    journal = "Phys. Rev. D",
    volume = "76",
    pages = "055013",
    year = "2007"
}

@article{Lenz:2020awd,
    author = "Lenz, Alexander and Wilkinson, Guy",
    title = "{Mixing and CP Violation in the Charm System}",
    eprint = "2011.04443",
    archivePrefix = "arXiv",
    primaryClass = "hep-ph",
    doi = "10.1146/annurev-nucl-102419-124613",
    journal = "Ann. Rev. Nucl. Part. Sci.",
    volume = "71",
    pages = "59--85",
    year = "2021"
}

@article{HeavyFlavorAveragingGroup:2022wzx,
    author = "Amhis, Yasmine Sara and others",
    collaboration = "Heavy Flavor Averaging Group, HFLAV",
    title = "{Averages of b-hadron, c-hadron, and \ensuremath{\tau}-lepton properties as of 2021}",
    eprint = "2206.07501",
    archivePrefix = "arXiv",
    primaryClass = "hep-ex",
    doi = "10.1103/PhysRevD.107.052008",
    journal = "Phys. Rev. D",
    volume = "107",
    number = "5",
    pages = "052008",
    year = "2023"
}

@article{Branco:2011iw,
    author = "Branco, G. C. and Ferreira, P. M. and Lavoura, L. and Rebelo, M. N. and Sher, Marc and Silva, Joao P.",
    title = "{Theory and phenomenology of two-Higgs-doublet models}",
    eprint = "1106.0034",
    archivePrefix = "arXiv",
    primaryClass = "hep-ph",
    doi = "10.1016/j.physrep.2012.02.002",
    journal = "Phys. Rept.",
    volume = "516",
    pages = "1--102",
    year = "2012"
}

@article{Faro:2019vcd,
    author = "Faro, Francisco S. and Ivanov, Igor P.",
    title = "{Boundedness from below in the $U(1)\times U(1)$ three-Higgs-doublet model}",
    eprint = "1907.01963",
    archivePrefix = "arXiv",
    primaryClass = "hep-ph",
    reportNumber = "CFTP-19-021, CFTP/19-022",
    doi = "10.1103/PhysRevD.100.035038",
    journal = "Phys. Rev. D",
    volume = "100",
    number = "3",
    pages = "035038",
    year = "2019"
}

@article{Altmannshofer:2017uvs,
    author = "Altmannshofer, Wolfgang and Gori, Stefania and Robinson, Dean J. and Tuckler, Douglas",
    title = "{The Flavor-locked Flavorful Two Higgs Doublet Model}",
    eprint = "1712.01847",
    archivePrefix = "arXiv",
    primaryClass = "hep-ph",
    doi = "10.1007/JHEP03(2018)129",
    journal = "JHEP",
    volume = "2018",
    number = "03",
    pages = "129",
    year = "2018"
}

@article{Altmannshofer:2016zrn,
    author = "Altmannshofer, Wolfgang and Eby, Joshua and Gori, Stefania and Lotito, Matteo and Martone, Mario and Tuckler, Douglas",
    title = "{Collider Signatures of Flavorful Higgs Bosons}",
    eprint = "1610.02398",
    archivePrefix = "arXiv",
    primaryClass = "hep-ph",
    reportNumber = "FERMILAB-PUB-16-499-PPD",
    doi = "10.1103/PhysRevD.94.115032",
    journal = "Phys. Rev. D",
    volume = "94",
    number = "11",
    pages = "115032",
    year = "2016"
}

@article{Altmannshofer:2018bch,
    author = "Altmannshofer, Wolfgang and Maddock, Brian",
    title = "{Flavorful Two Higgs Doublet Models with a Twist}",
    eprint = "1805.08659",
    archivePrefix = "arXiv",
    primaryClass = "hep-ph",
    doi = "10.1103/PhysRevD.98.075005",
    journal = "Phys. Rev. D",
    volume = "98",
    number = "7",
    pages = "075005",
    year = "2018"
}

@article{Altmannshofer:2012az,
    author = "Altmannshofer, Wolfgang and Straub, David M.",
    title = "{Cornering New Physics in $b \to s$ Transitions}",
    eprint = "1206.0273",
    archivePrefix = "arXiv",
    primaryClass = "hep-ph",
    reportNumber = "FERMILAB-PUB-12-257-T",
    doi = "10.1007/JHEP08(2012)121",
    journal = "JHEP",
    volume = "2012",
    number = "08",
    pages = "121",
    year = "2012"
}

@article{Altmannshofer:2017wqy,
    author = "Altmannshofer, Wolfgang and Niehoff, Christoph and Straub, David M.",
    title = "{$B_s\to\mu^+\mu^-$ as current and future probe of new physics}",
    eprint = "1702.05498",
    archivePrefix = "arXiv",
    primaryClass = "hep-ph",
    doi = "10.1007/JHEP05(2017)076",
    journal = "JHEP",
    volume = "2017",
    number = "05",
    pages = "076",
    year = "2017"
}

@article{Buchalla:1995vs,
    author = "Buchalla, Gerhard and Buras, Andrzej J. and Lautenbacher, Markus E.",
    title = "{Weak decays beyond leading logarithms}",
    eprint = "hep-ph/9512380",
    archivePrefix = "arXiv",
    reportNumber = "SLAC-PUB-7009, SLAC-PUB-95-7009, MPI-PH-95-104, TUM-T31-100-95, FERMILAB-PUB-95-305-T",
    doi = "10.1103/RevModPhys.68.1125",
    journal = "Rev. Mod. Phys.",
    volume = "68",
    pages = "1125--1144",
    year = "1996"
}

@article{Buras:2010mh,
    author = "Buras, Andrzej J. and Carlucci, Maria Valentina and Gori, Stefania and Isidori, Gino",
    title = "{Higgs-mediated FCNCs: Natural Flavour Conservation vs. Minimal Flavour Violation}",
    eprint = "1005.5310",
    archivePrefix = "arXiv",
    primaryClass = "hep-ph",
    reportNumber = "TUM-HEP-761-10, MPP-2010-57",
    doi = "10.1007/JHEP10(2010)009",
    journal = "JHEP",
    volume = "2010",
    number = "10",
    pages = "009",
    year = "2010"
}

@article{Buras:2012fs,
    author = "Buras, Andrzej J. and Girrbach, Jennifer",
    title = "{Complete NLO QCD Corrections for Tree Level Delta F = 2 FCNC Processes}",
    eprint = "1201.1302",
    archivePrefix = "arXiv",
    primaryClass = "hep-ph",
    reportNumber = "TUM-HEP-824-12, FLAVOUR(267104)-ERC-7",
    doi = "10.1007/JHEP03(2012)052",
    journal = "JHEP",
    volume = "2012",
    number = "03",
    pages = "052",
    year = "2012"
}

@article{Buras:2013rqa,
    author = "Buras, Andrzej J. and De Fazio, Fulvia and Girrbach, Jennifer and Knegjens, Robert and Nagai, Minoru",
    title = "{The Anatomy of Neutral Scalars with FCNCs in the Flavour Precision Era}",
    eprint = "1303.3723",
    archivePrefix = "arXiv",
    primaryClass = "hep-ph",
    reportNumber = "FLAVOUR(267104)-ERC-38, BARI-TH-13-671, NIKHEF-2013-008, UT-13-09",
    doi = "10.1007/JHEP06(2013)111",
    journal = "JHEP",
    volume = "2013",
    number = "06",
    pages = "111",
    year = "2013"
}

@article{ETM:2013jap,
    author = "Carrasco, N. and others",
    collaboration = "ETM",
    title = "{B-physics from $N_f$ = 2 tmQCD: the Standard Model and beyond}",
    eprint = "1308.1851",
    archivePrefix = "arXiv",
    primaryClass = "hep-lat",
    reportNumber = "CERN-PH-TH-2013-173, IFIC-13-47, LPT-ORSAY-13-59, LTH-982, MITP-13-045, RM3-TH-13-6, ROM2F-2013-12",
    doi = "10.1007/JHEP03(2014)016",
    journal = "JHEP",
    volume = "2014",
    number = "03",
    pages = "016",
    year = "2014"
}

@article{Aebischer:2020dsw,
    author = "Aebischer, Jason and Bobeth, Christoph and Buras, Andrzej J. and Kumar, Jacky",
    title = "{SMEFT ATLAS of $\Delta$F = 2 transitions}",
    eprint = "2009.07276",
    archivePrefix = "arXiv",
    primaryClass = "hep-ph",
    doi = "10.1007/JHEP12(2020)187",
    journal = "JHEP",
    volume = "2020",
    number = "12",
    pages = "187",
    year = "2020"
}

@article{FermilabLattice:2016ipl,
    author = "Bazavov, A. and others",
    collaboration = "Fermilab Lattice, MILC",
    title = "{$B^0_{(s)}$-mixing matrix elements from lattice QCD for the Standard Model and beyond}",
    eprint = "1602.03560",
    archivePrefix = "arXiv",
    primaryClass = "hep-lat",
    reportNumber = "FERMILAB-PUB-16-030-T",
    doi = "10.1103/PhysRevD.93.113016",
    journal = "Phys. Rev. D",
    volume = "93",
    number = "11",
    pages = "113016",
    year = "2016"
}

@article{LHCb:2017rmj,
    author = "Aaij, Roel and others",
    collaboration = "LHCb",
    title = "{Measurement of the $B^0_s\to\mu^+\mu^-$ branching fraction and effective lifetime and search for $B^0\to\mu^+\mu^-$ decays}",
    eprint = "1703.05747",
    archivePrefix = "arXiv",
    primaryClass = "hep-ex",
    reportNumber = "CERN-EP-2017-041, LHCB-PAPER-2017-001",
    doi = "10.1103/PhysRevLett.118.191801",
    journal = "Phys. Rev. Lett.",
    volume = "118",
    number = "19",
    pages = "191801",
    year = "2017"
}

@article{Altmannshofer:2021qrr,
    author = "Altmannshofer, Wolfgang and Stangl, Peter",
    title = "{New physics in rare B decays after Moriond 2021}",
    eprint = "2103.13370",
    archivePrefix = "arXiv",
    primaryClass = "hep-ph",
    doi = "10.1140/epjc/s10052-021-09725-1",
    journal = "Eur. Phys. J. C",
    volume = "81",
    number = "10",
    pages = "952",
    year = "2021"
}

@article{FlavourLatticeAveragingGroupFLAG:2021npn,
    author = "Aoki, Y. and others",
    collaboration = "Flavour Lattice Averaging Group (FLAG)",
    title = "{FLAG Review 2021}",
    eprint = "2111.09849",
    archivePrefix = "arXiv",
    primaryClass = "hep-lat",
    reportNumber = "CERN-TH-2021-191, JLAB-THY-21-3528, FERMILAB-PUB-21-620-SCD-T",
    doi = "10.1140/epjc/s10052-022-10536-1",
    journal = "Eur. Phys. J. C",
    volume = "82",
    number = "10",
    pages = "869",
    year = "2022"
}

@article{AltmannshoferToner:2025,
    author = "Altmannshofer, Wolfgang and Toner, Kevin",
    title = "{Flavor Constraints in a Generational Three Higgs Doublet Model}",
    eprint = "2502.04579",
    archivePrefix = "arXiv",
    primaryClass = "hep-ph",
    doi = "10.1103/PhysRevD.111.075009",
    journal = "Phys. Rev. D",
    volume = "111",
    number = "7",
    pages = "075009",
    year = "2025"
}

@article{Yang:2018fvw,
    author = "Yang, Jin-Lei and Feng, Tai-Fu and Zhang, Hai-Bin and Ning, Guo-Zhu and Yang, Xiu-Yi",
    title = "{Top quark decays with flavor violation in the B-LSSM}",
    eprint = "1806.01476",
    archivePrefix = "arXiv",
    primaryClass = "hep-ph",
    doi = "10.1140/epjc/s10052-018-5919-5",
    journal = "Eur. Phys. J. C",
    volume = "78",
    pages = "438",
    year = "2018"
}

@article{Ge:2024fdx,
    author = "Ge, Zhao-Feng and Yang, Jin-Lei",
    title = "{Top quark decays in the flavor-dependent $U(1)_X$ model}",
    doi = "10.1140/epjc/s10052-024-13507-w",
    journal = "Eur. Phys. J. C",
    volume = "84",
    pages = "1189",
    year = "2024"
}

@misc{Ma:2026G3HDMbs,
    author = "Ma, Hao-Ran and Hou, Ti-Bin and Yang, Jin-Lei and Feng, Tai-Fu",
    title = "{The Higgs boson decay $h\to bs$ in the Generational Three-Higgs-Doublet Model}",
    eprint = "2608.15162",
    archivePrefix = "arXiv",
    primaryClass = "hep-ph",
    year = "2026"
}

@misc{Hou:2026Type1B,
    author = "Hou, Ti-Bin and Li, Zheng and Peng, Yu-Ju and Yang, Jin-Lei and Ma, Hao-Ran and Feng, Tai-Fu",
    title = "{Flavor-Violating Higgs and Top Decays in the Type 1B Flavorful Two-Higgs-Doublet Model with a Twist}",
    eprint = "2609.00692",
    archivePrefix = "arXiv",
    primaryClass = "hep-ph",
    year = "2026"
}

@misc{Ma:2026G3HDMEDM,
    author = "Ma, Hao-Ran and Hou, Ti-Bin and Yang, Jin-Lei and Feng, Tai-Fu",
    title = "{Electric Dipole Moments in the CP-violating Generational Three-Higgs-Doublet Model}",
    eprint = "2609.00916",
    archivePrefix = "arXiv",
    primaryClass = "hep-ph",
    year = "2026"
}

@article{Altmannshofer:2012azMFV2HDM,
    author = "Altmannshofer, Wolfgang and Gori, Stefania and Kribs, Graham D.",
    title = "{A Minimal Flavor Violating 2HDM at the LHC}",
    eprint = "1210.2465",
    archivePrefix = "arXiv",
    primaryClass = "hep-ph",
    doi = "10.1103/PhysRevD.86.115009",
    journal = "Phys. Rev. D",
    volume = "86",
    pages = "115009",
    year = "2012"
}

@article{Altmannshofer:2016oaq,
    author = "Altmannshofer, Wolfgang and Carena, Marcela and Crivellin, Andreas",
    title = "{A $L_\mu-L_\tau$ theory of Higgs flavor violation and $(g-2)_\mu$}",
    eprint = "1604.08221",
    archivePrefix = "arXiv",
    primaryClass = "hep-ph",
    doi = "10.1103/PhysRevD.94.095026",
    journal = "Phys. Rev. D",
    volume = "94",
    pages = "095026",
    year = "2016"
}

@article{Altmannshofer:2020shb,
    author = "Altmannshofer, Wolfgang and Gori, Stefania and Hamer, Nick and Patel, Hiren H.",
    title = "{Electron EDM in the complex two-Higgs-doublet model}",
    eprint = "2009.01258",
    archivePrefix = "arXiv",
    primaryClass = "hep-ph",
    doi = "10.1103/PhysRevD.102.115042",
    journal = "Phys. Rev. D",
    volume = "102",
    pages = "115042",
    year = "2020"
}

@article{Altmannshofer:2024edmmeg,
    author = "Altmannshofer, Wolfgang and Assi, Bastian and Brod, Joachim and Hamer, Nick and Julio, Joao and Uttayarat, Patipan and Volkov, Dmitry",
    title = "{Electron EDM and $\Gamma(\mu\to e\gamma)$ in the 2HDM}",
    eprint = "2410.17313",
    archivePrefix = "arXiv",
    primaryClass = "hep-ph",
    doi = "10.1007/JHEP06(2025)156",
    journal = "JHEP",
    volume = "2025",
    number = "06",
    pages = "156",
    year = "2025"
}

@article{Aebischer:2019mlg,
    author = "Aebischer, Jason and Altmannshofer, Wolfgang and Guadagnoli, Diego and Reboud, M{\'e}ril and Stangl, Peter and Straub, David M.",
    title = "{$B$-decay discrepancies after Moriond 2019}",
    eprint = "1903.10434",
    archivePrefix = "arXiv",
    primaryClass = "hep-ph",
    doi = "10.1140/epjc/s10052-020-7817-x",
    journal = "Eur. Phys. J. C",
    volume = "80",
    pages = "252",
    year = "2020"
}

@article{Staub:2013tta,
    author = "Staub, Florian",
    title = "{SARAH 4: A tool for (not only SUSY) model builders}",
    eprint = "1309.7223",
    archivePrefix = "arXiv",
    primaryClass = "hep-ph",
    doi = "10.1016/j.cpc.2014.02.018",
    journal = "Comput. Phys. Commun.",
    volume = "185",
    pages = "1773--1790",
    year = "2014"
}

@article{Porod:2003um,
    author = "Porod, Werner",
    title = "{SPheno, a program for calculating supersymmetric spectra, SUSY particle decays and SUSY particle production at $e^+e^-$ colliders}",
    eprint = "hep-ph/0301101",
    archivePrefix = "arXiv",
    doi = "10.1016/S0010-4655(03)00222-4",
    journal = "Comput. Phys. Commun.",
    volume = "153",
    pages = "275--315",
    year = "2003"
}

@article{Porod:2011nf,
    author = "Porod, Werner and Staub, Florian",
    title = "{SPheno 3.1: Extensions including flavour, CP-phases and models beyond the MSSM}",
    eprint = "1104.1573",
    archivePrefix = "arXiv",
    primaryClass = "hep-ph",
    doi = "10.1016/j.cpc.2012.05.021",
    journal = "Comput. Phys. Commun.",
    volume = "183",
    pages = "2458--2469",
    year = "2012"
}

@article{Hahn:1998yk,
    author = "Hahn, Thomas and Perez-Victoria, Manuel",
    title = "{Automatized one-loop calculations in four and $D$ dimensions}",
    eprint = "hep-ph/9807565",
    archivePrefix = "arXiv",
    doi = "10.1016/S0010-4655(98)00173-8",
    journal = "Comput. Phys. Commun.",
    volume = "118",
    pages = "153--165",
    year = "1999"
}

@misc{UTfit:2025NP,
    author = "UTfit Collaboration",
    title = "{New Physics Fit Results: Summer 2025}",
    year = "2025",
    url = "https://www.utfit.org/foswiki/bin/view/UTfit/ResultsSummer2025NP",
    note = "Accessed 30 August 2026"
}

@article{ATLAS:2024mih,
    author = "Aad, Georges and others",
    collaboration = "ATLAS",
    title = "{Search for flavour-changing neutral-current couplings between the top quark and the Higgs boson in multi-lepton final states in 13 TeV $pp$ collisions with the ATLAS detector}",
    eprint = "2404.02123",
    archivePrefix = "arXiv",
    primaryClass = "hep-ex",
    doi = "10.1140/epjc/s10052-024-12994-1",
    journal = "Eur. Phys. J. C",
    volume = "84",
    number = "7",
    pages = "757",
    year = "2024"
}

@article{ATLAS:2023mvd,
    author = "Aad, Georges and others",
    collaboration = "ATLAS",
    title = "{Searches for lepton-flavour-violating decays of the Higgs boson into $e\tau$ and $\mu\tau$ in $\sqrt{s}=13$ TeV $pp$ collisions with the ATLAS detector}",
    eprint = "2302.05225",
    archivePrefix = "arXiv",
    primaryClass = "hep-ex",
    doi = "10.1007/JHEP07(2023)166",
    journal = "JHEP",
    volume = "2023",
    number = "07",
    pages = "166",
    year = "2023"
}

@article{CMS:2023pte,
    author = "Hayrapetyan, Aram and others",
    collaboration = "CMS",
    title = "{Search for the lepton-flavor violating decay of the Higgs boson and additional Higgs bosons in the $e\mu$ final state in proton-proton collisions at $\sqrt{s}=13$ TeV}",
    eprint = "2305.18106",
    archivePrefix = "arXiv",
    primaryClass = "hep-ex",
    doi = "10.1103/PhysRevD.108.072004",
    journal = "Phys. Rev. D",
    volume = "108",
    number = "7",
    pages = "072004",
    year = "2023"
}

@article{MEGII:2025gzr,
    author = "Afanaciev, K. and others",
    collaboration = "MEG II",
    title = "{New limit on the $\mu^+\to e^+\gamma$ decay with the MEG II experiment}",
    eprint = "2504.15711",
    archivePrefix = "arXiv",
    primaryClass = "hep-ex",
    doi = "10.1140/epjc/s10052-025-14906-3",
    journal = "Eur. Phys. J. C",
    volume = "85",
    number = "10",
    pages = "1177",
    year = "2025",
    note = "[Erratum: Eur. Phys. J. C 85, 1317 (2025)]"
}

@article{SINDRUMII:2006,
    author = "Bertl, W. H. and others",
    collaboration = "SINDRUM II",
    title = "{A Search for muon to electron conversion in muonic gold}",
    doi = "10.1140/epjc/s2006-02582-x",
    journal = "Eur. Phys. J. C",
    volume = "47",
    pages = "337--346",
    year = "2006"
}

@article{Mu2e:2023RunI,
    author = "Abdi, F. and others",
    collaboration = "Mu2e",
    title = "{Mu2e Run I Sensitivity Projections for the Neutrinoless $\mu^-\to e^-$ Conversion Search in Aluminum}",
    eprint = "2210.11380",
    archivePrefix = "arXiv",
    primaryClass = "hep-ex",
    doi = "10.3390/universe9010054",
    journal = "Universe",
    volume = "9",
    number = "1",
    pages = "54",
    year = "2023"
}

@article{COMET:2020PhaseI,
    author = "Abramishvili, R. and others",
    collaboration = "COMET",
    title = "{COMET Phase-I Technical Design Report}",
    eprint = "1812.09018",
    archivePrefix = "arXiv",
    primaryClass = "physics.ins-det",
    doi = "10.1093/ptep/ptz125",
    journal = "PTEP",
    volume = "2020",
    number = "3",
    pages = "033C01",
    year = "2020"
}

\end{document}